\documentclass[twocolumn]{aastex631}

\usepackage{lineno}
\graphicspath{{./}{}}

\begin{document}

\title{Supernova Siblings and Spectroscopic Host-Galaxy Properties}
 
\author[0000-0001-5473-6871]{Laura Salo}
\affiliation{School of Physics and Astronomy, University of Minnesota, 116 Church Street S.E., Minneapolis, MN 55455}
\affiliation{Minnesota Institute for Astrophysics, University of Minnesota, 116 Church Street S.E., Minneapolis, MN 55455}
\author[0000-0003-0031-9241]{Rui Zhou}
\affiliation{School of Physics and Astronomy, University of Minnesota, 116 Church Street S.E., Minneapolis, MN 55455}
\affiliation{Minnesota Institute for Astrophysics, University of Minnesota, 116 Church Street S.E., Minneapolis, MN 55455}
\affiliation{School of Statistics, University of Minnesota, 224 Church Street S.E., Minneapolis, MN 55455}
\author{Samuel Johnson}
\affiliation{College of Science and Engineering, University of Minnesota, 117 Pleasant St., Minneapolis, MN 55455}
\author[0000-0002-0786-7307]{Patrick Kelly}
\affiliation{School of Physics and Astronomy, University of Minnesota, 116 Church Street S.E., Minneapolis, MN 55455}
\affiliation{Minnesota Institute for Astrophysics, University of Minnesota, 116 Church Street S.E., Minneapolis, MN 55455}
\author[0000-0002-6869-6855]{Galin L. Jones}
\affiliation{School of Statistics, University of Minnesota, 224 Church Street S.E., Minneapolis, MN 55455}

\begin{abstract}
Over the past century, supernova (SN) searches have detected multiple supernovae (SNe) in hundreds of individual galaxies. So-called SN siblings discovered in the same galaxy present an opportunity to constrain the dependence of the properties of  SNe on those of their host galaxies. To investigate whether there is a connection between sibling SNe in galaxies that have hosted multiple SNe and the properties of galaxies, we have acquired integrated optical spectroscopy of 59 galaxies with multiple core-collapse SNe. %When we measure the integrated strong emission lines from the galaxies, 
Perhaps surprisingly, a strong majority of host-galaxy spectra fall within the composite region of the Baldwin-Phillips-Terlevich (BPT) diagram. We find a statistically significant difference (KS test p-value = 0.044) between the distributions of the [NII] $\lambda 6583$/H$\alpha$ of galaxies that have hosted a majority SN Ibc and those that have hosted a majority SN II, where the majority SN Ibc galaxies have, on average, higher ratios.  
The difference between the distributions of [NII] $\lambda 6583$/H$\alpha$ may arise from either increased contribution from AGN or LINERs in SN Ibc host galaxies, greater metallicity for SN Ibc host galaxies, or both. When comparing the inferred oxygen abundance and the ionization parameter for the galaxies in the Star-Forming region on the BPT diagram, we find statistically significant differences between the distributions for SN Ibc hosts and SN II hosts (p=0.008 and p=0.001, respectively), as well as SN Ib hosts and SN II hosts (p=0.030 and p=0.006, respectively). We also compare the H$\alpha$ equivalent width distributions, also integrated across the galaxies, and find no significant difference. 

\end{abstract}

\section{Introduction} \label{sec:intro}
Supernovae (SNe) are classified according to their optical spectra into several primary categories. 
%SNe whose spectra lack hydrogen are classified as Type I, while those that exhibit hydrogen features are Type II.  
Type Ia SNe, the thermonuclear explosions of white dwarf stars \citep{1960HoyleFowler}, have spectra that are characterized by a prominent blueshifted Si II absorption feature near 6150\,\AA. All other major spectroscopic types of SNe are core-collapse supernovae (CCSNe), which are the fatal explosions of stars whose initial masses exceed $\sim$8 $M_\odot$. SNe II are the explosions of massive stars that have retained their outer hydrogen envelope. SNe Ib (helium-rich) and SNe Ic (helium-poor) are so-called ``stripped-envelope'' SNe, since their progenitors have instead lost their outer hydrogen envelope to winds or a binary companion.  SNe IIb are a transitional class whose early optical spectra exhibit hydrogen features that subsequently fade, which indicates that  their progenitors retain only a small fraction of their hydrogen envelope at the time of explosion. As SN IIb evolve, helium lines become evident. 

Gravitational-wave detectors, including the Laser Interferometric Gravitational-Wave Observatory (LIGO), have now detected almost one hundred mergers of compact objects \citep[e.g.,][]{LIGO}. Many of the binary systems of neutron stars and black holes detected during their inspiral must have originated from close massive stellar binaries whose member stars have exploded as CCSNe.
Improved understanding of the formation of binary systems of neutron stars and black holes, including how their rates of formation depend on their host galaxies' properties, should increase our ability to interpret the growing population of gravitational wave events.

The Milky Way is expected to  host approximately one SN each century on average \citep{1991SNRates}. More highly star-forming galaxies, however, can be much more prolific. NGC 6946, for example, has hosted ten detected SNe: SN\,1917A (SN\,II), SN\,1939C (SN\,I), SN\,1948B (SN\,IIP), SN\,1968D (SN\,II), SN\,1969P (unknown), SN\,1980K (SN\,IIL), SN\,2002hh (SN\,IIP), SN\,2004et (SN II), SN\,2008S (SN\,IIn-pec), and SN\,2017eaw (SN\,IIP). %, and NGC6946-BH1 \citep{NGC6946}. 

\cite{AndersonSoto} (hereafter A\&S) examined the relative fractions of the spectroscopic SN types found in galaxies that hosted multiple SNe through 2012. They found that, if a galaxy had hosted more than one recorded SN, subsequent SNe may be more likely by about 10\% to be of the same spectroscopic type as the first. They also found evidence with modest statistical significance ($\sim$1 $\sigma$) that SNe Ibc account for an elevated fraction of SNe in galaxies that have hosted multiple SNe. Zhou et al. (submitted) analyzed SN discoveries that span an additional decade through 2022, and applied importance sampling to mitigate biases that arise from differences among the luminosity functions of the SN types.  Zhou et al. obtains 1.4$\sigma$ evidence that SNe Ib are overrepresented in galaxies that have hosted multiple SNe and 2.0$\sigma$ evidence that SNe Ib are more likely to occur consecutively.

 \citet{2020NGC2770} examined the four SNe discovered in NGC 2770 (SN 1999eh, SN 2007uy, SN 2008D, SN 2015bh), which include three SNe Ib.  They found that the SN rate in this galaxy is significantly higher than expected given its star formation rate (SFR) measured from fitting of its broadband spectral energy distribution. The SN rate is, however, consistent with its dust-corrected H$\alpha$ luminosity, which can be expected to reflect the galaxy's recent star-formation history over the past approximately 10 Myrs.

The gas phase oxygen abundance of a galaxy provides an easily measured indicator of the metallicity of its massive stars, given their brief lifetimes. Moreover, the opacity of a stellar atmosphere depends upon its metallicity; and therefore metallicity will affect a star's evolution through both the rate of wind-driven mass loss and, through the radius of the photosphere, potential interaction with binary companions.

Empirical efforts have examined the relationship between the spectroscopic type of core-collapse SNe and stellar metallicity. These consist of studies that have measured the properties of the entire host galaxies (or the nuclear region), and those that have analysed the local environment of the SN.
\citet{Prieto} analysed 2$''$-diameter fiber spectra taken by the Sloan Digital Sky Survey (SDSS) where the fiber was, in most cases, positioned on the nucleus of the host galaxy. They reported evidence (Kolmogorov-Smirnov (KS) p-value = 0.05) that SNe Ibc occur in galaxies with higher oxygen abundances than in galaxies that hosted SNe II. \citet{Arcavi} found evidence for a higher fraction of SNe Ib, SNe Ic-BL, and SNe IIb in dwarf galaxies, among discoveries by the Palomar Transient Factory (PTF). In their study, SNe Ic, in contrast, were found only in massive galaxies.  
%They attribute this difference to the low metallicity in the dwarf galaxies.  
Their interpretation was that metallicity should reduce wind-driven mass loss, which may in turn cause massive stars that otherwise would have exploded as SNe Ic to retain H and He and explode as SNe Ib or SNe IIb.  
\citet{patspaper} also analyzed SDSS fiber spectroscopy of SN host galaxies, and examined whether fibers were positioned at similar distances from the galaxy nucleus as the SN location. %This is relevant particularly for CCSNe since their progenitors are relatively short-lived and therefore explode near to where they were born.  
\citet{patspaper} found evidence that SNe Ic-BL originate in lower-metallicity host galaxy environments than SNe Ic, while the host-galaxy environments of SNe IIb had lower oxygen abundances than those of SNe Ib. 

\citet{Modjaz} obtained long-slit spectroscopy of environments and found evidence that SNe Ic exploded in environments that had higher oxygen abundances ($\sim$0.2\,dex) than in those of SNe Ib. 
%, central spectra, and a metallicity gradient to find this result. 
\citet{Anderson2010} and \citet{Leloudas}, however, find no significant difference between the metallicity at SNe Ib and SNe Ic events.  
\citet{GalbanyI} and \citet{GalbanyII} use integral field spectroscopy to measure gas and stellar population properties at the location of the SNe. 
%\citet{Stanishev} found no correlation between the properties of the host galaxy and the SN spectroscopic type. Pat says -- the Stanishev paper considered only SNe Ia
\citet{GalbanyI} found that SNe Ibc and SNe IIb tend to explode closer to regions with high SFR, and \citet{GalbanyII} found no significant difference in the average gas-phase oxygen abundance at the locations of SNe Ia, SNe Ibc, and SNe II.  

We examine integrated optical spectroscopy population of galaxies that have hosted multiple core-collapse SNe. %We  these galaxies to establish whether there is a connection between the principal spectroscopic types of the SNe in these galaxies and the host-galaxy properties. 
We measure the strengths of the strong nebular lines in the optical, and constrain the locations of the galaxy spectra on the Balwin-Phillips-Terlevich (BPT; \citealt{baldwinphillipsterlevich81}) diagram.  The [NII] $\lambda 6583$/H$\alpha$ ratio has a strong dependence both on metallicity and the ionizing spectrum, and therefore reflects the contribution of Active Galactic Nuclei (AGN) or low-ionization nuclear emission-line regions (LINERs).  
In Section~\ref{sec:data}, we present our optical spectroscopic observations of galaxies that have hosted multiple SNe.
The methods that we employ in our reduction and analysis of the data are discussed in Section~\ref{sec:methods}.
We then discuss, in Section~\ref{sec:results}, the signal-to-noise cuts we apply, the distribution of stellar masses of the galaxies in our sample, as well as the results of our analysis. In Section~\ref{sec:conclusions}, we discuss our conclusions.

\begin{figure*}
  \includegraphics[width=\linewidth]{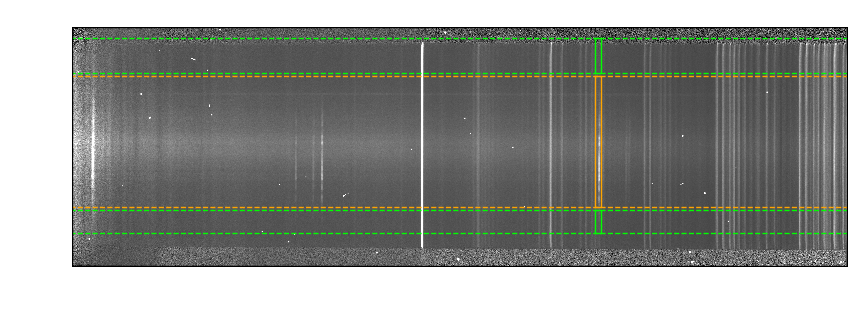}
  \caption{Spectrum of NGC 4490 acquired with the Boller \& Chivens spectrograph mounted on the Bok telescope. During each observation, we panned the slit repeatedly across the target galaxy. The orange dashed lines delineate the spectral extraction region, and the green dashed lines indicate the background region. As shown by the orange (extraction) and green (background) solid boxes, we designed the extraction region to bracket the H$\alpha$ emission detected along the spatial axis of the slit.}
  \label{ds9_regions}
\end{figure*}

\begin{figure*}
  \includegraphics[width=\linewidth]{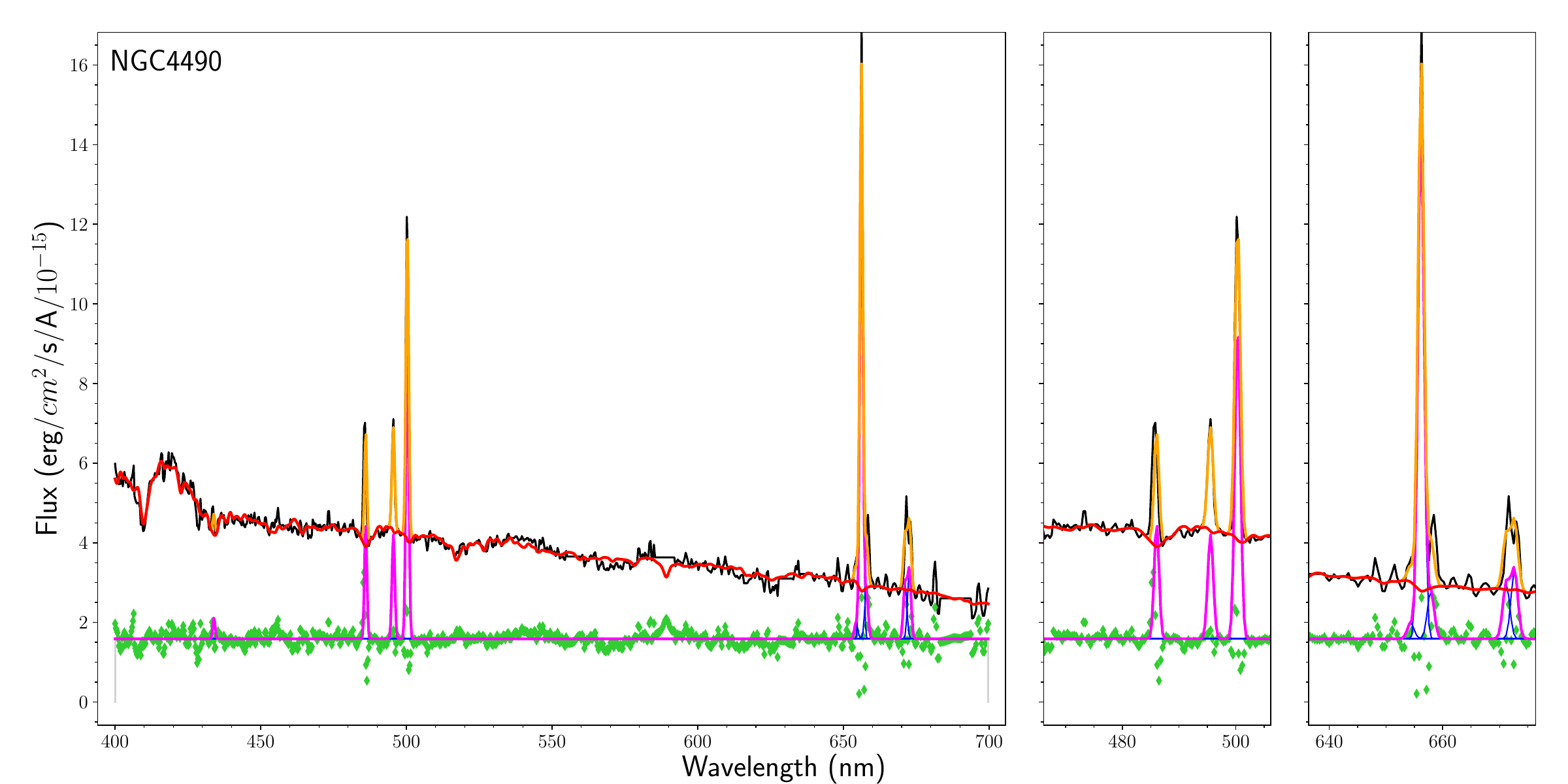}
  \caption{Best-fitting {\tt pPXF} model of both nebular emission and stellar continuum for our integrated spectrum of galaxies NGC 4490, NGC 2276, and NGC 4666. The left plot shows the full spectrum. The middle plot show a zoom in around H$\beta$, and the right plot shows a zoom in around H$\alpha$. The black line is the flux density of the observed spectrum, the red line is the pPXF model for the stellar component, and the orange line is the model of the nebular emission lines and the stellar continuum. The magenta line shows the nebular line model alone, and the blue line shows the individual lines the contribute to the nebular emission model. The green dots are the fit residuals.}%In the upper panel, the measured spectrum is plotted in black, the model stellar continuum in red, the model emission line strengths in yellow, and the residuals in green.  The bottom panel plots the posterior distribution of the metallicity ([M/H]) and age 3 of the stellar populations that contribute to the best fit to the stellar continuum. }
  \label{ppxf_fit}
\end{figure*}

\begin{figure*}
  \includegraphics[width=\linewidth]{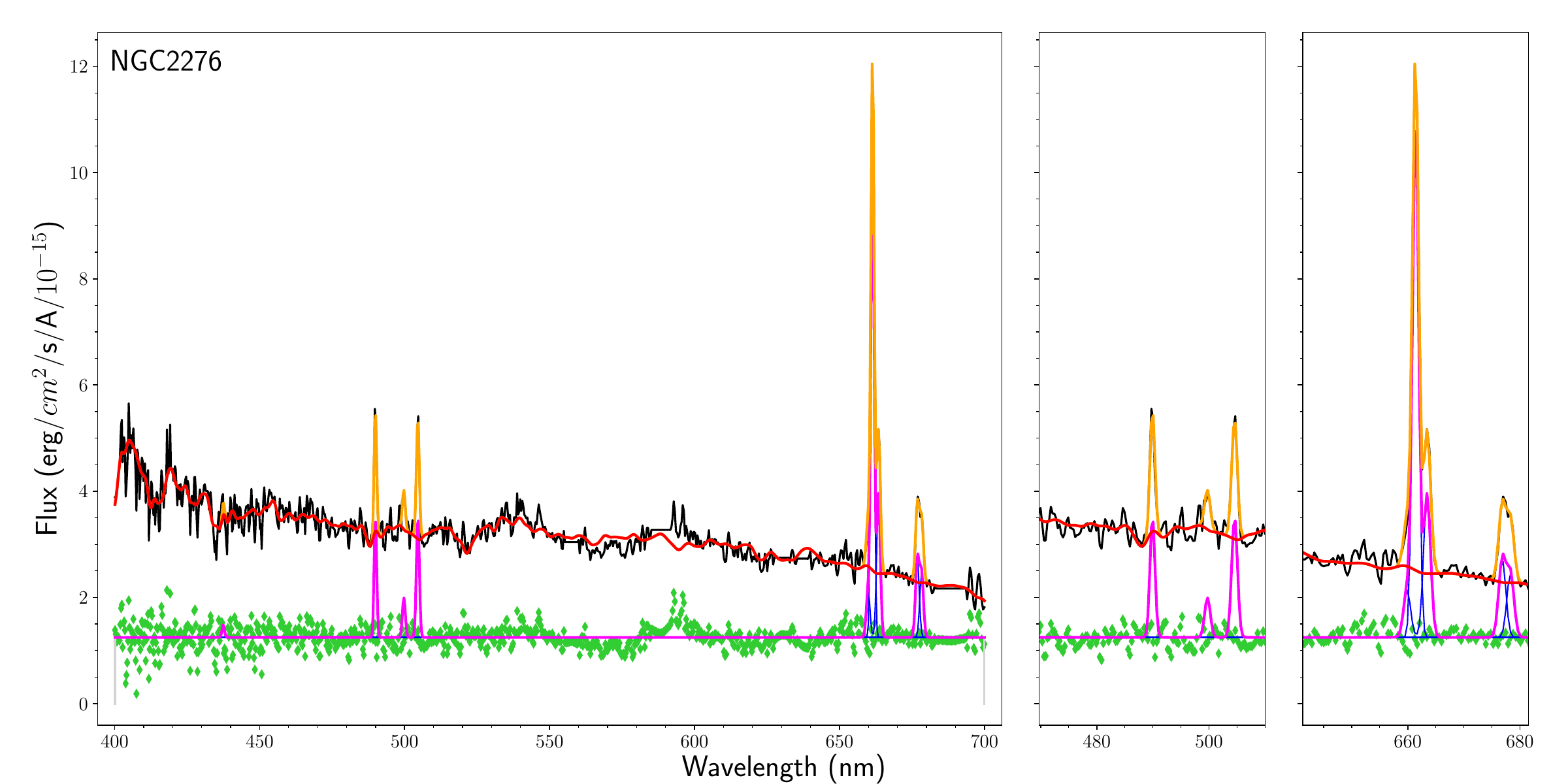}
  \includegraphics[width=\linewidth]{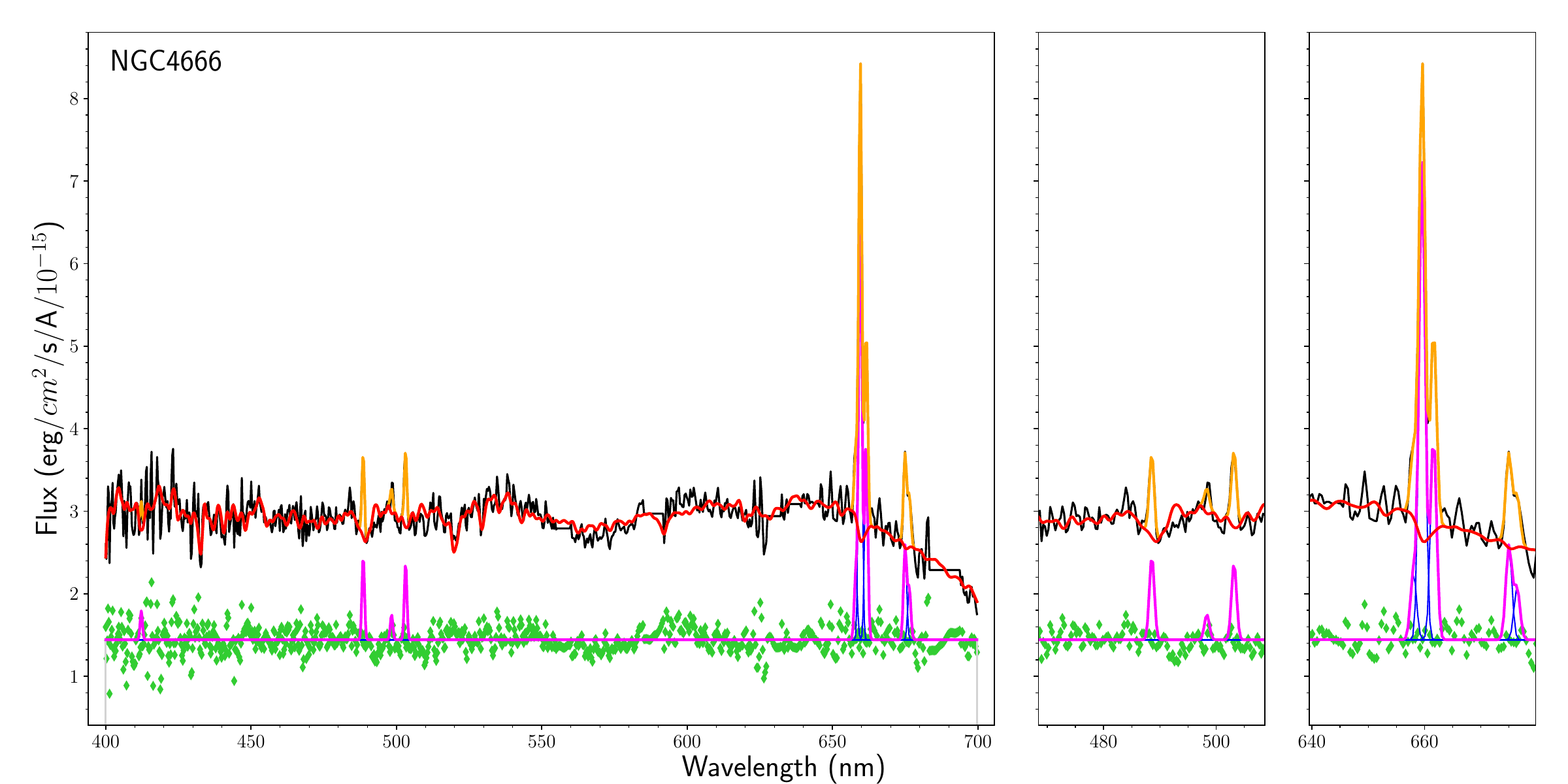}
    \caption{Continuation of Figure~\ref{ppxf_fit}.}
\end{figure*}

\begin{figure}
  \includegraphics[width=\linewidth]{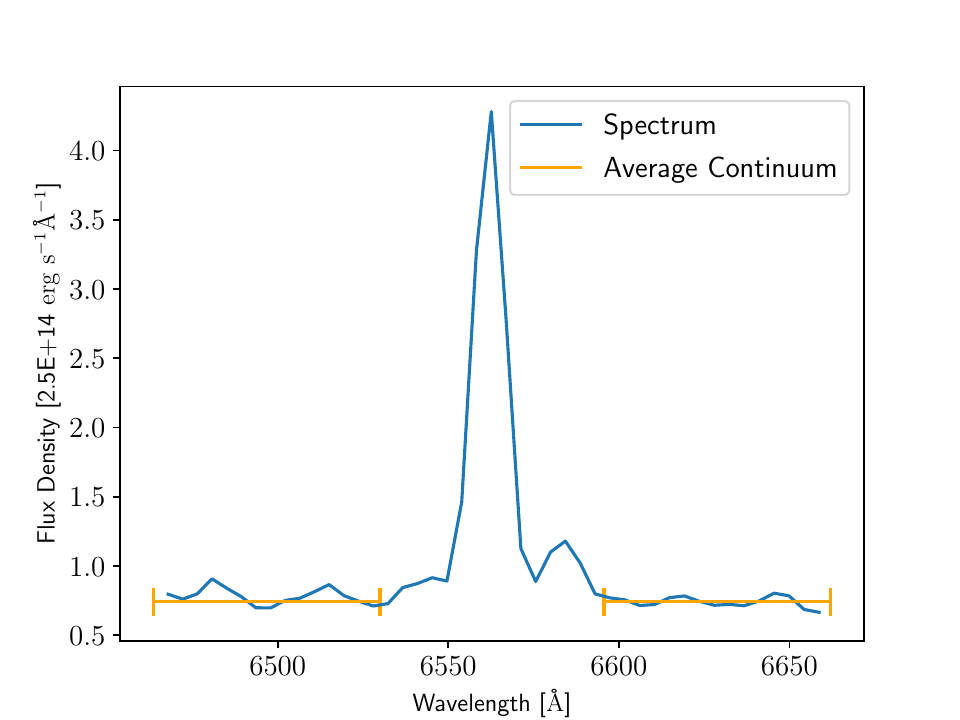}
  \includegraphics[width=\linewidth]{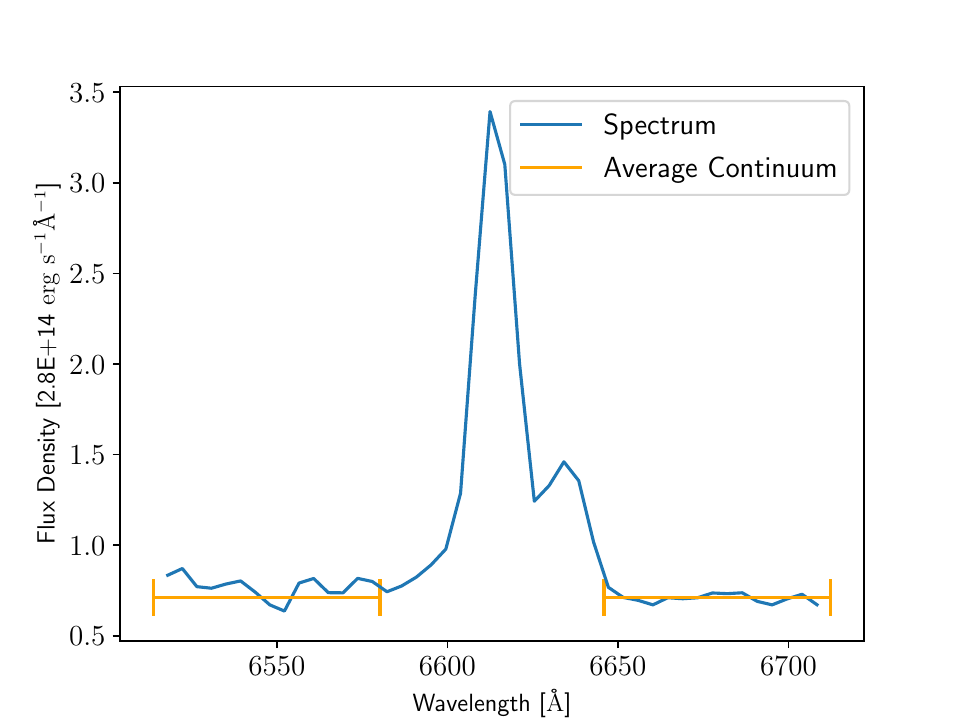}
  \includegraphics[width=\linewidth]{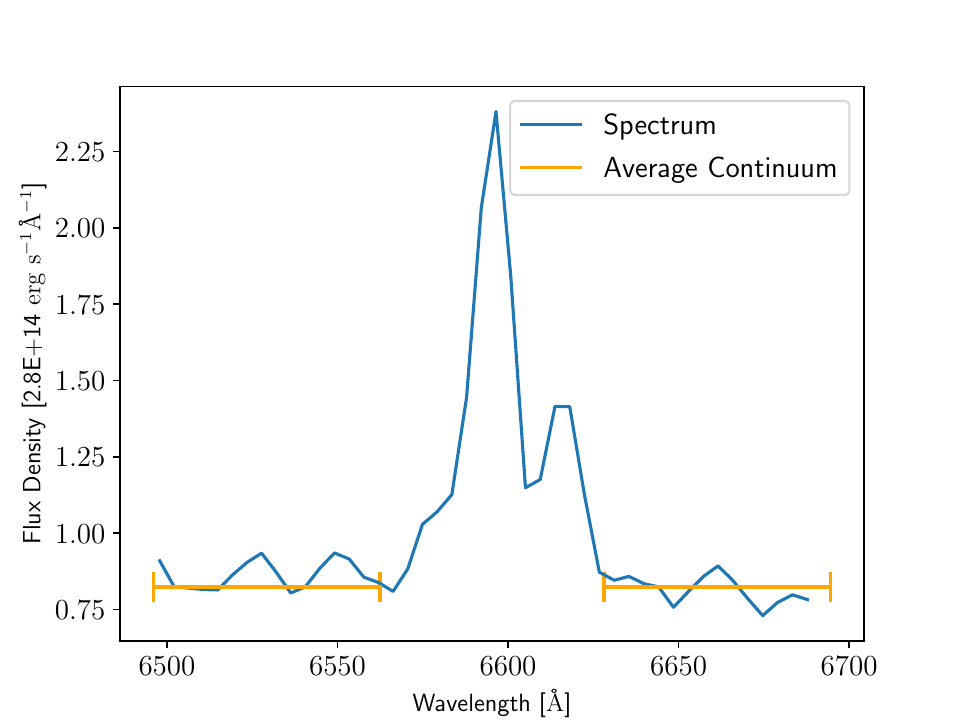}
  \caption{ Spectra of the H$\alpha$ and [NII]$\lambda\lambda$ 6548,6583 \AA\ emission lines for NGC 4490 (top), NGC 2276 (middle), and NGC 4666 (bottom). The blue line shows the integrated spectrum, and the orange lines show our measurement of the average of the stellar continuum adjacent to the H$\alpha$ and [NII]$\lambda\lambda$ 6548,6583 \AA\ doublet.  We measure the  continuum by computing the average of the flux density within the pair of intervals indicated by the orange lines (6463--6531 \AA\ and 6595--6663 \AA\ in the rest frame).}
  \label{h-alpha-NGC4490}
\end{figure}

\section{Data} \label{sec:data}

We identified galaxies that had hosted SNe from the list of SNe in the Open Supernova Catalog \citep{OpenSNeCat} through 11 January 2022. When selecting targets, we only chose galaxies that had hosted two or more SNe, and we assigned higher priority to galaxies that had hosted greater numbers of SNe. The smallest redshift of a host galaxy in our sample is z=0.0008 and the largest is z=0.0296.  

Of our list of potential galaxies to observe, we acquired integrated spectra of 55 galaxies that have hosted multiple SNe using the Boller \& Chivens Spectrograph mounted on the 2.3\,m Bok telescope on Kitt Peak, Arizona. We used a 2.5$''$ long-slit mask and the 300 lines mm$^{-1}$ grating with a central wavelength setting of 6693 \AA. To obtain an integrated spectrum of the galaxy, we repeatedly panned the slit continuously across the full extent of the host galaxy during each observation. The endpoints of each scan were determined from imaging of each target, and we verified that the slit and the scan both extended across the entire galaxy. Each full scan was completed every 60 to 120 seconds, depending upon the target, during the integrations. 
%, and ensured that light from all parts of the galaxy was included in the data. 
The galaxies that we observed are listed in Tab.~\ref{hosts_of_Ibs}, Tab.~\ref{hosts_of_IIs}, and Tab.~\ref{hosts_of_Ics} in the Appendix, and are grouped by spectroscopic type of SNe that account for a majority of SNe discovered in each galaxy. These tables also contain the line strengths observed from these galaxies. 

To complement our observations, we identified archival IFU spectroscopy of 4 galaxies (IC 701, NGC 5480, NGC 5630, and NGC 5888) that have hosted multiple SNe.  The galaxies are included in Data Release 3 of the Calar Alto Legacy Integral Field Area (CALIFA) survey, which was carried out with the 3.5\,m Calar Alto telescope \citep{2012Sanchez_Califa,2014Walcher_Califa,2016Sanchez_Califa}. The CALIFA galaxies were observed using PMAS \citep{2005Roth} in the PPak configuration \citep{2004Verheijen, 2006Keltz}. The PPak IFU has a field of view of 74$''$ x 64$''$ and contains 382 fibers. 331 of these fibers are science fibers. They are arranged in a hexagonal grid with each fiber covering a 2.7$''$ diameter circular aperture on the sky. An additional 36 fibers measure sky background.

Reduced datacubes were available for each galaxy target \citep{2016Sanchez_Califa}. Each datacube was prepared using semi-automatic python pipeline for the reduction of fiber-fed IFS data \citep{2006Sanchez}. In the reduction, the four different FITS files from the amplifiers of the detectors were combined into a single FITS file. This file was bias-subtracted, and cosmic rays were removed or masked (see \citet{Cosmicrays}). Next, a wavelength solution was applied and a stray-light map was subtracted. An extraction algorithm from \citet{extraction} was used to extract the spectra. Flux-calibration was done using the method described in \citet{flux_calibration}, and a correction for Galactic extinction was applied \citep{galactic_extinction, 1989Cardelli}. More details on the data reduction can be found in \citet{2016Sanchez_Califa}. 

For consistency with our analysis of the Bok long-slit spectroscopy where the slit moves across regions with low host-galaxy surface brightness, we do not apply a S/N cut on the spaxels. Likewise, for the purpose of consistency, we do not apply a cut to individual spaxels on the basis of their location on the BPT diagram. On average, the flux density of the CALIFA IFU spectra is a factor of 10--100 greater than that of the Bok long-slit spectra we acquired, since the long-slit admits only a small fraction of the galaxy's light during each scan.

\citet{GalbanyII} has led a CALIFA extension program focusing on environments of CCSN. 
Two of the galaxy spectra we have used from CALIFA, NGC 5480 and NGC 5630, are also used analyzed by \citet{GalbanyII}.  They measure oxygen abundances ($12 + \log({\rm O/H})$) in the \citet{marino13} calibration of the O3N2 index of 8.55 dex and 8.32 dex for NGC 5480 and NGC 5630, respectively. When we use the same \citet{marino13} calibration of the O3N2 index (elsewhere we adopt the PP04 calibration) together with our integrated emission-line measurements, we obtain, in excellent agreement, 8.53 dex and 8.33 dex for NGC 5480 and NGC 5630, respectively. 

\subsection{Reduction of Bok Spectroscopy}

We reduce the data we acquired with the Bok telescope using the {\tt Image Reduction and Analysis Facility} \citep[{\tt IRAF};][]{1986_IRAF}. 
 We subtract bias exposures from the data, corrected pixel-to-pixel sensitivity variations using quartz-lamp exposures, and, finally, remove large-scale gradients using flat-field exposures. 
Fig.~\ref{ds9_regions} shows, as an example, the extraction and background regions that we used to extract the integrated spectrum of NGC 4490.

\section{Methods} 
\label{sec:methods}
We measure the fluxes of the strong nebular lines, including H$\alpha$, H$\beta$, [NII] $\lambda 6583$, [OII] $\lambda $3727, 3729, and [O III] $\lambda 5007$. 
%, and  to find the ratio [NII] $\lambda 6583$/H$\alpha$. 
We first use the ratios [O III] $\lambda 5007$/H$\beta$ and [NII] $\lambda 6583$/H H$\beta$ to constrain the locations of the galaxies on the Baldwin-Phillips-Terlevich (BPT; \citealt{baldwinphillipsterlevich81}) diagram. 
Where these ratios are consistent with the star-forming sequence, we place constraints on the oxygen abundance, $12 + \log$(O/H), the gas pressure, and the ionization parameter, $q_{\rm ion}$, which describes the ratio of mean ionizing photon flux to mean atomic density. 

\subsection{Measurement of Emission-Line Strengths}
After extracting the Bok spectra, we correct for foreground Galactic extinction, using the \citet{1989Cardelli} extinction law, with $R_V = 3.1$. We do not apply this correction to the four galaxies with CALIFA IFU observations, as those data already have Galactic extinction removed.
%We next measured the emission lines and used the ratios of their strengths to constrain the properties of the galaxies' nebular gas.  
We model each spectrum as the sum of nebular emission lines and an underlying stellar continuum using the {\tt Penalized PiXel-Fitting} ({\tt pPXF}) code  \citep{2004Cappellari_ppxf,2017Cappellari_ppxf} and  the MILES stellar spectral library to model the continuum \citep{2006MILES,2011MILES}. The fluxes of the emission lines are obtained from these pPXF fits.

Fig.~\ref{ppxf_fit} shows three examples (NGC 4490, NGC 2276, and NGC 4666) of the {\tt pPXF} model of our integrated spectra. The black line is the flux density of the observed spectrum, the red line is the {\tt pPXF} model of the stellar component, the magenta line shows the best-fitting model for the nebular emission lines, the orange line is the sum of the model stellar continuum and nebular emission, the blue line shows the individual lines comprising the emission-line spectrum, and the green dots show the fit residuals. Our {\tt pPXF} fits to the entire set of host-galaxy spectra can be seen in Fig.~\ref{all_ppxf_fits} in the Appendix.

\subsection{Inferences from Nebular Emission-Line Strengths}
After correcting the emission-line fluxes for Galactic extinction, we %find the 
compute ratios of [NII] $\lambda 6583$/H$\alpha$, as well as [O III] $\lambda 5007$/H$\beta$. The respective pairs of lines are close to each other in wavelength, and therefore experience similar extinction due to dust. These ratios can be used to assess whether the nebular emission is consistent with ionization by massive stars or instead AGN / LINER. The flux ratio [NII] $\lambda 6583$/H$\alpha$ also varies monotonically with the oxygen abundance. For instance, the \citet{PP04_paper} calibration of the N2 ([NII] $\lambda 6583$/H$\alpha$) diagnostic of the oxygen abundance (hereafter {\tt PP04 N2}) is $12 + \log$(O/H)$ = 8.90 + 0.57 * N2$ to calculate the oxygen abundance, $12 + \log$(O/H). \citet{PP04_paper} also provides a calibration of the O3N2 diagnostic of oxygen abundance, where O3N2$ \equiv $log$\{($[OIII]$ \lambda 5007/$H$\beta)/$[NII]$ \lambda 6583/$H$\alpha\}^2$ and $12 + \log$(O/H)$ = 8.73 - 0.32 * $O3N2 (hereafter {\tt PP04 O3N2}).

We use the {\tt NebulaBayes} software package \citep{2018NebulaBayes} to constrain the properties of nebular gas from the fluxes of the strong emission line.
Before employing {\tt NebulaBayes}, we correct the emission-line fluxes we measure for host-galaxy extinction using the Balmer decrement under the assumption of Case B recombination, using the \citet{1989Cardelli} extinction law with $R_V=3.1$.
 %\textbf{Different from the foreground Galactic extinction we corrected the entire spectrum for earlier, this correction accounts for the extinction of the emission lines from the host galaxy and thus is applied only to the emission lines.} 
We infer values of the oxygen abundance, $12 + \log$(O/H), the ionization parameter, $\log$(U), and the gas pressure, $\log$(P/k). For galaxies that fall within the star-forming main sequence, {\tt NebulaBayes} uses the \citet{Sutherland&Dopita} photoionization grid. For galaxies in the Composite and AGN regions, we also apply a photoionization model appropriate for narrow line regions (NLRs) using {\tt NebulaBayes}. We use the OXAF ionizing spectrum \citep{NBNLR} when modeling ionizing radiation from NLRs.

Finally, the equivalent widths of the Balmer emission lines are indicators of the age of the massive stellar population. %is the interval along the continuum one must integrate to reach the full strength of the H$\alpha$ line.  
To constrain the equivalent width of H$\alpha$, we estimated the average continuum value adjacent to the line.  Fig.~\ref{h-alpha-NGC4490} shows the H$\alpha$ and [NII]$\lambda\lambda$ 6548,6583 \AA\ emission lines and adjacent stellar continuum. We compute an average continuum strength by calculating the mean of the flux density within a pair of rest-frame wavelength intervals (6463--6531 \AA\ and 6595--6663 \AA) that bracket the [NII] doublet. The equivalent width of H$\alpha$ is calculated by dividing the integrated flux of H$\alpha$ emission estimated using the {\tt pPXF} code by the average flux density of the adjacent stellar continuum. We note that our estimate of the H$\alpha$ flux accounts for absorption by the stellar continuum, since pPXF simultaneously models the contributions from nebular emission and the underlying stellar continuum.

\begin{figure*}
  \includegraphics[width=0.5\linewidth]{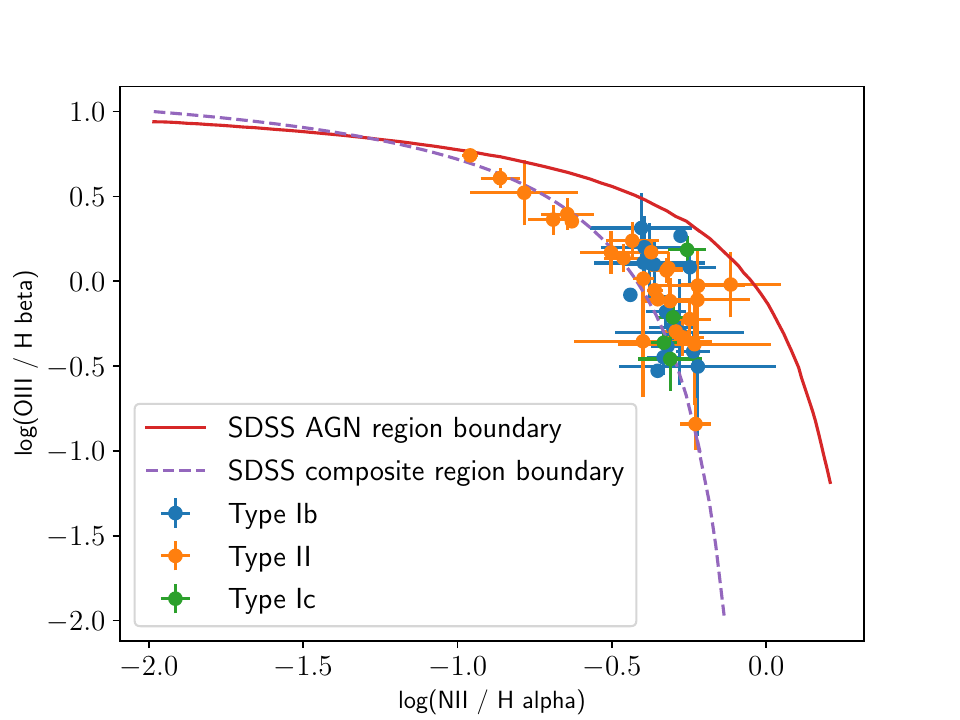}
  \includegraphics[width=0.5\linewidth]{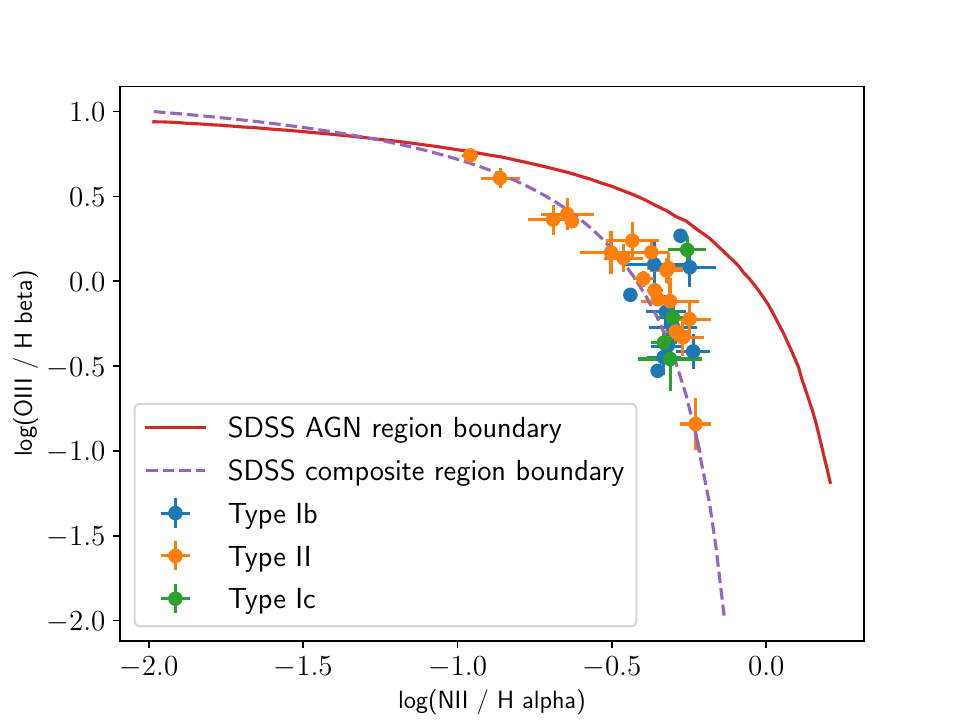}
  \begin{center}
  \includegraphics[width=0.5\linewidth]
  {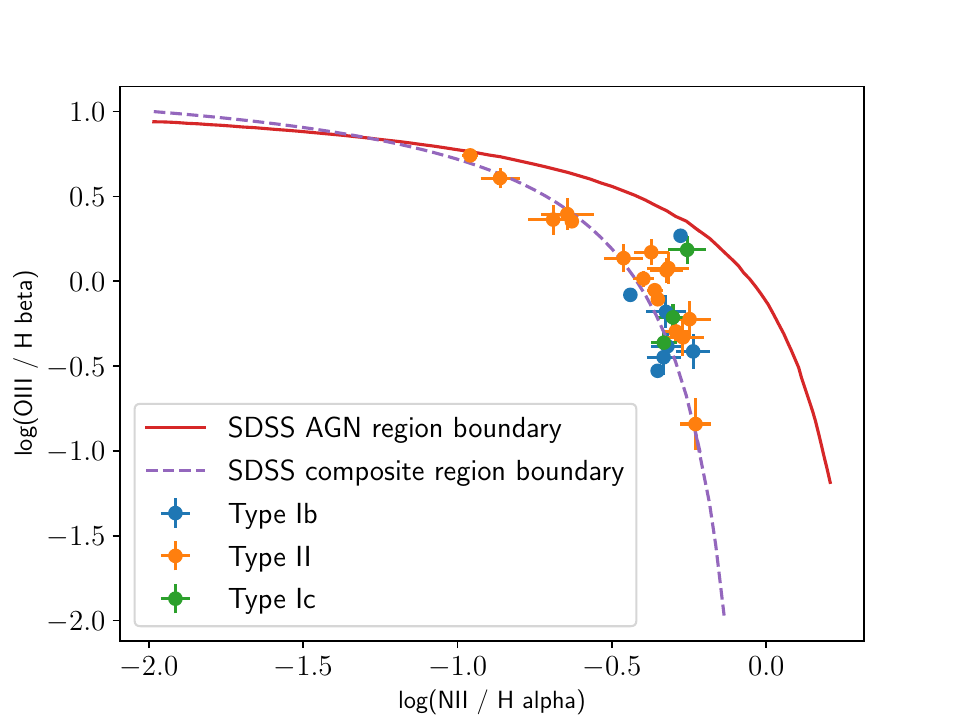}
  \end{center}
  \caption{Baldwin-Phillips-Terlevich (BPT; \citealt{baldwinphillipsterlevich81}) diagram of the galaxies in our sample. The solid magenta line \citep{SDSS_solid_line} separates the AGN region (the upper-most portion). The dashed purple line \citep{SDSS_dashed_line} separates the composite region (middle portion) and the star-forming region (lower-most portion). In the upper left plot, we include only galaxies that have H$\beta$ signal-to-noise (S/N) ratio exceeding 1.5. There are 46 galaxies with S/N $>$ 1.5, 13 of which fall in the Star-Forming region and 33 of which fall in the Composite region. The upper right plot shows the 36 galaxies with S/N $>$ 3 (11 in Star-Forming region; 25 in Composite region). The lower plot shows the 29 galaxies with S/N $>$ 5 (9 in Star-Forming region; 20 in Composite region). }
  \label{BPT_divid}
\end{figure*}

\section{Results}
\label{sec:results}
%\section{Sample} \label{sec:sample}
%\textbf{In this section, we detail the ways we looked at our data and show what results we found. While most of our results showed no difference in hosts of SN Ibc and hosts of SN II, we did find a distinction in the [NII] $\lambda 6583$/H$\alpha$ ratio, which indicates either a higher metallicity or greater AGN or LINER contributions to the ionizing flux.}

\subsection{Signal-to-Noise and AGN/LINER Contamination Cuts}
In order to be able to determine the location of galaxies on the BPT diagram, we eliminate galaxies for which our measurement of H$\beta$ has a signal-to-noise ratio (S/N) of less than 1.5, which eliminates 13 galaxies from our sample of 59, leaving a sample of 46 galaxies.
In Fig.~\ref{BPT_divid}, we plot the BPT diagram for the 46 host galaxies in our sample with H$\beta$ S/N of at least 1.5. 

\begin{figure}
    \centering
    \includegraphics[width=\linewidth]{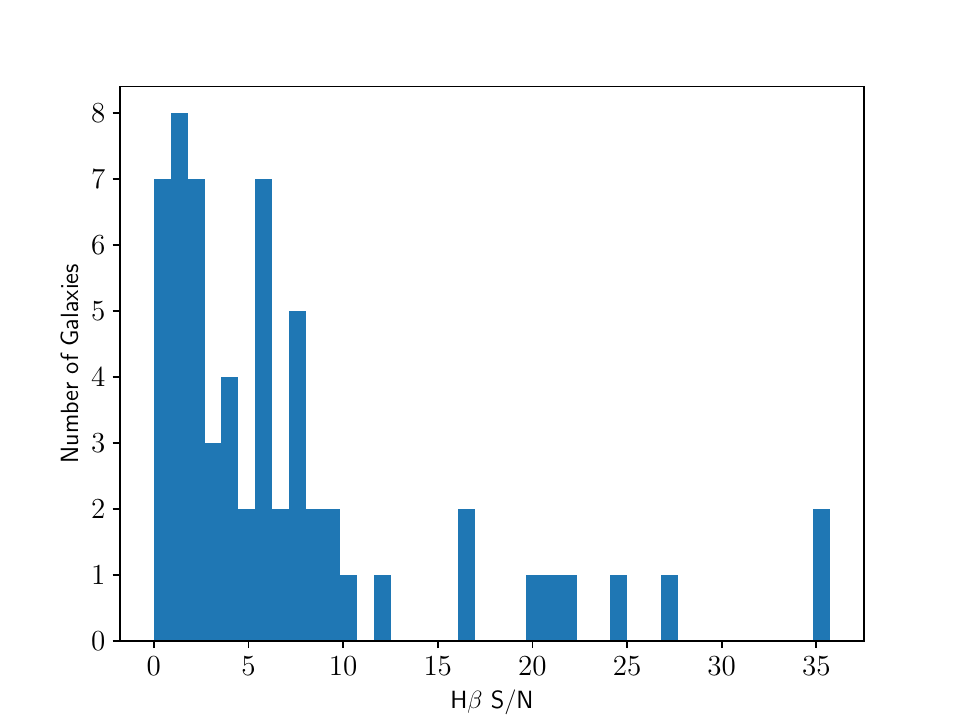}
    \caption{Histogram of the S/N ratio of the H$\beta$ emission-line flux for all the host galaxies in our sample. %Almost all galaxies, except for 11 galaxies ($19\%$ of the sample), have S/N $< 10$.
    }
    \label{SNhist}
\end{figure}

Between the solid \citet{SDSS_solid_line} and dashed \citet{SDSS_dashed_line} lines in Fig.~\ref{BPT_divid} is the Composite region, where the ionizing spectrum may include contribution from AGN. Below the dashed line is the Star-Forming region. We also implement a higher S/N cut of 3 and 5 to see if these more stringent cuts affect the proportion of galaxies found in the Composite region. There are 36 galaxies with S/N greater than 3 and 29 galaxies with S/N greater than 5. We can see from the plots that, even with higher S/N cuts, a majority of galaxies still fall in the Composite region. To be exact, for the 36 galaxies with S/N greater than 3, 11 galaxies fall in the Star-Forming region and 25 galaxies fall in the Composite region. For the 29 galaxies with S/N greater than 5, 9 galaxies fall in the Star-Forming region and 20 galaxies fall in the Composite region. We also present a histogram of H$\beta$ S/N (see Fig.~\ref{SNhist}).

We refer to the galaxies in the composite region as {\tt Sample Composite} (hereafter {\tt Sample Comp.}) and the galaxies in the star-forming region as {\tt Sample Star-Forming} (hereafter {\tt Sample SF}). {\tt Sample Comp.} consists of 13 SN Ib, 18 SN II, and 2 SN Ic, for a total of 33 galaxies. {\tt Sample SF} consists of 4 SN Ib, 7 SN II, and 2 SN Ic, for a total of 13 galaxies.  Tab.~\ref{BPT_ratios} lists the ratios of nebular spectra classified as AGN or composite to star forming galaxies ({\tt Sample SF}) for galaxies that have hosted a majority SN Ib, SN Ic, or SN II. Tab.~\ref{sample_galaxies} in the Appendix lists all galaxies in our sample and specifies whether we classify their spectra as star-forming or composite. 

\begin{table}[hbt!]
\centering
\begin{tabular}{ |c|c|c|c| } 
\hline
SN Hosts & Composite & Star-Forming & Ratio\\
\hline
SN Ib & 17.0 $\pm$ 4.1 & 4.0 $\pm$ 2.0 & 4.3 $\pm$ 2.4 \\
\hline
SN Ic & 4.0 $\pm$ 2.0 & 2.0 $\pm$ 1.4 & 2.0 $\pm$ 1.7\\
\hline
SN II & 25.0 $\pm$ 5.0 & 7.0 $\pm$ 2.6 & 3.6 $\pm$ 1.5 \\
\hline
\end{tabular}
\caption{The hosts of SNe and the corresponding number of galaxies falling in the AGN or Composite region and in the star-forming region. These numbers are from after the S/N cut on H$\beta$. }
\label{BPT_ratios}
\end{table}

%In some of the host-galaxy spectra, the [O III] line or the [N II] line was only faintly detected at less than one sigma, giving these galaxies very large errors on the BPT diagram.  We determined that these galaxies lie within the star-forming region of the BPT diagram, but they are not well localised.  

\begin{figure}[h!]
    \includegraphics[width=\linewidth]{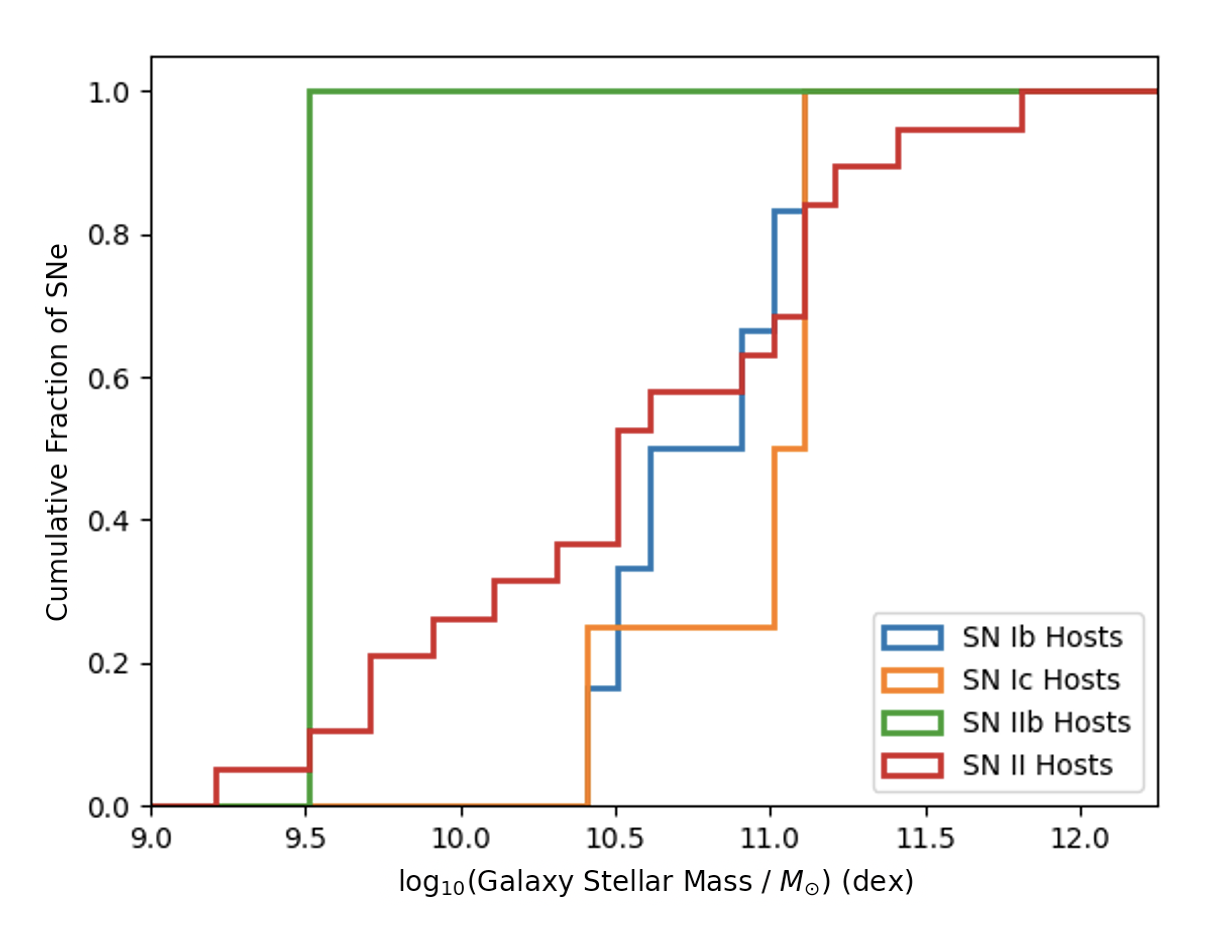}
  \caption{Cumulative distributions of stellar mass for galaxies that hosted majority SN Ib, Ic, IIb, and II in our sample.  The KS test does not yield significance differences among the distributions.}
  \label{masses_multi_types}
\end{figure}

\begin{figure*}
\centering
\includegraphics[width=\linewidth]{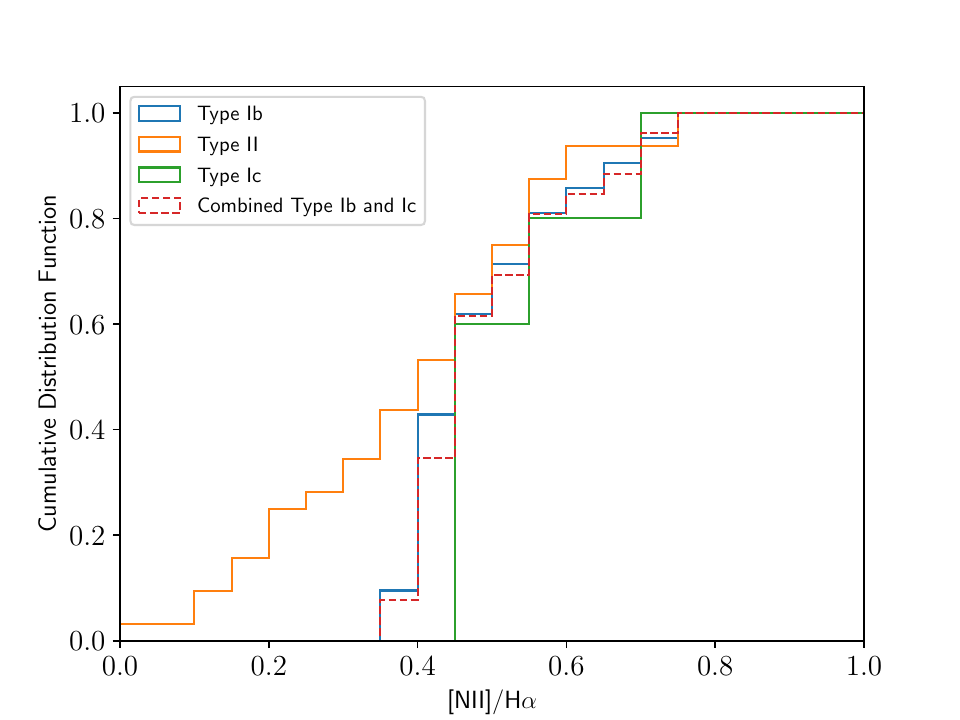}
\caption{This plot shows the distributions of the ratio [NII] $\lambda 6583$/H$\alpha$ from the entire sample. We find a statistically significant difference between the distributions for the host galaxies of SN Ibc (red dashed line) and those of SN II (orange line), with a KS test p-value of 0.044. }
\label{cdf_niihalpha}
\end{figure*}

\subsection{Host-Galaxy Stellar Mass Distributions}
We next examine whether the galaxies for which we obtained spectroscopy, which have hosted multiple SNe, have distributions of stellar masses that are representative of those of SNe discovered at low redshift.
We identify galaxies in our sample with stellar masses measured by \citet{patspaper}. We compare the distributions of stellar masses of the multiple-SN host galaxies that we have followed-up to the full sample of SN host galaxies studied by \citet{patspaper} for a sample of $z < 0.04$ SNe reported through 2012.  As shown in Fig.~\ref{masses_all}, the host-galaxy masses of the sample we have observed %are more massive by $\sim$0.5 dex, on average. 
have a median value of $10^{11.0}$ $M_{\odot}$, while the host-galaxy masses of the entire sample used by \citet{patspaper} have a median value of $10^{10.3}$ $M_{\odot}$. The median stellar masses of the multiple-SN host galaxies we observed are more massive than the median of the \citet{patspaper} by 0.65 dex, or a factor of $\sim$4.4. The difference may arise in part from the fact that more massive galaxies have higher star-formation rates, and therefore greater rates of core-collapse SNe.

We also plot the host-galaxy stellar masses of the SN host galaxies in our sample with masses measured by \citet{patspaper} in Fig.~\ref{masses_multi_types}, and compute the KS statistics for pairs of samples of the spectroscopic types of SNe in Tab.~\ref{stellar_mass_distributions}, which do not yield statistically significant differences.

\begin{table}[h!]
\begin{center}
\begin{tabular}{|c|c|c|} 
\tableline
\multicolumn{2}{|c|}{SN Samples} & $p$-value\\
\tableline
SN II & SN Ib & 0.53 \\ 
SN II & SN Ic & 0.75 \\ 
SN II & SN Ibc & 0.37 \\ 
SN Ib & SN Ic & 0.55 \\
\tableline
\end{tabular}
    \caption{Comparison between stellar mass distributions of the multiple-SN host galaxies in our sample with stellar mass measurements in \citet{patspaper}. The $p$-values from the KS test yield no evidence for statistically significant differences. \label{stellar_mass_distributions}}
\end{center}
\end{table}

\subsection{[NII] $\lambda 6583$/H$\alpha$ Ratio and Nebular Emission-Lines}
In Fig.~\ref{cdf_niihalpha}, we plot the ratio [NII] $\lambda 6583$/H$\alpha$ for the entire sample, consisting of 21 SN Ib hosts, 5 SN Ic hosts, and 33 SN II hosts. Host galaxies are categorized according to the type of SNe that accounts for half or more of all the SNe the galaxy has hosted. When we compare the hosts of SN Ibc and SN II, the KS test yields a statistically significant p-value of 0.044. When SN Ib hosts are compared to SN II hosts separately, the p-value is 0.089. When comparing SN Ic hosts and SN II hosts, the KS test yields a p-value of 0.102, which corresponds to no statistically significant difference. We note, however, that the SN Ic sample includes only 5 galaxies. 
%The comparison between SN Ic hosts and SN II hosts may suffer from low number statistics. 

\subsection{Nebular Gas Properties}

We next restrict the sample to galaxies that fall in the \citet{SDSS_dashed_line} star-forming region, and
examine whether there are statistically significant differences between the distributions of the nebular gas properties of the galaxies that have primarily hosted SNe Ib, SNe Ic, SNe Ibc, and SNe II, in our sample.

\begin{figure}
  \includegraphics[width=\linewidth]{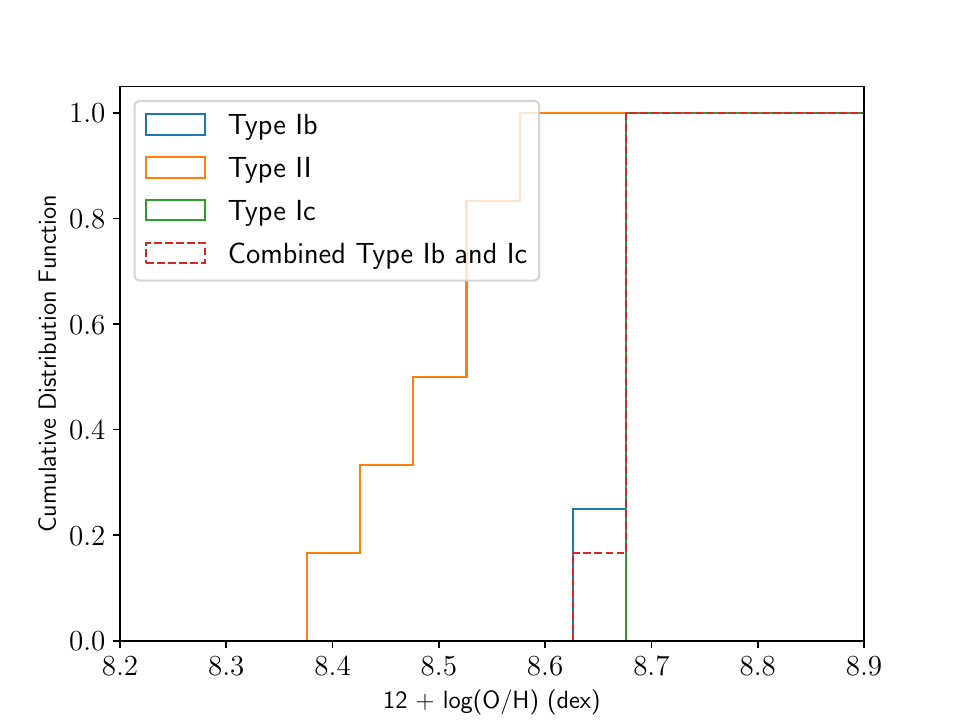}
  \caption{Cumulative distribution function of the {\tt PP04 N2} oxygen abundance for {\tt Sample SF}, which contains 13 host galaxies. There are 4 SN Ib host galaxies, 2 SN Ic host galaxies, and 7 SN II host galaxies. We find a statistically significant difference between the SN Ibc hosts and SN II hosts (p=0.008). Comparing these separately, we find the difference between the SN Ib hosts and SN II hosts to still be statistically significant (p=0.030), but the difference between the SN Ic hosts and SN II hosts to not be statistically significant (p=0.167). The number of SN Ic hosts also suffers from low number statistics, having only two galaxies.}
  \label{cdf_pp04}
\end{figure}

In Fig.~\ref{cdf_pp04}, we plot the distributions of oxygen abundance inferred using the {\tt PP04 N2} indicator for {\tt Sample SF}. Small samples (4 SN Ib hosts, 2 SN Ic host, 7 SN II hosts) remain after applying the H$\beta$ S/N cut and selecting spectra consistent with the star-forming region. The KS test, however, yields a statistically significant p-value of 0.008 for the comparison between SN Ibc hosts and SN II hosts. Moreover, there is significant evidence for a difference between the distributions for SN Ib and SN II hosts (p=0.030). The comparison between the two SN Ic hosts and seven SN II hosts is not significant (p=0.167).

\begin{figure}[h!]
  \includegraphics[width=\linewidth]{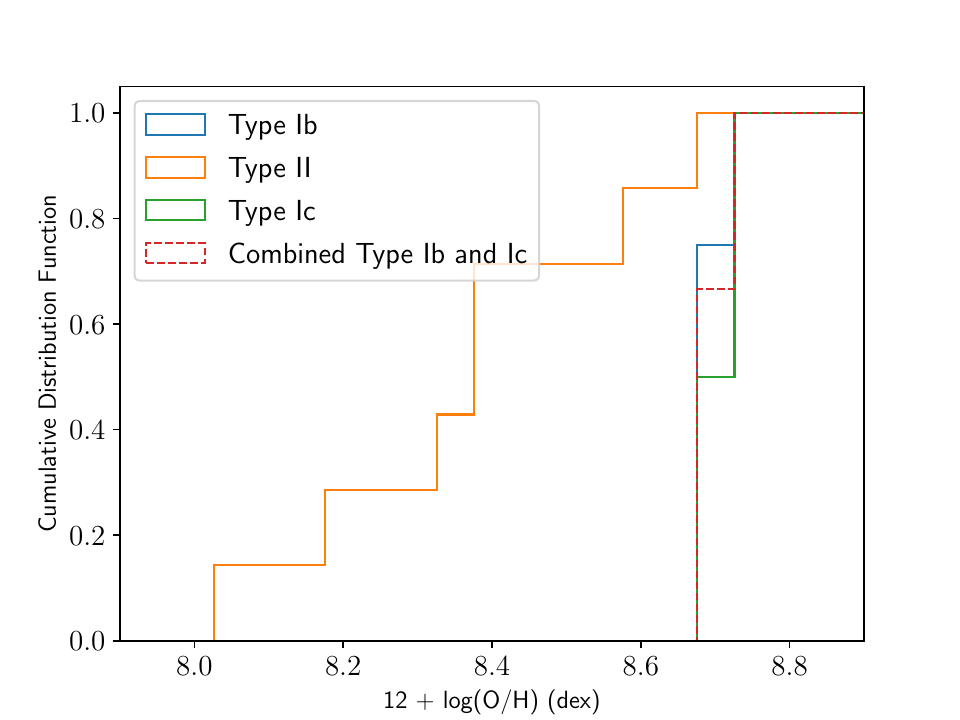}
  \caption{This plot shows the cumulative distribution function of {\tt PP04 O3N2} oxygen abundance for {\tt Sample SF}. There is a statistically significant difference between the SN Ibc hosts and the SN II hosts (p=0.008). Furthermore, there is also a significant difference between the SN Ib hosts and the SN II hosts (p=0.030), but there is no significant difference between the SN Ic hosts and the SN II hosts (p=0.056).}
  \label{cdf_pp04_o3n2}
\end{figure}

We also plot the distributions of oxygen abundance using the {\tt PP04 O3N2} indicator (see Fig.~\ref{cdf_pp04_o3n2}). Again, this is for {\tt Sample SF}, so our sample size is small. The KS test, though, yields a p-value of 0.008, which is statistically significant, for the comparison between SN Ibc hosts and SN II hosts. There is also a significant difference between the distributions of SN Ib and SN II hosts (p=0.030), but not between the distributions of SN Ic and SN II hosts (p=0.056).

To further explore the metallicity and other gas-phase properties of these galaxies, we next use the photoionization models from {\tt NebulaBayes} to model the combined {\tt Sample Comp.} and {\tt Sample SF} samples. In order to examine whether any differences are robust to assumptions about the ionizing spectrum, we apply two prescriptions: (a) model both samples using the \citet{Sutherland&Dopita} photoionization grid for an HII region, and (b) use the \citet{NBNLR} OXAF ionizing spectrum for a NLR when modelling the {\tt Sample Comp.} While our objective is to interpret the difference we observe in the [NII]/Halpha ratios in Fig.~\ref{cdf_niihalpha}, we note that analysis of the integrated spectra of the host galaxies using {\tt NebulaBayes} models for either an HII region or a NLR represents a substantial simplification.

\begin{figure*}[h!]
\centering
\includegraphics[width=3.25in]{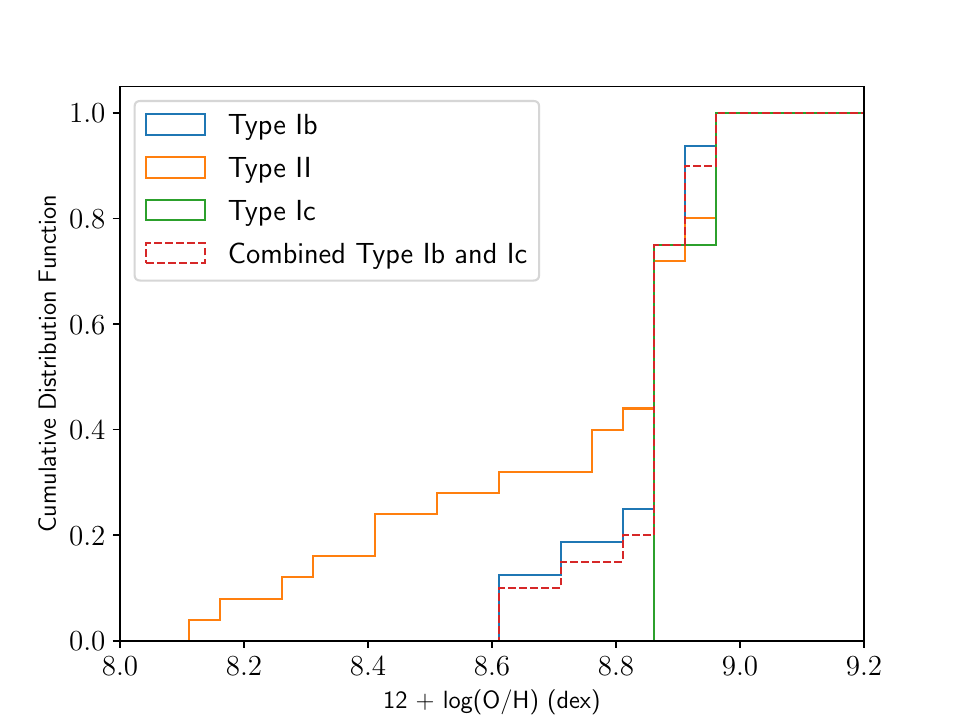}
  \includegraphics[width=3.25in]{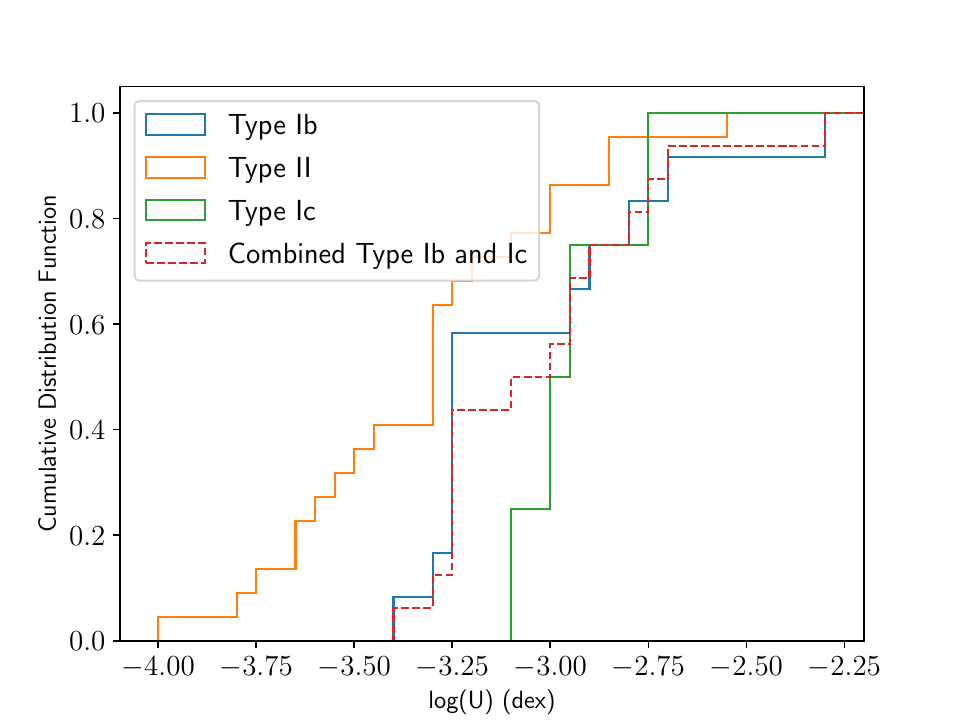}
  \includegraphics[width=3.25in]{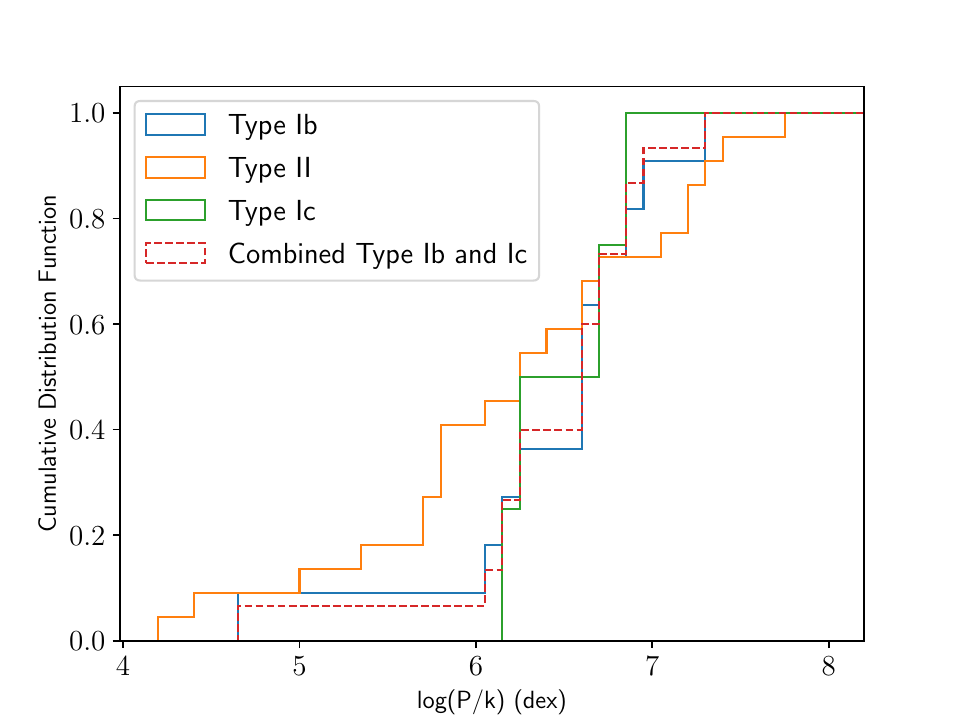}
  \caption{Cumulative distribution functions of the oxygen abundance, ionization, and gas pressure for host-galaxies in {\tt Sample SF} and {\tt Sample Comp.} found from {\tt NebulaBayes}. Both samples are modelled as an HII regions. Each host galaxy is assigned to a sample (Type Ib, Ic, and II) according to the spectroscopic type of SN that accounts for a majority of discoveries in the galaxy.}
  \label{cdf_plots_HII_samp1+2}
\end{figure*}

In Fig.~\ref{cdf_plots_HII_samp1+2}, we plot the posterior distributions from {\tt NebulaBayes} on the oxygen abundance, the ionization parameter, and gas pressure for the combination of {\tt Sample Comp.} and {\tt Sample SF}. While our sample includes galaxies in the Composite region, for this figure we model the emission-line fluxes using the ionizing spectrum for an HII region. When comparing the majority of SN Ibc and the majority of SN II samples, the KS statistic yields p-values of 0.27, 0.009, and 0.17 for oxygen abundance, ionization parameter, and gas pressure, respectively. KS test p-values for these properties considering the SN Ib and Ic subtypes can be found in Tab.~\ref{Sample2_table}.

\begin{figure*}[h!]
\centering
\includegraphics[width=3.25in]{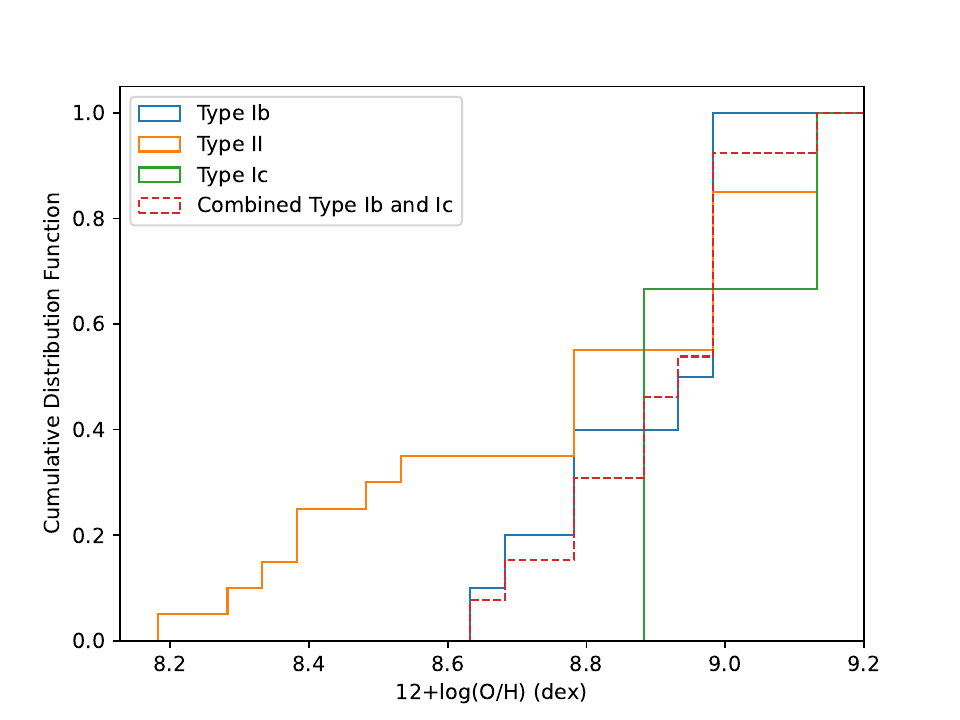}
  \includegraphics[width=3.25in]{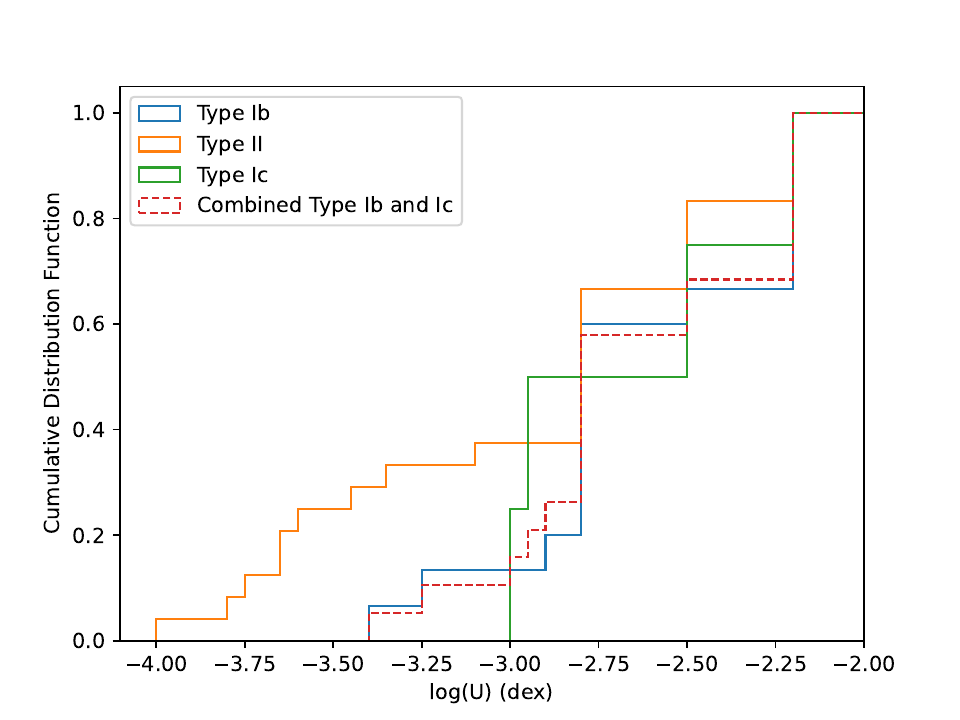}
  \includegraphics[width=3.25in]{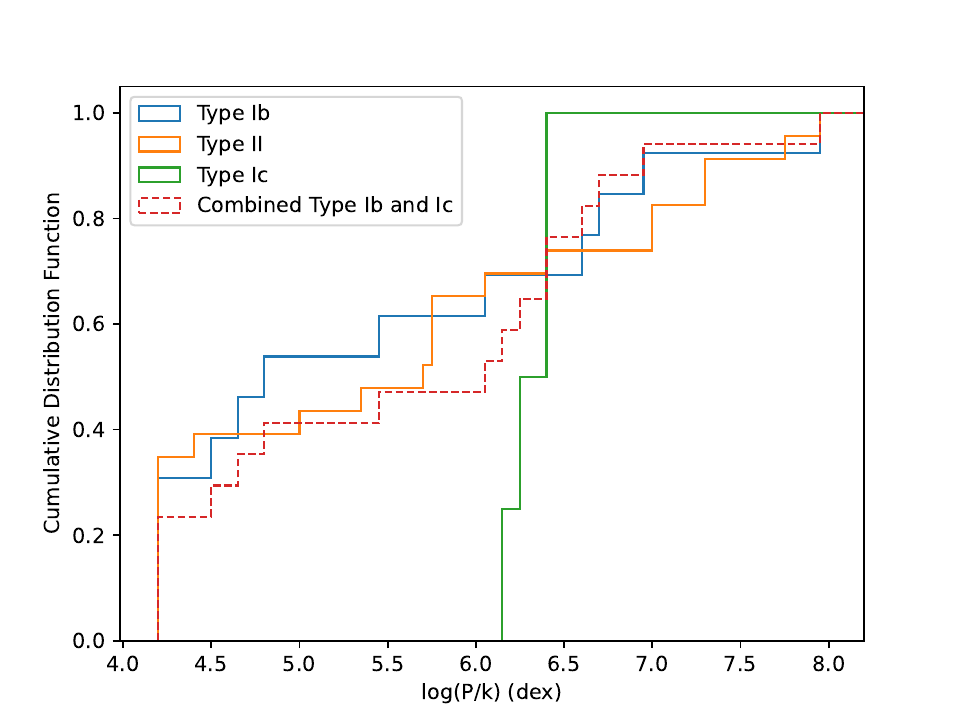}
  \caption{Cumulative distribution functions of the oxygen abundance, ionization, and gas pressure for host-galaxies in {\tt Sample SF} and {\tt Sample Comp.} found from {\tt NebulaBayes}. {\tt Sample SF} is modelled as an HII region while {\tt Sample Comp.} is modelled as an NLR region. Each host galaxy is assigned to a sample (Type Ib, Ic, and II) according to the spectroscopic type of SN that accounts for a majority of discoveries in the galaxy. }
  \label{cdf_plots_NLR_samp1+2}
\end{figure*}

\begin{figure*}[h!]
  \includegraphics[width=\linewidth]{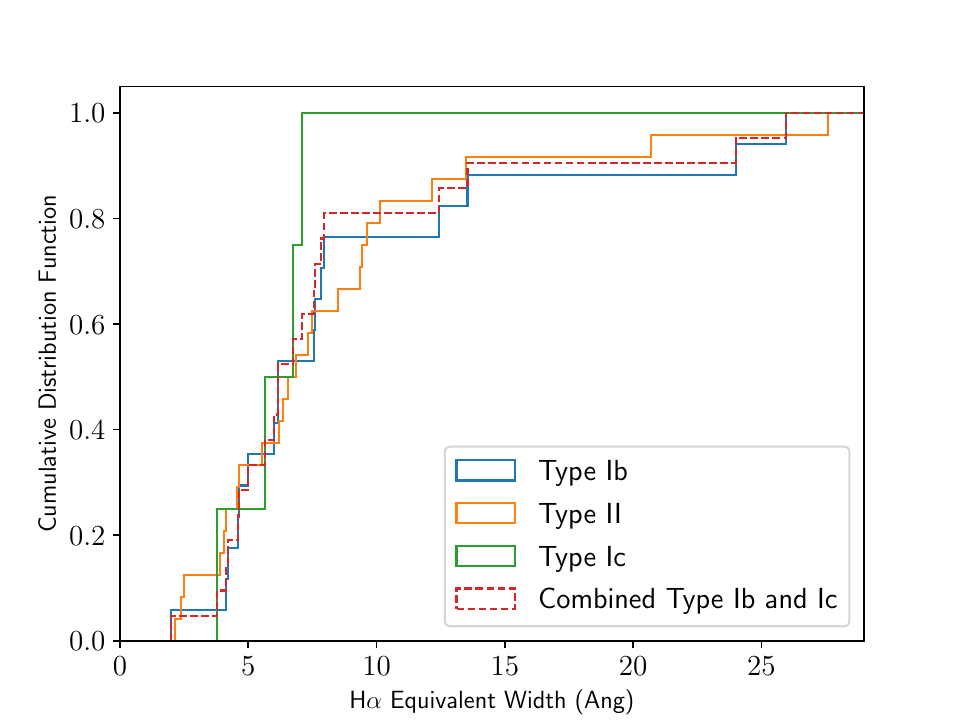}
  \caption{Cumulative distribution functions of the equivalent width of H$\alpha$ emission line in the host-galaxy spectra from the {\tt Full Sample}. Each host galaxy is assigned to a sample (Type Ib, Ic, and II) according to the spectroscopic type of SN that accounts for a majority of discoveries in the galaxy. This plot shows no statistically significant difference between the samples.}
  \label{h-alpha-IbIIb}
\end{figure*}

\begin{table*}
\centering
 \begin{tabular}{ |c|c|c|c| } 
  \hline
    & SN Ib vs. SN II & SN Ic vs. SN II & SN Ibc vs. SN II \\
  \hline
  12 + log(O/H) & 0.34 & 0.40 & 0.27 \\ 
   \hline
  log(U) & 0.027 & 0.074 & 0.009 \\
  \hline
  log(P/k) & 0.26 & 0.52 & 0.17 \\  
  \hline
 \end{tabular}
 \caption{The gas phase metallicity, ionization, and gas pressure are compared between different groupings of galaxies from {\tt Sample SF} and {\tt Sample Comp.}, where galaxies in both samples are modelled as HII regions. These groupings are SN Ib hosts compared to SN II hosts (SN Ib vs. SN II), SN Ic hosts compared to SN II hosts (SN Ic vs. SN II), and SN Ibc hosts compared to SN II hosts (SN Ibc vs. SN II). The values presented are the p-values of the KS test for these comparisons. }
 \label{Sample2_table}
\end{table*}

We next repeat the analysis of the gas-phase properties of {\tt Samp. Comp} and {\tt Samp. SF} using {\tt NebulaBayes}, where we continue modelling {\tt Samp SF} as an HII region but model {\tt Samp. Comp.} with a NLR model. For the NLR model, we use an OXAF ionizing spectrum \citep{NBNLR} with three parameters in {\tt NebulaBayes}. Fig.~\ref{cdf_plots_NLR_samp1+2} shows the posterior distributions from {\tt NebulaBayes} for  oxygen abundance, ionization parameter, and gas pressure. When comparing hosts of SN Ibc to hosts of SN II, the KS statistic yields p-values of 0.21, 0.27, and 0.55 for the oxygen abundance, ionization parameter, and gas pressure distributions, respectively. The full set of KS p-values can be found in Tab.~\ref{Sample1_table}. 
We conclude that we are not able to identify statistically significant differences among the distributions of oxygen abundance, ionization parameter, and gas pressure that are robust to the choice of ionizing source.

\begin{table*}
\centering
 \begin{tabular}{ |c|c|c|c| } 
  \hline
    & SN Ib vs. SN II & SN Ic vs. SN II & SN Ibc vs. SN II \\
  \hline
  12 + log(O/H) & 0.34 & 0.40 & 0.21 \\ 
   \hline
  log(U) & 0.34 & 0.65 & 0.27 \\
  \hline
  log(P/k) & 0.85 & 0.074 & 0.55 \\  
  \hline
 \end{tabular}
 \caption{These values are the p-values of the KS tests comparing different groupings of host galaxies.} The gas phase metallicity, ionization, and gas pressure are compared between different groupings of galaxies from {\tt Sample SF} and {\tt Sample Comp.}, where galaxies in {\tt Sample SF} are modelled as an HII region and galaxies in {\tt Sample Comp.} are modelled as NLRs. These groupings are SN Ib hosts compared to SN II hosts (SN Ib vs. SN II), SN Ic hosts compared to SN II hosts (SN Ic vs. SN II), and SN Ibc hosts compared to SN II hosts (SN Ibc vs. SN II). 
 \label{Sample1_table} 
\end{table*}

\begin{figure*}
\centering
\includegraphics[width=3.25in]{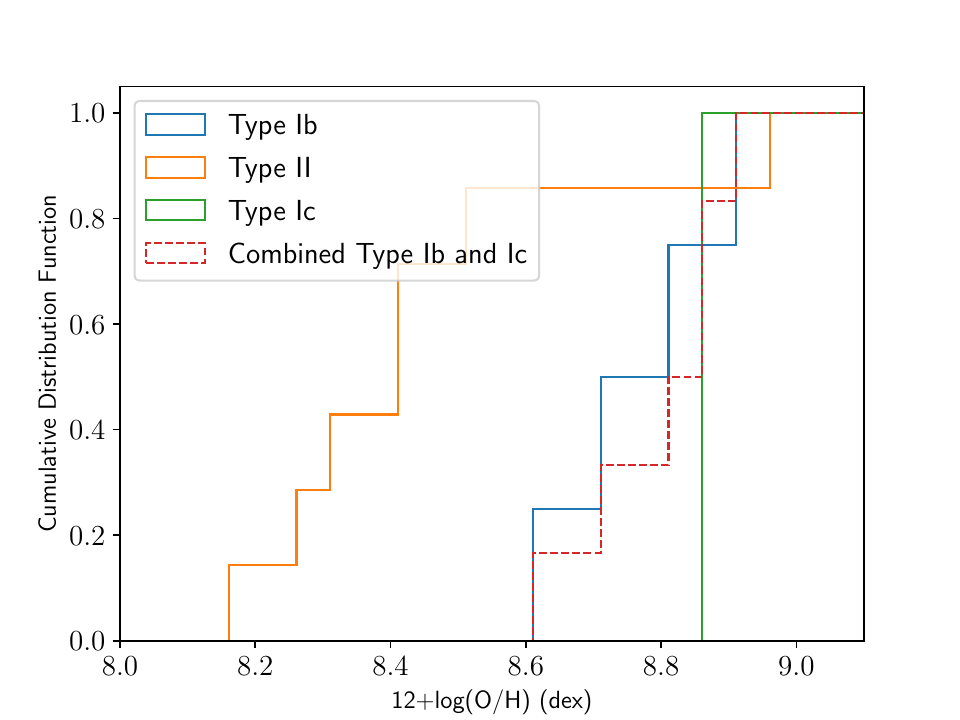}
  \includegraphics[width=3.25in]{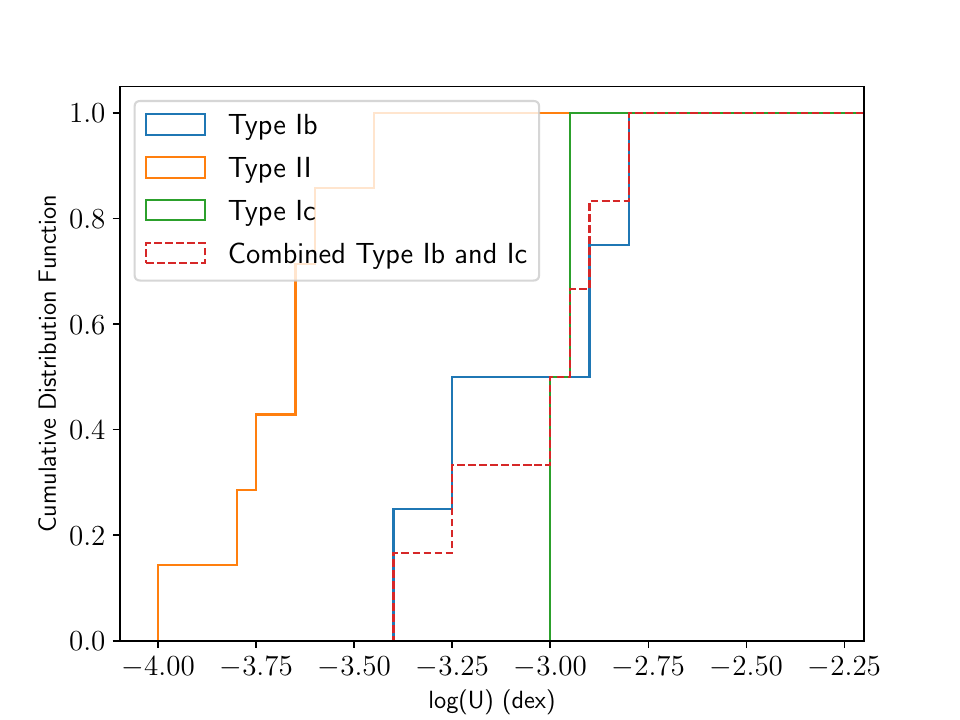}
  \includegraphics[width=3.25in]{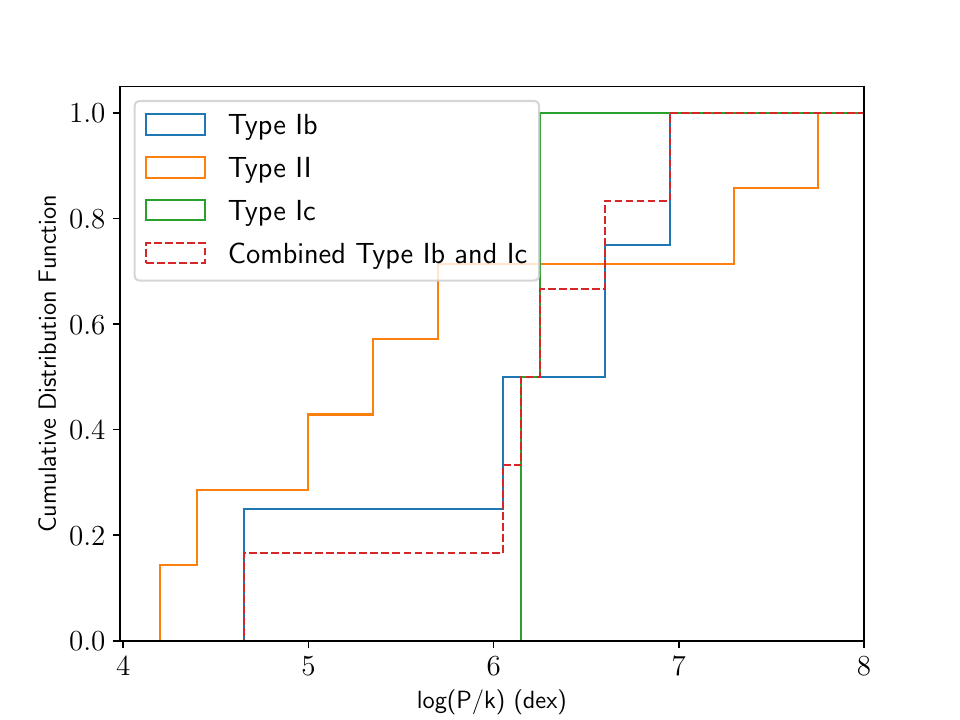}
  \caption{Cumulative distribution functions of the oxygen abundance, ionization, and gas pressure for host-galaxies in {\tt Sample SF} found from {\tt NebulaBayes}. Each host galaxy is assigned to a sample (Type Ib, Ic, and II) according to the spectroscopic type of SN that accounts for a majority of discoveries in the galaxy.}
  \label{cdf_plots}
\end{figure*}

We next analyse the gas-phase properties of the host galaxies by examining just the {\tt Sample SF} galaxies. Fig.~\ref{cdf_plots} shows the posterior distributions from {\tt NebulaBayes} on the oxygen abundance, the ionization parameter, and the gas pressure for this sample. Using the KS test to compare the distribution of SN Ibc hosts to SN II hosts, we find a significant difference for the oxygen abundance (p=0.008) and the ionization (p=0.001) but not for the gas pressure (p=0.212). 
All KS statistics for this sample are listed in Tab.~\ref{Sample2_table}.  

\begin{table*}[h!]
\centering
 \begin{tabular}{ |c|c|c|c| } 
  \hline
    & SN Ib vs. SN II & SN Ic vs. SN II & SN Ibc vs. SN II \\
  \hline
  12 + log(O/H) & 0.030 & 0.167 & 0.008 \\ 
   \hline
  log(U) & 0.006 & 0.055 & 0.001 \\
  \hline
  log(P/k) & 0.536 & 0.333 & 0.212 \\  
  \hline
 \end{tabular}
 \caption{The gas phase metallicity, ionization, and gas pressure are compared between different groupings of galaxies from {\tt Sample SF}. These groupings are SN Ib hosts compared to SN II hosts (SN Ib vs. SN II), SN Ic hosts compared to SN II hosts (SN Ic vs. SN II), and SN Ibc hosts compared to SN II hosts (SN Ibc vs. SN II). The values presented are the p-values of the KS test for these comparisons. The KS test shows no statistically significant differences. }
 \label{Sample2_table} 
\end{table*}

\subsection{H$\alpha$ Equivalent Width}
Fig.~\ref{h-alpha-IbIIb} shows the distributions of H$\alpha$ equivalent widths that we measure for the samples of SN host galaxies. 
For a hypothetical stellar population where all stars are formed at a single time, the H$\alpha$ equivalent width is a strong indicator of the time since the stars were formed. While we have acquired integrated spectra of the galaxies, the equivalent width may provide a useful indicator of the recent star-formation history. A KS test does not find evidence that the SN Ibc and SN II distributions are drawn from distinct underlying distributions (p=0.61). 

Tab.~\ref{Halph_eq_widths_table} lists the H$\alpha$ equivalent widths of the galaxies in our entire sample. The galaxies that have the greatest H$\alpha$ equivalent widths are NGC 2993, NGC 3690, NGC 7714, and MCG-02-07-10. NGC 2993, NGC 3690, and MCG-02-07-10 are host to a majority of type II SNe, while NGC 7714 has hosted primarily SN Ibc.

\section{Conclusions}
\label{sec:conclusions}

We have analyzed integrated optical spectroscopy of a sample of 59 galaxies that have hosted multiple SNe. The properties of these galaxies provide a new opportunity to examine potential connections between the spectroscopic type of SNe and their host-galaxy environments. Surprisingly, we find that a strong majority of the galaxies that have hosted multiple core-collapse SNe fall within the Composite region of the BPT diagram. From our complete sample of 59 galaxies, we remove 13 due to low S/N measurements of H$\beta$. The cut leaves us with 33 galaxies in the Composite region of the BPT diagram and 13 galaxies in the Star-Forming region of the BPT diagram. We refer to these two groupings of galaxies as {\tt Sample Comp.} for the galaxies in the composite region and {\tt Sample SF} for the galaxies in the Star-Forming region.

We first find, when we consider the full sample, that  galaxies that have hosted a majority SNe Ibc have higher [NII] $\lambda$6584 / H$\alpha$ ratios than those that have hosted SNe II (KS test p-value = 0.044). Greater ratios of [NII] $\lambda$6584 / H$\alpha$ can arise from either higher metallicity, or from AGN or LINER contributions to the ionizing flux.

We further examine the 13 host galaxies in our {\tt Sample SF}. Given the galaxies' classification as star-forming, we are able to make confident inferences about their oxygen abundance. We find a significant difference in oxygen abundance inferred from the \citet{PP04_paper} N2 diagnostic between the distributions for the SN Ibc hosts and the SN II hosts (p=0.008). We also examine oxygen abundance inferred from the 03N2 index and again find a significance between the distributions of SN Ibc hosts and SN II hosts (p=0.008).

We also find a significant difference for {\tt Sample SF} when using the \citet{Sutherland&Dopita} photoionization grid and the {\tt NebulaBayes} software package. When comparing the distributions of SN Ibc hosts and SN II hosts, we find a significant difference in oxygen abundance (p=0.008) and ionization parameter (p=0.001), but not in gas pressure (p=0.212).  Overall, we find significant differences in oxygen abundance from three different oxygen abundance diagnostics for our small {\tt Sample SF}.

 Increased oxygen abundance may  play a role in determining the relative fraction of massive stars that explode as SNe Ibc or SNe II, although not all studies have found evidence for a metallicity dependence \citep[e.g.,][]{Anderson2010,Leloudas}. 
 A recent study has found evidence for higher abundances of Wolf-Rayet stars in association with high-ionization nebular emission lines \citep{kauffmannmillancrowther24}. Since Wolf-Rayet stars are expected to explode as SNe Ibc, the pattern we identify may have connection with the Wolf-Rayet population. 

Our findings motivate spatially resolved spectroscopy of the host galaxies of multiple SNe. The positions of individual HII regions as well as diffuse emission on the BPT diagram would enable constraints on the origin of the high [NII] / H$\alpha$ ratios we observed for galaxies that have primarily hosted SNe Ibc.  Spectra of HII regions consistent with stellar ionizing sources would enable oxgyen abundance measurements, and make it possible to disentangle the contributions of the ionizing source and metallicity to the galaxies' [NII] / H$\alpha$ ratios.

\begin{acknowledgements}
L.S., R.Z., and S.J. were supported by the NSF through NRT-WoU 1922512 (PI: Mandic). P.K. has been supported by NSF grants AST-1908823 and AST-2308051 and a McKnight Land-Grant Professorship. G.J. is supported by NSF grant DMS-2152746. his study uses data provided by the Calar Alto Legacy Integral Field Area (CALIFA) survey (https://califa.caha.es/). Based on observations collected at the Central Astron\'omico Hispano Alem\'an (CAHA) at Calar Alto, operated jointly by the Max-Planck-Institut f\H ur Astronomie and the Instituto de Astrof\'isica de Andaluc\'ia (CSIC).
\end{acknowledgements}

\software{Open Supernova Catalog, Astroquery, Numpy, Pandas, Matplotlib, IRAF}

\bibliography{other_files/references}
\bibliographystyle{aasjournal}

\pagebreak
\renewcommand\thefigure{A\arabic{figure}}   
\setcounter{figure}{0} 
\renewcommand\thetable{A\arabic{table}}   
\setcounter{table}{0} 
\appendix

\begin{figure}[h!]
  \includegraphics[width=\linewidth]{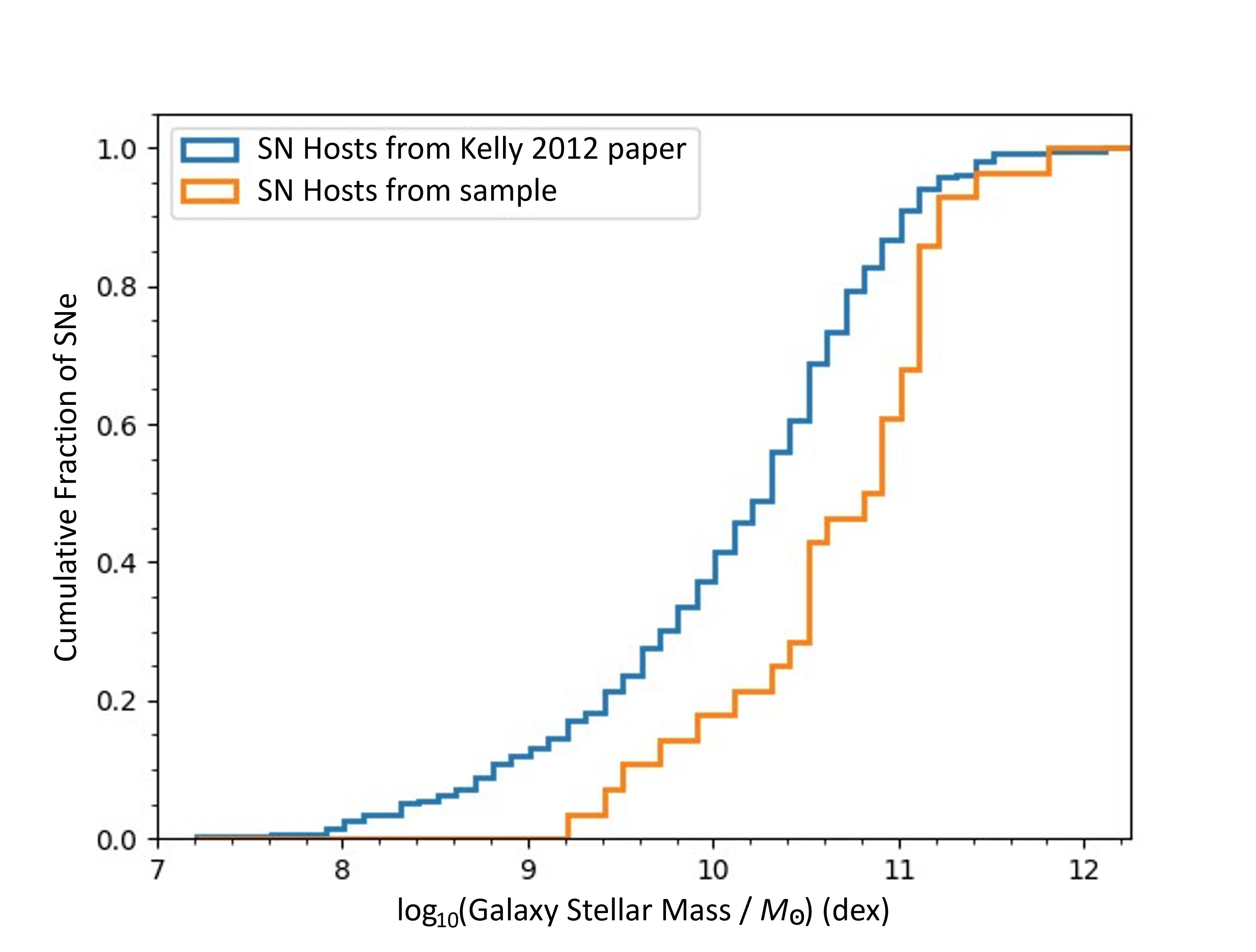}
  \caption{Comparison between the distributions of SN host-galaxy stellar masses. Here we use the stellar masses of low-redshift SN host galaxies measured by \citet{patspaper}. The blue curve plots the full $z<0.04$ sample from \citet{patspaper}, while the orange curve shows the overlapping objects between the galaxies we analyze in this paper, which have hosted multiple SNe, and the \citet{patspaper}. Our sample of galaxies has greater stellar masses, on average, than the \citet{patspaper}. More massive galaxies have higher average star-formation rates, and therefore core-collapse SN rates.}
  \label{masses_all}
\end{figure}

\pagebreak

\section{Observation Details}

\begin{longtable*}{c c c}
 \hline
 Galaxy Name & Redshift & SNe \\
 \hline\hline
  IC701 & 0.0205 & 13gl(Ib), 04gj(IIb)\\
  NGC7479 & 0.0079 & 09jf(Ib), 90U(Ib)\\
  NGC2770 & 0.0065 & 07uy(Ib pec), 08D(Ib), 99eh(Ib), 15bh(IIn)\\
  NGC7714 & 0.0093 & 07fo(Ib), 99dn(Ib)\\
  NGC3690 & 0.0104 & 10P(Ib/IIb), 99D(II), 05U(IIb), 98T(Ib), 92bu(nan), 10O(Ib), 93G(IIL)\\
  NGC5480 & 0.0064 & 88L(Ib), 17iro(Ib)\\
  NGC6118 & 0.0052 & 04dk(Ib), 20hvp(Ib)\\
  UGC7020 & 0.0205 & 01cf(IIb), 06C(II)\\
  IC1099 & 0.0296 & 40C(II), 10av(Ib/c)\\
  IC1216 & 0.0254 & 97da(II), 19klq(Ib)\\
  NGC5278 & 0.0252 & 01ai(Ib), 19cec(II)\\
  NGC3147 & 0.0093 & 08fv(Ia), 97bq(Ia), 72H(I), 06gi(Ib) \\
  NGC3506 & 0.0214 & 03L(Ic), 17dfq(Ia) \\
  NGC5806 & 0.0045 & 04dg(IIP), 12P(Ib/c) \\
  NGC7624 & 0.0143 & 10kc(Ib), 08ea(IIP)\\
  IC900 & 0.0236 & 15at(II), 09gl(IIb)\\
  KUG1313+309 & 0.0189 & 16jdw(Ib), 19uwd(II)\\
  NGC4666 & 0.0050 & 65H(II), 19yvr(Ib)\\
  NGC5605 & 0.0113 & 22bn(Ib), 22ec(II)\\
  UGC4671 & 0.0122 & 16bme(II), 00dv(Ib)\\
  NGC1187 & 0.0046 & 82R(Ib), 07Y(Ib)\\
  NGC5630 & 0.0089 & 05dp(II), 06am(IIn)\\
  NGC5888 & 0.0291 & 10fv(II), 07Q(II)\\
  MCG-02-33-20 & 0.0080 & 08aq(IIb), 16aai(IIP), 15P(IIP) \\
  M+07-20-73 & 0.0160 & 89U(II), 14eb(IIP), 01er(Ia) \\
  NGC2532 & 0.0175 & 02hn(Ic), 99gb(IIn), 16gil(II) \\
  NGC2939 & 0.0111 & 11cf(II), 09ao(IIP) \\
  NGC4017 & 0.0115 & 07an(II), 02hm(II), 06st(II)\\
  NGC4303 & 0.0052 & 64F(II), 26A(IIL), 08in(IIP), 06ov(IIP), 14dt(Ia pec), 61I(II), 99gn(IIP)\\
  UGC5434 & 0.0186 & 06bx(II), PSN J10051374+2127231(II) \\
  NGC1084 & 0.0047 & 96an(II), 08al(II), 98dl(IIP), 63P(Ia), 12ec(IIP), 09H(II), \\
  NGC1643 & 0.0163 & 99et(II), 95G(IIn)\\
  NGC3034 & 0.0009 & 04am(IIP), 14J(Ia), 08iz(II)\\
  NGC337 & 0.0055 & 11dq(IIP), 14cx(IIP)\\
  NGC2993 & 0.0077 & 17ejx(IIP), 03ao(IIP)\\
  NGC4141 & 0.0064 & 09E(IIP), 08X(IIP) \\
  NGC3340 & 0.0183 & 05O(Ib), 07fp(II), 99O(II)\\
  NGC3627 & 0.0024 & 09ii(II), 07bb(IIn)\\
  NGC4490 & 0.0019 & 82F(IIP), 08ax(IIb), 08ic(II)\\
  NGC7316 & 0.0185 & 16hvu(IIP), 06cx(II), 11he(Ia pec)\\
  UGC11946 & 0.0184 & 08dn(II), 21rfs(Ib)\\
  NGC772 & 0.0082 & 03hl(II), 03iq(II)\\
  NGC2276 & 0.0081 & 05dl(II), 68W(nan), 93X(II), 62Q(nan), 68V(II), 16gfy(II)\\
  NGC3184 & 0.0019 & 99gi(IIP), 10dn(nan), 21B(II), 21C(I), 16bkv(II), 37F(IIP)\\
  NGC3631 & 0.0039 & 96bu(IIn), 65L(IIP), 16bau(Ib), 64A(II pec)\\
  NGC4258 & 0.0015 & 14bc(IIP), 81K(II)\\
  NGC5457 & 0.0008 & 11fe(Ia), 70G(IIL), 09A(IIP), 51H(II)\\
  NGC493 & 0.0078 & 04br(Ia pec), 94M(Ia) \\
  MCG-02-07-10 & 0.0070 & 85S(II), 21suk(II)\\
  UGC5695 & 0.0098 & 93N(IIn), 94N(II)\\
  NGC4157 & 0.0026 & 55A(nan), 03J(IIP), 37A(IIP)\\
  NGC3938 & 0.0027 & 05ay(IIP), 17ein(Ic pec), 61U(IIL), 10kn(Ia), 64L(Ic) \\
  NGC4273 & 0.0079 & 36A(IIP), 08N(II)\\
  NGC5020 & 0.0112 & 15D(IIP), 91J(II)\\
  NGC4051 & 0.0023 & 83I(Ic), 10br(Ib/c), 03ie(IIP) \\
  NGC3810 & 0.0033 & 97dq(Ic), 00ew(Ic) \\
  NGC4568 & 0.0074 & 04cc(Ic), 90B(Ic) \\
  NGC5374 & 0.0145 & 16P(Ic-BL), 03bl(II), 10do(Ic) \\
  UGC4798 & 0.0267 & 05mf(Ic), 13V(Ia)\\
  \hline
  \caption{All host galaxies are listed here, along with their redshifts and the SNe they have hosted. The name of each SN is listed with the type of SN indicated in parentheses. To be included in our sample, a galaxy must have hosted a minimum of two SNe, with preference given to those galaxies that have hosted more SNe.}
  \label{redshiftsandsne_galaxies}
 \end{longtable*}

\begin{longtable*}{c c c c}
 \hline
 Galaxy Name & Date Observed & Exposure Time (s) & Galaxy Diameter ($''$) [Passband] \\
 \hline\hline
  NGC7479 & 22 Dec 2020 & 1200 & 264.00 [POSS1 103a-O]\\
  NGC2770 & 19 Dec 2020 & 1200 & 255.52 [r (SDSS Isophotal)]\\
  NGC7714 & 21 Dec 2020 & 1200 & 150.00 [POSS1 103a-O]\\
  NGC3690 & 22 Dec 2020 & 1200 & 170.40 [K\_s (LGA/2MASS ``total'')]\\
  NGC6118 & 24 May 2022 & 1800 & 307.70 [RC3 D\_0 (blue)]\\
  UGC7020 & 24 May 2022 & 1800 & 128.30 [RC3 D\_25, R\_25 (blue)]\\
  IC1099 & 25 May 2022 & 1200 & 89.80 [K\_s (2MASS ``total'')]\\
  IC1216 & 25 May 2022 & 1200 & 66.00 [POSS1 103a-O]\\
  NGC5278 & 25 May 2022 & 1200 & 80.90 [RC3 D\_25, R\_25 (blue)]\\
  NGC3147 & 26 May 2022 & 1200 & 282.00 [POSS1 103a-O]\\
  NGC3506 & 26 May 2022 & 1200 & 78.00 [POSS1 103a-O]\\
  NGC5806 & 26 May 2022 & 1200 & 189.70 [RC3 D\_0 (blue)]\\
  NGC7624 & 26 May 2022 & 1200 & 95.50 [K\_s (2MASS ``total'')]\\
  IC900 & 27 May 2022 & 1200 & 110.60 [K\_s (2MASS ``total'')]\\
  KUG1313+309 & 27 May 2022 & 1200 & 68.20 [K\_s (2MASS ``total'')]\\
  NGC4666 & 27 May 2022 & 1000 & 429.40 [r (SDSS Isophotal)]\\
  NGC5605 & 27 May 2022 & 1200 & 126.10 [K\_s (2MASS ``total'')]\\
  UGC4671 & 27 May 2022 & 1200 & 96.00 [POSS1 103a-O]\\
  NGC1187 & 20 Jan 2023 & 1200 & 495.50 [ESO-LV ``Quick Blue'' IIa-O]\\
  MCG-02-33-20 & 17 Feb 2020 & 900 & 210.00 [POSS1 103a-O]\\
  M+07-20-73 & 17 Feb 2020 & 900 & 138.00 [POSS1 103a-O]\\
  NGC2532 & 17 Feb 2020 & 900 & 134.30 [RC3 D\_0 (blue)]\\
  NGC2939 & 17 Feb 2020 & 900 & 156.00 [POSS1 103a-O]\\
  NGC4017 & 22 Dec 2020 & 1200 & 112.50 [K\_s (2MASS ``total'')]\\
  NGC4303 & 17 Feb 2020 & 900 & 399.20 [K\_s (LGA/2MASS ``total'')]\\
  UGC5434 & 19 Dec 2020 & 1200 & 94.40 [K\_s (2MASS ``total'')]\\
  NGC1084 & 19 Dec 2020 & 1080 & 68.90 [RC3 D\_0 (blue)]\\
  NGC1643 & 20 Dec 2020 & 1200 & 67.30 [RC3 D\_0 (blue)]\\
  NGC3034 & 20 Dec 2020 & 1200 & 780.00 [POSS1 103a-O]\\
  NGC337 & 20 Dec 2020 & 1200 & 197.40 [K\_s (LGA/2MASS ``total'')]\\
  NGC2993 & 21 Dec 2020 & 1200 & 84.80 [RC3 D\_0 (blue)]\\
  NGC4141 & 22 Dec 2020 & 1200 & 84.00 [POSS1 103a-O]\\
  NGC3340 & 5 Jun 2021 & 1200 & 85.30 [K\_s (2MASS ``total'')]\\
  NGC3627 & 5 Jun 2021 & 1200 & 560.00 [RC3 D\_0 (blue)]\\
  NGC4490 & 5 Jun 2021 & 1200 & 425.80 [K\_s (LGA/2MASS ``total'')]\\
  NGC7316 & 7 Jun 2021 & 1200 & 75.40 [K\_s (2MASS ``total'')]\\
  UGC11946 & 26 May 2022 & 1200 & 90.40 [K\_s (2MASS ``total'')]\\
  NGC772 & 21 Dec 2020 & 1200 & 480.00 [POSS1 103a-O]\\
  NGC2276 & 19 Jan 2023 & 1800 & 170.10 [RC3 D\_0 (blue)]\\
  NGC3184 & 19 Jan 2023 & 1800 & 552.20 [K\_s (LGA/2MASS ``total'')]\\
  NGC3631 & 19 Jan 2023 & 1800 & 380.60 [K\_s (LGA/2MASS ``total'')]\\
  NGC4258 & 19 Jan 2023 & 1200 & 1320.00 [POSS1 103a-O]\\
  NGC5457 & 19 Jan 2023 & 1200 & 2400.00 [POSS1 103a-O]\\
  NGC493 & 20 Jan 2023 & 1200 & 258.00 [POSS1 103a-O]\\
  MCG-02-07-10 & 20 Jan 2023 & 1000 & 114.80 [K\_s (2MASS ``total'')]\\
  UGC5695 & 20 Jan 2023 & 1200 & 90.86 [r (SDSS Isophotal)]\\
  NGC4157 & 20 Jan 2023 & 1200 & 552.20 [K\_s (LGA/2MASS ``total'')]\\
  NGC3938 & 20 Jan 2023 & 1200 & 380.60 [K\_s (LGA/2MASS ``total'')]\\
  NGC4273 & 20 Jan 2023 & 1200 & 150.00 [POSS1 103a-O]\\
  NGC5020 & 20 Jan 2023 & 1200 & 198.00 [POSS1 103a-O]\\
  NGC4051 & 20 Dec 2020 & 1200 & 405.00 [K\_s (LGA/2MASS ``total'')]\\
  NGC3810 & 21 Dec 2020 & 1200 & 261.90 [RC3 D\_0 (blue)]\\
  NGC4568 & 21 Dec 2020 & 1200 & 357.00 [K\_s (LGA/2MASS ``total'')]\\
  NGC5374 & 25 May 2022 & 1200 & 104.80 [K\_s (2MASS ``total'')]\\
  UGC4798 & 19 Jan 2023 & 1600 & 66.00 [POSS1 103a-O]\\
  \hline
  \caption{Details of observations of host galaxies. The date of observation, the exposure time, and the airmass are included. The galaxies in this list were observed with the Bok telescope. Data for four galaxies, IC 701, NGC 5480, NGC 5630, and NGC 5888, was obtained from CALIFA. The physical galaxy diameter was obtained from the \citet{https://doi.org/10.26132/ned1}. Galaxies too large to be scanned by the slit of the Bok telescope were skipped. For some larger galaxies, the slit was rotated to the shortest diameter of the galaxy and scanned across the larger diameter of the galaxy. Extraction regions were determined by visual inspection of the data. }
  \label{redshiftsandsne_galaxies}
 \end{longtable*}

\begin{longtable*}{c c c c}
 \hline
 Galaxy Name & Date Observed & Exposure Time (s) & Galaxy Diameter ($''$) [Passband] \\
 \hline\hline
IC701 & 19 Dec 2015 & 900 & 92.00 [K\_s (2MASS ``total'')]\\
NGC5480 & 14 Aug 2013 & 900 & 134.40 [K\_s (2MASS ``total'')]\\
NGC5630 & 22 Jun 2012 & 900 & 150.00 [POSS1 103a-O]\\
NGC5888 & 6 Jun 2012 & 900 & 109.20 [K\_s (2MASS ``total'')]\\
\hline
  \caption{Details of observations of host galaxies. The date of observation, the exposure time, and the airmass are included. The galaxies in this list were observed in the CALIFA surveys. The physical galaxy diameter was obtained from the \citet{https://doi.org/10.26132/ned1}.}
  \label{redshiftsandsne_galaxies_CALIFA}
 \end{longtable*}

\section{Emission-Line Strengths of Host Galaxies}

\begin{longtable*}{c c c c c c c c} 
 \hline
 \multicolumn{8}{c}{Hosts of Galaxies that Have Hosted Majority Type Ib SNe} \\
 \hline
 Galaxy Name & H$\beta$ & H$\alpha$ & [SII]6716 & [SII]6731 & [OIII]5007 & [NII]6583 \\ [0.5ex] 
 \hline\hline
IC701 & $39.68 \pm 0.70$ & $108.88 \pm 0.68$ & $24.49 \pm 0.69$ & $14.34 \pm 0.69$ & $32.99 \pm 0.89$ & $39.52 \pm 0.87$\\
NGC7479 & $3.42 \pm 1.03$ & $12.40 \pm 0.99$ & $1.69 \pm 0.92$ & $2.06 \pm 0.92$ & $1.08 \pm 1.21$ & $7.45 \pm 1.15$\\
NGC2770 & $4.49 \pm 0.89$ & $30.34 \pm 0.84$ & $4.00 \pm 0.84$ & $5.40 \pm 0.84$ & $9.25 \pm 1.11$ & $11.95 \pm 1.06$\\
NGC7714 & $106.25 \pm 2.02$ & $237.00 \pm 1.91$ & $28.72 \pm 1.94$ & $19.92 \pm 1.94$ & $259.67 \pm 2.55$ & $95.75 \pm 2.43$\\
NGC3690 & $51.72 \pm 1.14$ & $197.51 \pm 1.08$ & $33.33 \pm 1.16$ & $29.17 \pm 1.16$ & $95.91 \pm 1.49$ & $104.23 \pm 1.44$\\
NGC5480 & $149.91 \pm 1.81$ & $442.54 \pm 1.77$ & $81.30 \pm 1.74$ & $56.33 \pm 1.74$ & $44.45 \pm 2.26$ & $196.86 \pm 2.20$\\
NGC6118 & $0.53 \pm 0.25$ & $3.89 \pm 0.24$ & $0.95 \pm 0.21$ & $0.38 \pm 0.21$ & $0.43 \pm 0.27$ & $1.58 \pm 0.26$\\
UGC7020 & $0.66 \pm 0.12$ & $1.67 \pm 0.12$ & $0.00 \pm 0.12$ & $0.01 \pm 0.12$ & $0.84 \pm 0.16$ & $0.70 \pm 0.15$\\
IC1099 & $0.62 \pm 0.14$ & $1.35 \pm 0.13$ & $0.14 \pm 0.13$ & $0.21 \pm 0.13$ & $0.31 \pm 0.17$ & $0.71 \pm 0.16$\\
IC1216 & $1.08 \pm 0.15$ & $2.60 \pm 0.15$ & $0.00 \pm 0.15$ & $0.00 \pm 0.15$ & $1.72 \pm 0.19$ & $1.05 \pm 0.18$\\
NGC5278 & $3.94 \pm 0.16$ & $9.58 \pm 0.15$ & $0.03 \pm 0.15$ & $0.23 \pm 0.15$ & $11.96 \pm 0.20$ & $7.42 \pm 0.19$\\
NGC3147 & $0.09 \pm 0.48$ & $9.36 \pm 0.46$ & $0.72 \pm 0.47$ & $0.00 \pm 0.47$ & $1.47 \pm 0.61$ & $6.80 \pm 0.59$\\
NGC3506 & $1.90 \pm 0.13$ & $7.30 \pm 0.13$ & $0.00 \pm 0.13$ & $0.00 \pm 0.13$ & $1.25 \pm 0.17$ & $3.45 \pm 0.16$\\
NGC5806 & $0.06 \pm 0.40$ & $8.85 \pm 0.38$ & $1.91 \pm 0.37$ & $0.67 \pm 0.37$ & $0.90 \pm 0.49$ & $5.88 \pm 0.47$\\
NGC7624 & $3.54 \pm 0.20$ & $14.45 \pm 0.19$ & $1.19 \pm 0.19$ & $1.85 \pm 0.19$ & $1.47 \pm 0.26$ & $6.91 \pm 0.24$\\
IC900 & $1.04 \pm 0.12$ & $4.96 \pm 0.11$ & $0.00 \pm 0.11$ & $0.00 \pm 0.11$ & $1.30 \pm 0.14$ & $2.15 \pm 0.14$\\
KUG1313+309 & $0.53 \pm 0.07$ & $1.73 \pm 0.07$ & $0.00 \pm 0.07$ & $0.00 \pm 0.07$ & $0.68 \pm 0.09$ & $0.69 \pm 0.09$\\
NGC4666 & $3.48 \pm 0.34$ & $18.04 \pm 0.33$ & $3.50 \pm 0.33$ & $1.91 \pm 0.33$ & $4.21 \pm 0.43$ & $10.20 \pm 0.41$\\
NGC5605 & $1.55 \pm 0.14$ & $6.74 \pm 0.13$ & $0.56 \pm 0.13$ & $0.42 \pm 0.13$ & $0.83 \pm 0.17$ & $3.36 \pm 0.16$\\
UGC4671 & $1.71 \pm 0.10$ & $5.94 \pm 0.10$ & $0.62 \pm 0.10$ & $0.37 \pm 0.10$ & $0.61 \pm 0.13$ & $2.76 \pm 0.12$\\
NGC1187 & $9.04 \pm 0.52$ & $27.90 \pm 0.50$ & $4.92 \pm 0.52$ & $3.52 \pm 0.52$ & $3.49 \pm 0.67$ & $16.18 \pm 0.65$\\
\hline
 \caption{Line strengths of galaxies that have hosted a majority Type Ib SNe. Line strengths are given in $erg/cm^2/sec/A/arcsec^2$. These six lines, H$\beta$, H$\alpha$, [SII]6716, [SII]6731, [OIII]5007, and [NII]6583, are the lines used in our analysis. Galaxies IC 701, NGC 5480, and NGC 5888 are from CALIFA while the rest of the galaxies are from the Bok telescope. The galaxies from CALIFA have higher fluxes than those observed with the Bok because the CAlIFA data is IFU data and observes the entire galaxy at once. The Bok data is of the entire galaxy, but is taken by sweeping a slit across the galaxy for the duration of the observation. This leads to the Bok having less overall light flux than CALIFA observations. }
 \label{hosts_of_Ibs}
 \end{longtable*}

\begin{longtable*}{c c c c c c c c} 
 \hline
 \multicolumn{8}{c}{Hosts of Galaxies that Have Hosted Majority Type II SNe} \\
 \hline
 Galaxy Name & H$\beta$ & H$\alpha$ & [SII]6716 & [SII]6731 & [OIII]5007 & [NII]6583 \\ [0.5ex] 
 \hline\hline
NGC5630 & $107.41 \pm 1.72$ & $298.48 \pm 1.67$ & $66.29 \pm 1.76$ & $45.78 \pm 1.76$ & $242.41 \pm 2.26$ & $70.01 \pm 2.22$\\
NGC5888 & $25.69 \pm 1.19$ & $49.01 \pm 1.18$ & $9.77 \pm 1.08$ & $6.30 \pm 1.08$ & $3.70 \pm 1.37$ & $28.92 \pm 1.35$\\
MCG-02-33-20 & $4.18 \pm 0.30$ & $12.11 \pm 0.30$ & $1.82 \pm 0.34$ & $1.58 \pm 0.34$ & $10.41 \pm 0.40$ & $2.74 \pm 0.40$\\
M+07-20-73 & $1.71 \pm 0.13$ & $7.57 \pm 0.13$ & $1.95 \pm 0.12$ & $0.14 \pm 0.12$ & $0.80 \pm 0.16$ & $4.06 \pm 0.15$\\
NGC2532 & $4.48 \pm 0.26$ & $19.34 \pm 0.26$ & $1.93 \pm 0.31$ & $0.00 \pm 0.31$ & $5.19 \pm 0.36$ & $9.20 \pm 0.36$\\
NGC2939 & $0.50 \pm 0.29$ & $6.04 \pm 0.28$ & $0.49 \pm 0.29$ & $0.48 \pm 0.29$ & $0.23 \pm 0.36$ & $4.57 \pm 0.35$\\
NGC4017 & $4.20 \pm 0.27$ & $13.04 \pm 0.26$ & $2.36 \pm 0.26$ & $1.62 \pm 0.26$ & $5.75 \pm 0.34$ & $4.50 \pm 0.32$\\
NGC4303 & $9.43 \pm 0.43$ & $37.35 \pm 0.42$ & $6.06 \pm 0.43$ & $3.68 \pm 0.43$ & $4.77 \pm 0.54$ & $19.02 \pm 0.53$\\
UGC5434 & $3.95 \pm 1.35$ & $15.62 \pm 1.29$ & $3.99 \pm 1.42$ & $0.00 \pm 1.42$ & $6.85 \pm 1.81$ & $2.99 \pm 1.75$\\
NGC1084 & $84.08 \pm 3.12$ & $342.30 \pm 2.95$ & $60.01 \pm 2.97$ & $39.83 \pm 2.97$ & $87.06 \pm 3.91$ & $136.90 \pm 3.73$\\
NGC1643 & $2.80 \pm 0.24$ & $7.93 \pm 0.23$ & $1.30 \pm 0.23$ & $0.00 \pm 0.23$ & $4.85 \pm 0.30$ & $2.92 \pm 0.29$\\
NGC3034 & $4.26 \pm 0.85$ & $83.09 \pm 0.80$ & $13.74 \pm 0.81$ & $13.44 \pm 0.81$ & $4.07 \pm 1.06$ & $63.74 \pm 1.01$\\
NGC337 & $5.07 \pm 0.36$ & $15.86 \pm 0.34$ & $3.25 \pm 0.34$ & $2.26 \pm 0.34$ & $11.72 \pm 0.45$ & $3.24 \pm 0.43$\\
NGC2993 & $5.05 \pm 0.10$ & $12.02 \pm 0.09$ & $1.65 \pm 0.09$ & $1.23 \pm 0.09$ & $3.95 \pm 0.12$ & $5.35 \pm 0.12$\\
NGC4141 & $0.58 \pm 0.20$ & $3.38 \pm 0.19$ & $0.36 \pm 0.19$ & $0.63 \pm 0.19$ & $2.57 \pm 0.25$ & $1.00 \pm 0.24$\\
NGC3340 & $0.79 \pm 0.08$ & $3.13 \pm 0.08$ & $0.00 \pm 0.08$ & $0.00 \pm 0.08$ & $0.61 \pm 0.10$ & $1.53 \pm 0.10$\\
NGC3627 & $1.36 \pm 0.42$ & $11.74 \pm 0.40$ & $1.93 \pm 0.40$ & $1.50 \pm 0.40$ & $0.58 \pm 0.52$ & $6.88 \pm 0.50$\\
NGC4490 & $9.65 \pm 0.49$ & $39.94 \pm 0.46$ & $4.71 \pm 0.51$ & $5.81 \pm 0.51$ & $39.16 \pm 0.65$ & $5.48 \pm 0.63$\\
NGC7316 & $0.99 \pm 0.07$ & $3.93 \pm 0.07$ & $0.05 \pm 0.07$ & $0.00 \pm 0.07$ & $1.19 \pm 0.09$ & $1.90 \pm 0.09$\\
UGC11946 & $0.70 \pm 0.14$ & $3.25 \pm 0.13$ & $0.05 \pm 0.16$ & $0.00 \pm 0.16$ & $0.54 \pm 0.20$ & $1.94 \pm 0.19$\\
NGC772 & $1.32 \pm 1.06$ & $10.16 \pm 1.11$ & $1.75 \pm 0.74$ & $0.30 \pm 0.74$ & $0.00 \pm 0.97$ & $0.00 \pm 0.98$\\ 
NGC2276 & $7.88 \pm 0.48$ & $30.64 \pm 0.45$ & $5.27 \pm 0.50$ & $4.31 \pm 0.50$ & $11.67 \pm 0.64$ & $12.99 \pm 0.62$\\
NGC3184 & $6.25 \pm 1.57$ & $19.56 \pm 1.53$ & $1.34 \pm 1.32$ & $2.54 \pm 1.32$ & $2.77 \pm 1.74$ & $7.80 \pm 1.66$\\ 
NGC3631 & $3.45 \pm 1.66$ & $19.08 \pm 1.59$ & $3.32 \pm 1.50$ & $1.50 \pm 1.50$ & $3.12 \pm 1.98$ & $10.31 \pm 1.89$\\
NGC4258 & $9.56 \pm 2.04$ & $32.25 \pm 2.00$ & $4.15 \pm 2.86$ & $20.22 \pm 2.86$ & $26.32 \pm 3.14$ & $40.39 \pm 3.15$\\
NGC5457 & $6.70 \pm 1.79$ & $10.57 \pm 1.69$ & $2.71 \pm 1.70$ & $0.04 \pm 1.70$ & $0.00 \pm 2.24$ & $4.92 \pm 2.14$\\
NGC493 & $3.05 \pm 0.52$ & $13.22 \pm 0.50$ & $2.51 \pm 0.53$ & $1.88 \pm 0.53$ & $10.13 \pm 0.67$ & $2.17 \pm 0.65$\\
MCG-02-07-10 & $33.82 \pm 0.42$ & $90.59 \pm 0.40$ & $9.51 \pm 0.44$ & $8.34 \pm 0.44$ & $186.71 \pm 0.56$ & $9.96 \pm 0.54$\\
UGC5695 & $2.63 \pm 0.25$ & $6.93 \pm 0.24$ & $1.76 \pm 0.24$ & $1.08 \pm 0.24$ & $3.88 \pm 0.31$ & $2.18 \pm 0.30$\\
NGC4157 & $5.74 \pm 1.01$ & $22.42 \pm 0.97$ & $3.40 \pm 1.02$ & $5.28 \pm 1.02$ & $5.39 \pm 1.30$ & $13.46 \pm 1.26$\\
NGC3938 & $5.48 \pm 0.78$ & $20.70 \pm 0.74$ & $3.02 \pm 0.78$ & $3.31 \pm 0.78$ & $5.16 \pm 1.01$ & $12.42 \pm 0.97$\\
NGC4273 & $9.17 \pm 0.24$ & $33.77 \pm 0.23$ & $6.30 \pm 0.24$ & $4.40 \pm 0.24$ & $8.11 \pm 0.31$ & $14.70 \pm 0.30$\\
NGC5020 & $7.26 \pm 0.50$ & $15.61 \pm 0.48$ & $1.52 \pm 0.50$ & $1.13 \pm 0.50$ & $4.34 \pm 0.65$ & $8.81 \pm 0.63$\\
\hline
 \caption{Line strengths of galaxies that have hosted a majority Type II SNe. Line strengths are given in $erg/cm^2/sec/A/arcsec^2$. These six lines, H$\beta$, H$\alpha$, [SII]6716, [SII]6731, [OIII]5007, and [NII]6583, are the lines used in our analysis. Galaxy NGC 5630 is from CALIFA while the rest of the galaxies are from the Bok telescope. The galaxies from CALIFA have higher fluxes than those observed with the Bok because the CAlIFA data is IFU data and observes the entire galaxy at once. The Bok data is of the entire galaxy, but is taken by sweeping a slit across the galaxy for the duration of the observation. This leads to the Bok having less overall light flux than CALIFA observations. }
 \label{hosts_of_IIs}
 \end{longtable*}

\begin{longtable*}{c c c c c c c c} 
 \hline
 \multicolumn{8}{c}{Hosts of Galaxies that Have Hosted Majority Type Ic SNe} \\
 \hline
 Galaxy Name & H$\beta$ & H$\alpha$ & [SII]6716 & [SII]6731 & [OIII]5007 & [NII]6583 \\ [0.5ex] 
 \hline\hline
NGC4051 & $4.20 \pm 0.29$ & $17.60 \pm 0.28$ & $2.50 \pm 0.29$ & $2.05 \pm 0.29$ & $6.44 \pm 0.37$ & $9.76 \pm 0.36$\\
NGC3810 & $8.85 \pm 0.39$ & $29.22 \pm 0.37$ & $4.36 \pm 0.37$ & $2.57 \pm 0.37$ & $3.85 \pm 0.49$ & $13.62 \pm 0.46$\\
NGC4568 & $3.61 \pm 0.43$ & $18.96 \pm 0.41$ & $2.79 \pm 0.41$ & $1.13 \pm 0.41$ & $1.26 \pm 0.54$ & $9.26 \pm 0.52$\\
NGC5374 & $2.77 \pm 0.15$ & $8.43 \pm 0.14$ & $0.76 \pm 0.14$ & $0.64 \pm 0.14$ & $1.70 \pm 0.18$ & $4.20 \pm 0.17$\\
UGC4798 & $0.52 \pm 0.19$ & $1.52 \pm 0.20$ & $0.00 \pm 0.40$ & $0.00 \pm 0.40$ & $1.69 \pm 0.35$ & $1.11 \pm 0.37$\\
\hline
 \caption{Line strengths of galaxies that have hosted a majority Type Ic SNe. Line strengths are given in $erg/cm^2/sec/A/arcsec^2$. These six lines, H$\beta$, H$\alpha$, [SII]6716, [SII]6731, [OIII]5007, and [NII]6583, are the lines used in our analysis.}
 \label{hosts_of_Ics}
 \end{longtable*}

\section{Sample Classification of Host Galaxies}

\begin{longtable*}{c c}
 \hline
 Galaxy Name & Sample \\
 \hline\hline
  IC701 & Sample SF \\
  NGC7479 & Low H$\beta$ S/N \\
  NGC2770 & Sample Comp. \\
  NGC7714 & Sample Comp. \\
  NGC3690 & Sample Comp. \\
  NGC5480 & Sample SF \\
  NGC6118 & Low H$\beta$ S/N \\
  UGC7020 & Sample Comp. \\
  IC1099 & Sample Comp. \\
  IC1216 & Sample Comp. \\
  NGC5278 & Sample Comp. \\
  NGC3147 & Low H$\beta$ S/N \\
  NGC3506 & Sample Comp. \\
  NGC5806 & Low H$\beta$ S/N \\
  NGC7624 & Sample SF \\
  IC900 & Sample Comp. \\
  KUG1313+309 & Sample Comp. \\
  NGC4666 & Sample Comp. \\
  NGC5605 & Sample Comp. \\
  UGC4671 & Sample SF \\
  NGC1187 & Sample Comp. \\
  NGC5630 & Sample SF \\
  NGC5888 & Sample Comp. \\
  MCG-02-33-20 & Sample SF \\
  M+07-20-73 & Sample Comp. \\
  NGC2532 & Sample Comp. \\
  NGC2939 & Low H$\beta$ S/N \\
  NGC4017 & Sample Comp. \\
  NGC4303 & Sample Comp. \\
  UGC5434 & Low H$\beta$ S/N \\
  NGC1084 & Sample Comp. \\
  NGC1643 & Sample Comp. \\
  NGC3034 & Sample Comp. \\
  NGC337 & Sample SF \\
  NGC2993 & Sample Comp. \\
  NGC4141 & Low H$\beta$ S/N \\
  NGC3340 & Sample Comp. \\
  NGC3627 & Low H$\beta$ S/N \\
  NGC4490 & Sample SF \\
  NGC7316 & Sample Comp. \\
  UGC11946 & Sample Comp. \\
  NGC772 & Low H$\beta$ S/N \\
  NGC2276 & Sample Comp. \\
  NGC3184 & Low H$\beta$ S/N \\
  NGC3631 & Low H$\beta$ S/N \\
  NGC4258 & Sample SF \\
  NGC5457 & Low H$\beta$ S/N \\
  NGC493 & Sample SF \\
  MCG-02-07-10 & Sample Comp. \\
  UGC5695 & Sample SF \\
  NGC4157 & Sample Comp. \\
  NGC3938 & Sample Comp. \\
  NGC4273 & Sample Comp. \\
  NGC5020 & Sample Comp. \\
  NGC4051 & Sample Comp. \\
  NGC3810 & Sample SF \\
  NGC4568 & Sample SF \\
  NGC5374 & Sample Comp. \\
  UGC4798 & Low H$\beta$ S/N \\
  \hline
  \caption{Host galaxies and classification of spectra from the BPT diagram. %Galaxies that are members of the {\tt Sample SF} and {\tt Sample Comp.} samples are also members of the {\tt Full Sample}. 
  Galaxies not in {\tt Sample SF} or {\tt Sample Comp.} are excluded due to low S/N on H$\beta$. There are 13 galaxies in the {\tt Sample SF}, 33 galaxies in the {\tt Sample Comp.}, and 13 galaxies with low H$\beta$ S/N, for a total of 59 galaxies in the {\tt Full Sample}.}
  \label{sample_galaxies}
 \end{longtable*}

%\pagebreak
\section{Gas-phase properties of Host Galaxies}

\begin{longtable*}{c c c c}
 \hline
 Galaxy Name & Oxygen Abundance & Ionization Parameter & Gas Pressure \\
 \hline\hline
IC701 & $8.65_{-0.06}^{+0.35}$ & $-3.37_{-0.26}^{+0.11}$ & $4.66_{-0.23}^{+1.16}$ \\
NGC5630 & $8.42_{-0.06}^{+0.12}$ & $-3.63_{-0.16}^{+0.11}$ & $4.20_{-0.35}^{+2.08}$ \\
NGC5888 & $8.89_{-0.12}^{+0.24}$ & $-3.26_{-0.21}^{+0.11}$ & $6.63_{-0.69}^{+0.35}$ \\
%NGC6373 & $8.42_{-0.12}^{+0.06}$ & $-3.68_{-0.16}^{+0.11}$ & $4.20_{-0.35}^{+2.08}$ \\
MCG-02-33-20 & $8.42_{-0.12}^{+0.59}$ & $-3.42_{-0.32}^{+0.26}$ & $5.71_{-1.04}^{+1.27}$ \\
M+07-20-73 & $8.89_{-0.18}^{+0.12}$ & $-3.00_{-0.21}^{+0.16}$ & $5.71_{-1.16}^{+0.35}$ \\
NGC2532 & $9.01_{-0.12}^{+0.18}$ & $-2.53_{-0.16}^{+0.37}$ & $6.63_{-0.69}^{+0.35}$ \\
NGC2939 & $9.24_{-1.77}^{+0.24}$ & $-2.95_{-0.68}^{+0.63}$ & $8.37_{-3.47}^{+0.35}$ \\
NGC4017 & $8.65_{-0.12}^{+0.41}$ & $-3.26_{-0.32}^{+0.16}$ & $5.82_{-1.27}^{+0.46}$ \\
NGC4303 & $8.89_{-0.12}^{+0.18}$ & $-3.05_{-0.16}^{+0.11}$ & $6.28_{-0.93}^{+0.35}$ \\
UGC5434 & $8.24_{-0.83}^{+0.59}$ & $-2.79_{-0.84}^{+0.53}$ & $5.13_{-0.35}^{+2.66}$ \\
NGC7479 & $9.01_{-0.89}^{+0.12}$ & $-3.37_{-0.37}^{+0.79}$ & $7.33_{-2.32}^{+0.58}$ \\
NGC1084 & $8.77_{-0.06}^{+0.24}$ & $-3.16_{-0.21}^{+0.11}$ & $5.82_{-1.27}^{+0.35}$ \\
NGC2770 & $8.65_{-0.18}^{+0.47}$ & $-3.21_{-0.47}^{+0.26}$ & $6.75_{-1.62}^{+0.93}$ \\
NGC1643 & $8.89_{-0.18}^{+0.18}$ & $-2.84_{-0.26}^{+0.26}$ & $5.82_{-1.16}^{+0.58}$ \\
NGC3034 & $9.01_{-0.41}^{+0.18}$ & $-3.26_{-0.42}^{+0.26}$ & $7.21_{-1.62}^{+0.69}$ \\
NGC337 & $8.36_{-0.12}^{+0.12}$ & $-3.63_{-0.21}^{+0.21}$ & $4.43_{-0.12}^{+2.08}$ \\
NGC4051 & $8.89_{-0.12}^{+0.30}$ & $-3.05_{-0.21}^{+0.16}$ & $6.75_{-0.58}^{+0.46}$ \\
NGC2993 & $8.89_{-0.12}^{+0.18}$ & $-3.00_{-0.16}^{+0.05}$ & $6.28_{-0.81}^{+0.35}$ \\
NGC3810 & $8.89_{-0.12}^{+0.18}$ & $-3.00_{-0.16}^{+0.11}$ & $6.17_{-1.04}^{+0.35}$ \\
NGC4568 & $8.89_{-0.18}^{+0.24}$ & $-2.95_{-0.21}^{+0.26}$ & $6.28_{-1.51}^{+0.46}$ \\
NGC7714 & $8.89_{-0.12}^{+0.18}$ & $-2.95_{-0.11}^{+0.11}$ & $6.17_{-1.16}^{+0.23}$ \\
NGC3690 & $8.89_{-0.06}^{+0.24}$ & $-3.21_{-0.16}^{+0.11}$ & $6.63_{-0.35}^{+0.23}$ \\
NGC4141 & $8.54_{-0.65}^{+0.53}$ & $-3.21_{-0.47}^{+0.74}$ & $6.40_{-1.51}^{+1.27}$ \\
NGC5480 & $8.83_{-0.12}^{+0.18}$ & $-3.21_{-0.16}^{+0.11}$ & $6.05_{-1.16}^{+0.23}$ \\
NGC3340 & $8.95_{-0.18}^{+0.12}$ & $-2.00_{-0.53}^{+0.05}$ & $8.60_{-1.04}^{inf}$ \\
NGC3627 & $8.95_{-0.77}^{+0.18}$ & $-3.26_{-0.42}^{+0.58}$ & $6.75_{-1.85}^{+1.04}$ \\
NGC4490 & $8.18_{-0.06}^{+0.12}$ & $-3.79_{-0.16}^{+0.26}$ & $7.33_{-0.81}^{+0.35}$ \\
NGC7316 & $8.95_{-0.12}^{+0.12}$ & $-2.00_{-0.47}^{+0.05}$ & $8.60_{-0.58}^{inf}$ \\
NGC6118 & $8.95_{-1.36}^{+0.06}$ & $-3.26_{-0.42}^{+0.89}$ & $6.05_{-1.27}^{+1.74}$ \\
UGC7020 & $8.89_{-1.06}^{+0.24}$ & $-2.26_{-0.95}^{+0.11}$ & $8.60_{-3.59}^{+0.46}$ \\
IC1099 & $8.95_{-0.71}^{+0.18}$ & $-3.26_{-0.42}^{+0.74}$ & $7.33_{-2.20}^{+0.69}$ \\
IC1216 & $8.89_{-0.41}^{+0.24}$ & $-2.00_{-0.95}^{+0.11}$ & $8.60_{-3.24}^{+0.35}$ \\
NGC5278 & $9.24_{-0.18}^{inf}$ & $-2.00_{-0.53}^{+0.05}$ & $8.60_{-0.58}^{inf}$ \\
NGC5374 & $9.01_{-0.18}^{+0.24}$ & $-2.74_{-0.21}^{+0.16}$ & $6.86_{-0.35}^{+0.46}$ \\
NGC3147 & $7.06_{0.18}^{+1.59}$ & $-2.00_{-1.47}^{+0.16}$ & $8.60_{-3.82}^{+0.58}$ \\
NGC3506 & $8.95_{-0.12}^{+0.06}$ & $-2.00_{-0.42}^{inf}$ & $8.60_{-0.58}^{inf}$ \\
NGC5806 & $8.12_{-0.71}^{+0.83}$ & $-3.00_{-0.68}^{+0.68}$ & $8.60_{-3.71}^{+0.69}$ \\
NGC7624 & $8.71_{-0.06}^{+0.47}$ & $-2.89_{-0.47}^{+0.11}$ & $6.98_{-0.12}^{+1.16}$ \\
UGC11946 & $9.01_{-1.12}^{+0.18}$ & $-2.00_{-1.00}^{+0.11}$ & $8.60_{-3.24}^{+0.23}$ \\
IC900 & $8.89_{-0.12}^{+0.12}$ & $-2.00_{-0.53}^{+0.05}$ & $8.60_{-0.93}^{inf}$ \\
KUG1313+309 & $8.89_{-0.24}^{+0.24}$ & $-2.00_{-0.74}^{+0.05}$ & $8.60_{-2.89}^{+0.23}$ \\
NGC4666 & $8.89_{-0.18}^{+0.18}$ & $-3.21_{-0.16}^{+0.21}$ & $6.28_{-1.39}^{+0.46}$ \\
NGC5605 & $9.01_{-0.24}^{+0.24}$ & $-2.68_{-0.21}^{+0.32}$ & $6.86_{-0.58}^{+0.69}$ \\
NGC772 & $7.59_{-0.35}^{+0.65}$ & $-2.00_{-1.63}^{+0.21}$ & $4.20_{-0.58}^{+3.71}$ \\
UGC4671 & $8.95_{-0.12}^{+0.24}$ & $-2.79_{-0.16}^{+0.21}$ & $6.63_{-0.81}^{+0.35}$ \\
NGC2276 & $8.77_{-0.12}^{+0.35}$ & $-3.21_{-0.26}^{+0.11}$ & $6.40_{-1.51}^{+0.23}$ \\
UGC4798 & $9.24_{-1.77}^{+0.18}$ & $-2.89_{-0.74}^{+0.63}$ & $8.60_{-3.71}^{+0.58}$ \\
NGC3184 & $8.77_{-0.71}^{+0.35}$ & $-3.00_{-0.58}^{+0.58}$ & $6.98_{-1.97}^{+0.93}$ \\
NGC3631 & $9.01_{-1.48}^{+0.06}$ & $-3.16_{-0.53}^{+0.79}$ & $6.05_{-1.16}^{+1.85}$ \\
NGC4258 & $9.01_{-0.47}^{+0.18}$ & $-4.00_{0.16}^{+1.11}$ & $7.79_{-1.85}^{+0.35}$ \\
NGC5457 & $8.89_{-1.30}^{+0.18}$ & $-3.26_{-0.42}^{+0.89}$ & $6.17_{-1.27}^{+1.74}$ \\
NGC493 & $8.30_{-0.18}^{+0.24}$ & $-3.74_{-0.16}^{+0.53}$ & $5.36_{-0.69}^{+1.62}$ \\
NGC1187 & $8.89_{-0.12}^{+0.24}$ & $-3.21_{-0.16}^{+0.16}$ & $6.63_{-0.69}^{+0.35}$ \\
MCG-02-07-10 & $8.12_{-0.06}^{+0.12}$ & $-3.53_{-0.11}^{+0.16}$ & $6.75_{-1.85}^{+0.23}$ \\
UGC5695 & $8.54_{-0.12}^{+0.47}$ & $-3.58_{-0.26}^{+0.26}$ & $5.01_{-0.46}^{+1.39}$ \\
NGC4157 & $8.89_{-0.24}^{+0.30}$ & $-3.47_{-0.37}^{+0.37}$ & $7.44_{-1.39}^{+0.46}$ \\
NGC3938 & $8.89_{-0.18}^{+0.30}$ & $-3.26_{-0.37}^{+0.26}$ & $7.09_{-1.16}^{+0.81}$ \\
NGC4273 & $8.83_{-0.12}^{+0.18}$ & $-3.26_{-0.16}^{+0.11}$ & $6.05_{-1.27}^{+0.23}$ \\
NGC5020 & $9.01_{-0.24}^{+0.24}$ & $-2.84_{-0.26}^{+0.37}$ & $7.21_{-0.58}^{+0.81}$ \\
 \hline
  \caption{Gas-phase properties of host galaxies measured from integrated spectra and assuming HII region ionizing spectrum.}
 \end{longtable*}

\begin{longtable*}{c c c c}
 \hline
 Galaxy Name & Oxygen Abundance & Ionization Parameter & Gas Pressure \\
 \hline\hline
NGC5888 & $8.98_{-0.32}^{+0.16}$ & $-2.77_{-0.29}^{+2.00}$ & $5.77_{-1.26}^{+0.63}$ \\
%NGC6373 & $8.34_{-0.16}^{+0.16}$ & $-3.63_{-0.29}^{+0.86}$ & $4.20_{0.31}^{+2.20}$ \\
M+07-20-73 & $9.14_{-0.32}^{inf}$ & $-2.49_{-0.29}^{+1.71}$ & $4.20_{-inf}^{+1.57}$ \\
NGC2532 & $9.30_{-0.16}^{inf}$ & $-2.20_{-0.29}^{+1.43}$ & $4.20_{-inf}^{+1.57}$ \\
NGC2939 & $7.86_{-0.48}^{+1.12}$ & $-2.20_{-1.14}^{+1.43}$ & $8.60_{-3.77}^{+-0.63}$ \\
NGC4017 & $8.82_{-0.32}^{+0.16}$ & $-2.77_{-0.29}^{+2.00}$ & $6.09_{-1.57}^{+0.63}$ \\
NGC4303 & $8.98_{-0.16}^{+0.16}$ & $-2.77_{0.00}^{+2.00}$ & $4.20_{-inf}^{+1.89}$ \\
UGC5434 & $8.18_{-0.80}^{+0.64}$ & $-2.20_{-1.14}^{+1.43}$ & $4.20_{0.63}^{+3.77}$ \\
NGC7479 & $8.98_{-1.12}^{+0.16}$ & $-2.77_{-0.57}^{+2.00}$ & $7.97_{-3.14}^{+0.31}$ \\
NGC1084 & $8.82_{-0.16}^{+0.16}$ & $-2.77_{-0.29}^{+2.00}$ & $5.77_{-1.26}^{+0.63}$ \\
NGC2770 & $8.82_{-0.32}^{+0.32}$ & $-2.77_{-0.57}^{+2.29}$ & $7.97_{-1.57}^{+0.31}$ \\
NGC1643 & $9.14_{-0.32}^{inf}$ & $-2.20_{-0.57}^{+1.43}$ & $4.20_{-inf}^{+2.20}$ \\
NGC3034 & $9.14_{-0.48}^{inf}$ & $-2.77_{-0.29}^{+2.00}$ & $7.03_{-2.20}^{+0.63}$ \\
NGC4051 & $9.14_{-0.32}^{inf}$ & $-2.49_{-0.29}^{+1.71}$ & $6.40_{-1.89}^{+0.31}$ \\
NGC2993 & $8.98_{-0.16}^{+0.16}$ & $-2.49_{-0.29}^{+1.71}$ & $4.20_{0.31}^{+2.20}$ \\
NGC7714 & $8.98_{-0.16}^{+0.16}$ & $-2.49_{-0.29}^{+1.71}$ & $4.51_{0.00}^{+1.57}$ \\
NGC3690 & $8.98_{-0.16}^{+0.16}$ & $-2.77_{-0.29}^{+2.00}$ & $6.71_{-0.94}^{+0.63}$ \\
NGC4141 & $8.82_{-1.12}^{+0.32}$ & $-0.20_{-3.14}^{+-0.29}$ & $8.60_{-3.46}^{+-0.31}$ \\
NGC3340 & $9.30_{-0.32}^{inf}$ & $-2.20_{-0.29}^{+1.14}$ & $8.60_{-3.14}^{inf}$ \\
NGC3627 & $8.98_{-1.12}^{+0.16}$ & $-2.77_{-0.57}^{+2.00}$ & $6.71_{-1.89}^{+1.26}$ \\
NGC7316 & $9.30_{-0.32}^{inf}$ & $-2.20_{-0.29}^{+0.57}$ & $8.60_{-1.89}^{inf}$ \\
NGC6118 & $8.66_{-1.28}^{+0.32}$ & $-2.49_{-1.14}^{+1.71}$ & $5.77_{-0.94}^{+2.20}$ \\
UGC7020 & $9.30_{-1.92}^{inf}$ & $-2.20_{-0.57}^{+1.43}$ & $4.20_{0.63}^{+3.77}$ \\
IC1099 & $8.98_{-0.96}^{+0.16}$ & $-2.77_{-0.57}^{+2.00}$ & $8.60_{-3.46}^{+-0.31}$ \\
IC1216 & $9.30_{-0.96}^{inf}$ & $-1.91_{-0.57}^{+1.14}$ & $4.20_{0.31}^{+3.77}$ \\
NGC5278 & $9.30_{-0.16}^{inf}$ & $-2.20_{-0.57}^{+0.29}$ & $8.60_{-0.63}^{inf}$ \\
NGC5374 & $9.30_{-0.32}^{inf}$ & $-2.20_{-0.57}^{+1.43}$ & $6.40_{-1.89}^{+0.63}$ \\
NGC3147 & $7.06_{0.16}^{+1.44}$ & $-1.63_{-1.43}^{+1.14}$ & $4.20_{0.63}^{+3.77}$ \\
NGC3506 & $9.30_{-0.32}^{inf}$ & $-2.20_{-0.29}^{+0.57}$ & $8.60_{-1.57}^{inf}$ \\
NGC5806 & $7.70_{-0.32}^{+1.28}$ & $-2.20_{-1.43}^{+1.43}$ & $8.60_{-3.77}^{+-0.63}$ \\
UGC11946 & $9.30_{-1.92}^{inf}$ & $-1.91_{-0.86}^{+1.14}$ & $4.20_{0.31}^{+3.77}$ \\
IC900 & $9.30_{-0.32}^{inf}$ & $-2.20_{-0.29}^{+1.14}$ & $8.60_{-3.77}^{+-0.31}$ \\
KUG1313+309 & $9.30_{-0.64}^{inf}$ & $-1.91_{-0.57}^{+1.14}$ & $4.20_{0.31}^{+3.77}$ \\
NGC4666 & $8.98_{-0.32}^{+0.16}$ & $-2.77_{-0.29}^{+2.00}$ & $4.83_{-0.31}^{+1.57}$ \\
NGC5605 & $9.30_{-0.32}^{inf}$ & $-2.20_{-0.57}^{+1.43}$ & $4.20_{0.31}^{+2.83}$ \\
NGC772 & $7.38_{-0.16}^{+0.96}$ & $-1.34_{-2.00}^{+0.86}$ & $4.20_{0.63}^{+3.77}$ \\
NGC2276 & $8.82_{-0.16}^{+0.32}$ & $-2.77_{-0.29}^{+2.00}$ & $6.40_{-1.57}^{+0.63}$ \\
UGC4798 & $7.06_{0.32}^{+1.92}$ & $-2.20_{-1.14}^{+1.43}$ & $8.60_{-3.77}^{+-0.63}$ \\
NGC3184 & $9.14_{-1.28}^{+0.00}$ & $-2.49_{-0.86}^{+1.71}$ & $8.60_{-3.46}^{+-0.31}$ \\
NGC3631 & $8.18_{-0.80}^{+0.80}$ & $-2.49_{-0.86}^{+1.71}$ & $4.20_{0.63}^{+3.77}$ \\
NGC5457 & $8.82_{-1.44}^{+0.16}$ & $-2.49_{-0.86}^{+1.71}$ & $4.20_{0.63}^{+3.77}$ \\
NGC1187 & $8.98_{-0.16}^{+0.16}$ & $-2.77_{-0.29}^{+2.00}$ & $5.46_{-0.94}^{+0.94}$ \\
MCG-02-07-10 & $8.50_{-0.32}^{+0.32}$ & $-3.34_{0.00}^{+2.86}$ & $7.03_{-0.94}^{+0.63}$ \\
NGC4157 & $8.98_{-0.48}^{inf}$ & $-3.06_{-0.29}^{+2.57}$ & $7.97_{-0.94}^{inf}$ \\
NGC3938 & $8.98_{-0.32}^{inf}$ & $-2.77_{-0.29}^{+2.00}$ & $7.34_{-2.20}^{+0.63}$ \\
NGC4273 & $8.82_{-0.16}^{+0.16}$ & $-2.49_{-0.57}^{+1.71}$ & $5.77_{-1.26}^{+0.63}$ \\
NGC5020 & $9.30_{-0.32}^{inf}$ & $-2.49_{-0.29}^{+1.71}$ & $4.20_{0.31}^{+3.14}$ \\
  \hline
  \caption{Gas-phase properties of host galaxies measured from integrated spectra and assuming NLR region ionizing spectrum. Galaxies in {\tt Sample SF} are not included in this table since they were only modelled as HII regions.}
 \end{longtable*}

\section{H$\alpha$ equivalent widths of host galaxies}

\begin{longtable*}{c c}
 \hline
 Galaxy Name & H$\alpha$ equivalent width \\
 \hline\hline
 IC701 & $13.55 \pm 0.29$ \\
 NGC5630 & $20.74 \pm 0.56$ \\
 NGC5888 & $2.50 \pm 0.08$ \\
 %NGC6373 & $12.98 \pm 0.38$ \\
 MCG-02-33-20 & $6.37 \pm 1.22$ \\
 M+07-20-73 & $2.42 \pm 0.77$ \\
 NGC2532 & $6.56 \pm 0.78$ \\
 NGC2939 & $2.33 \pm 0.92$ \\
 NGC4017 & $10.18 \pm 1.59$ \\
 NGC4303 & $6.86 \pm 0.43$ \\
 UGC5434 & $3.96 \pm 0.52$ \\
 NGC7479 & $2.28 \pm 0.41$ \\
 NGC1084 & $9.65 \pm 0.06$ \\
 NGC2770 & $4.18 \pm 0.28$ \\
 NGC1643 & $7.51 \pm 1.93$ \\
 NGC3034 & $6.21 \pm 0.15$ \\
 NGC337 & $9.46 \pm 1.22$ \\
 NGC4051 & $5.65 \pm 0.67$ \\
 NGC2993 & $27.61 \pm 4.69$ \\
 NGC3810 & $6.78 \pm 0.47$ \\
 NGC4568 & $3.80 \pm 0.41$ \\
 NGC7714 & $25.99 \pm 0.22$ \\
 NGC3690 & $24.02 \pm 0.25$ \\
 NGC4141 & $19.04 \pm 11.51$ \\
 NGC5480 & $12.49 \pm 0.23$ \\
 NGC3340 & $5.57 \pm 3.76$ \\
 NGC3627 & $2.76 \pm 0.49$ \\
 NGC4490 & $13.50 \pm 0.70$ \\
 NGC7316 & $8.52 \pm 4.45$ \\
 NGC6118 & $3.68 \pm 2.30$ \\
 UGC7020 & $4.22 \pm 5.21$ \\
 IC1099 & $2.03 \pm 3.09$ \\
 IC1216 & $7.85 \pm 6.24$ \\
 NGC5278 & $4.69 \pm 1.05$ \\
 NGC5374 & $7.11 \pm 1.74$ \\
 NGC3147 & $2.21 \pm 0.52$ \\
 NGC3506 & $7.62 \pm 2.16$ \\
 NGC5806 & $2.06 \pm 0.49$ \\
 NGC7624 & $7.56 \pm 1.08$ \\
 UGC11946 & $3.93 \pm 2.51$ \\
 IC900 & $6.18 \pm 2.57$ \\
 KUG1313+309 & $4.64 \pm 5.53$ \\
 NGC4666 & $6.19 \pm 0.72$ \\
 NGC5605 & $6.04 \pm 1.85$ \\
 NGC772 & $1.99 \pm 0.61$ \\
 UGC4671 & $7.95 \pm 2.76$ \\
 NGC2276 & $12.16 \pm 0.84$ \\
 UGC4798 & $2.97 \pm 4.45$ \\
 NGC3184 & $4.06 \pm 0.50$ \\
 NGC3631 & $5.80 \pm 0.67$ \\
 NGC4258 & $2.15 \pm 0.15$ \\
 NGC5457 & $2.35 \pm 0.46$ \\
 NGC493 & $7.35 \pm 1.18$ \\
 NGC1187 & $5.03 \pm 0.38$ \\
 MCG-02-07-10 & $29.85 \pm 0.68$ \\
 UGC5695 & $4.68 \pm 1.48$ \\
 NGC4157 & $4.06 \pm 0.40$ \\
 NGC3938 & $4.15 \pm 0.41$ \\
 NGC4273 & $9.38 \pm 0.59$ \\
 NGC5020 & $4.55 \pm 0.61$ \\
 \hline
  \caption{H$\alpha$ equivalent widths for all galaxies. }
  \label{Halph_eq_widths_table}
\end{longtable*}

\section{pPXF plots}

\begin{figure*}[h!]
\centering
\includegraphics[width=7in]{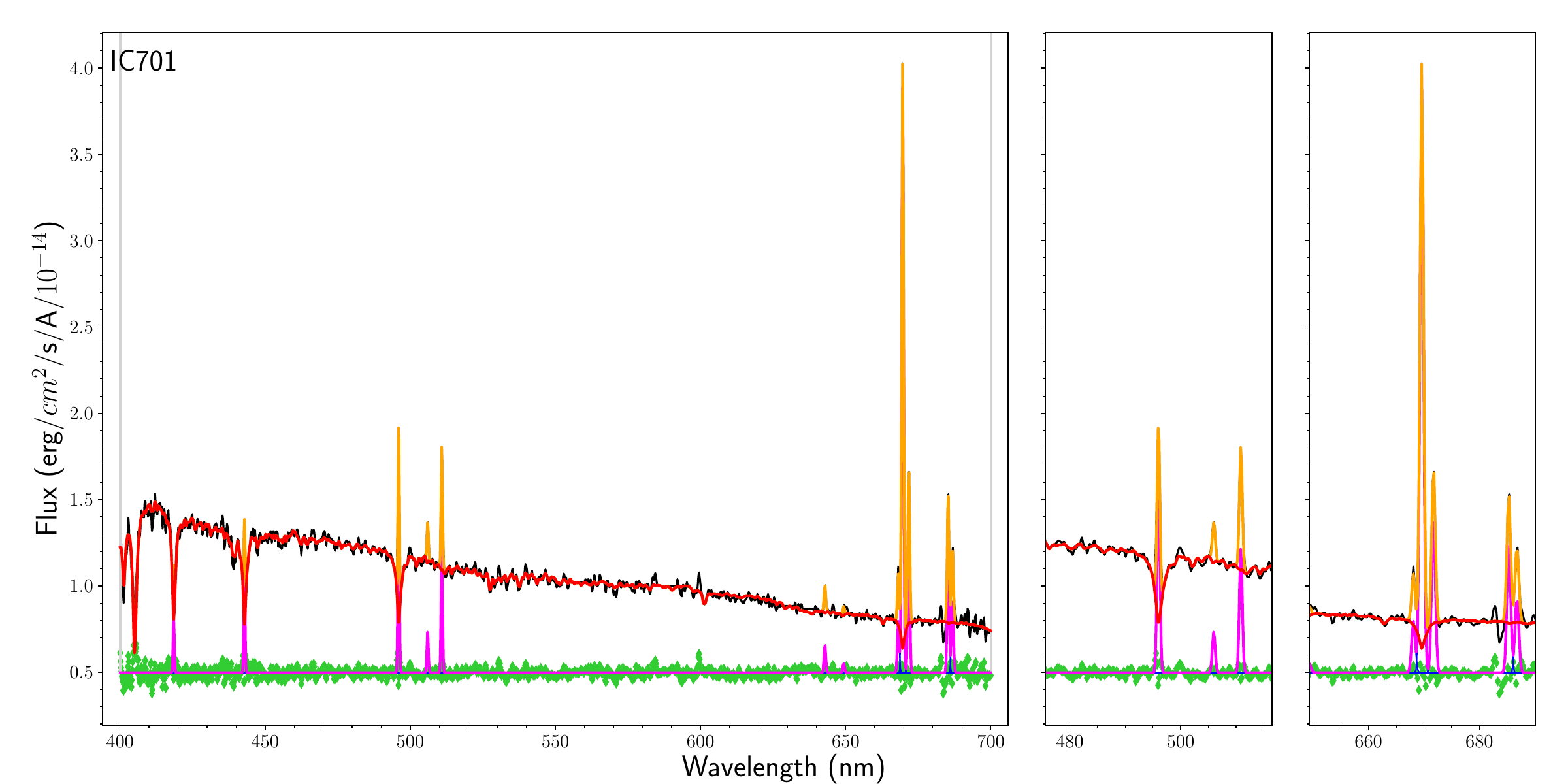}

\includegraphics[width=7in]{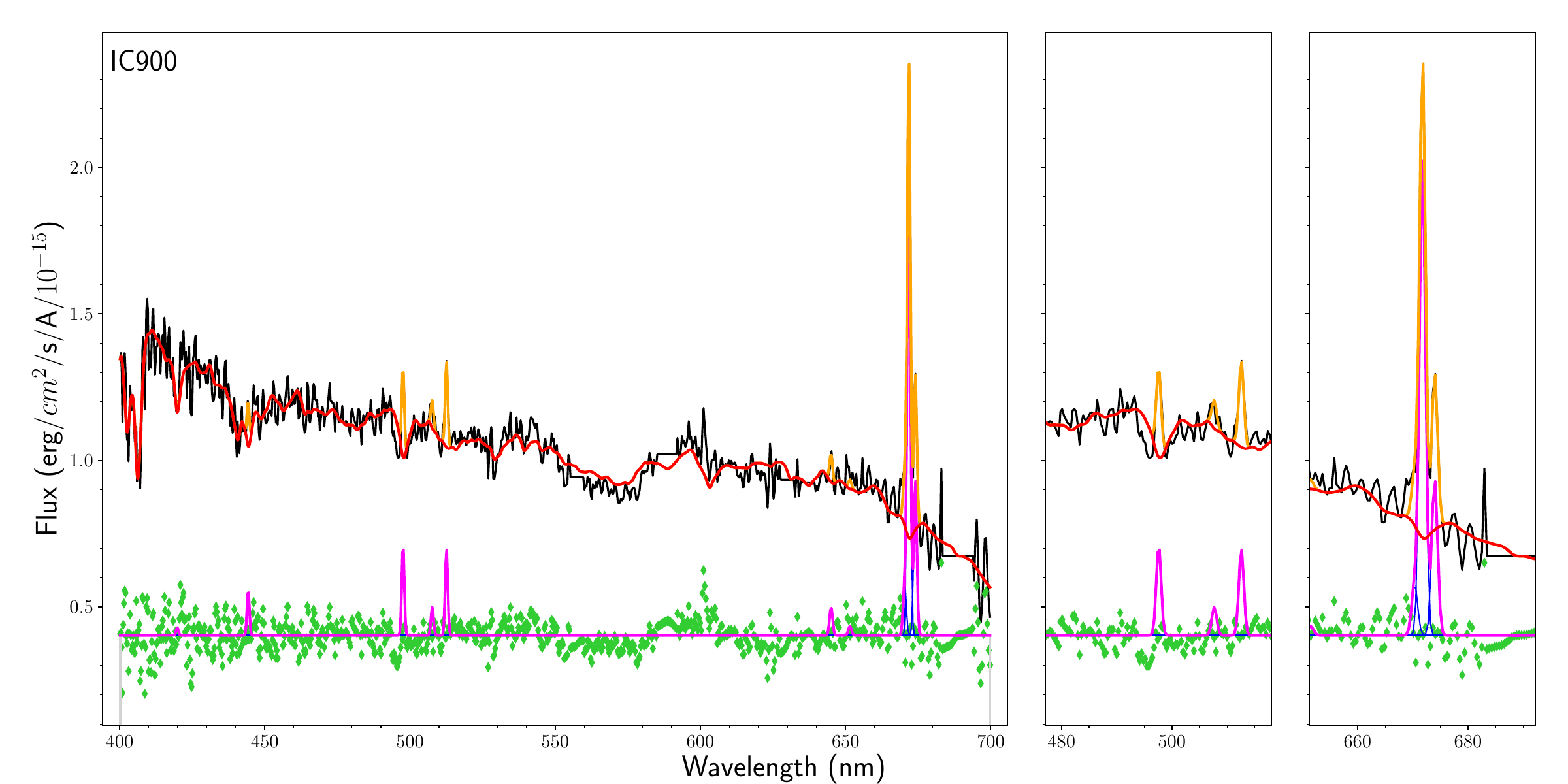}
\end{figure*}
\begin{figure*}
\includegraphics[width=7in]{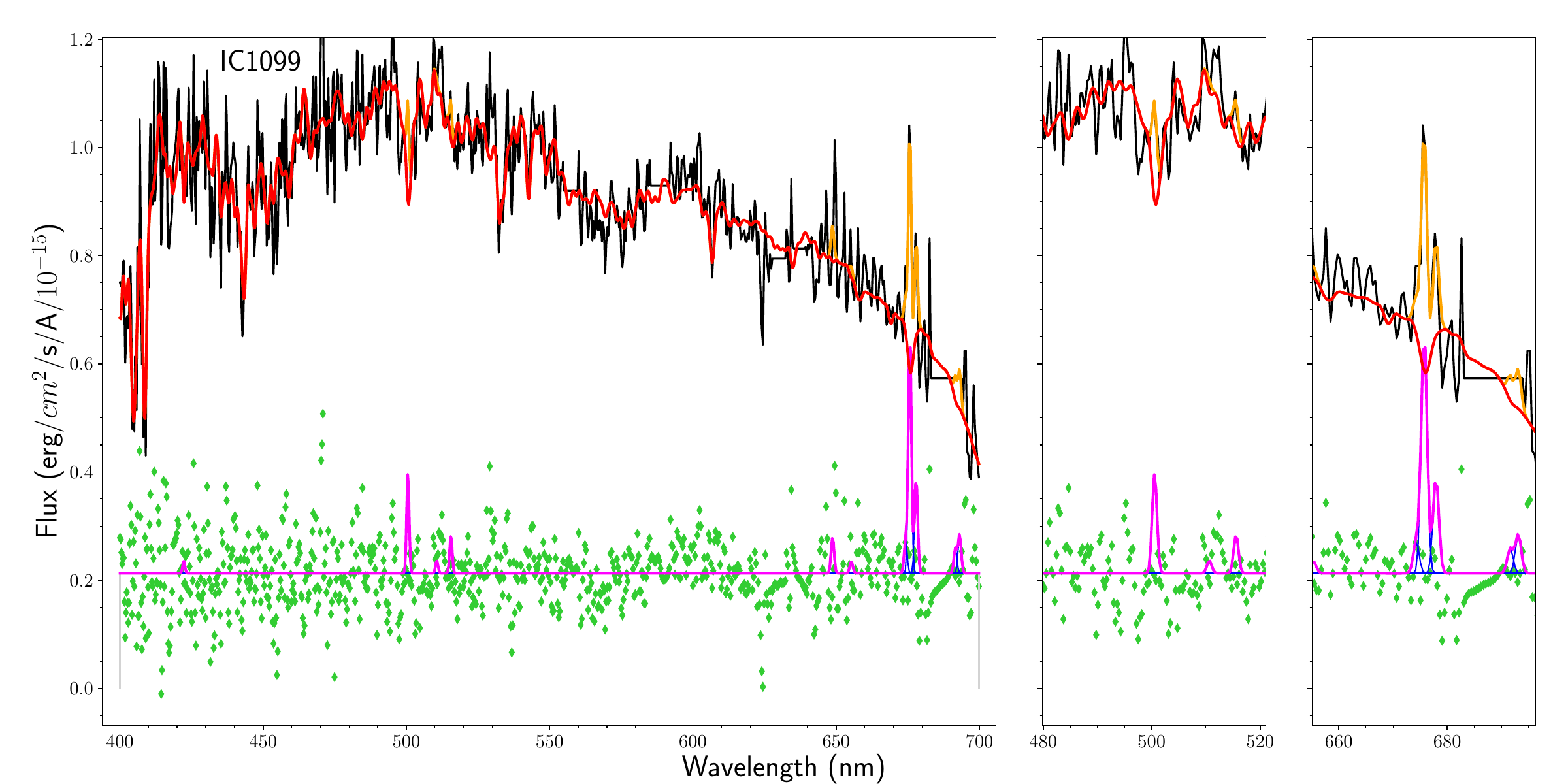}

\includegraphics[width=7in]{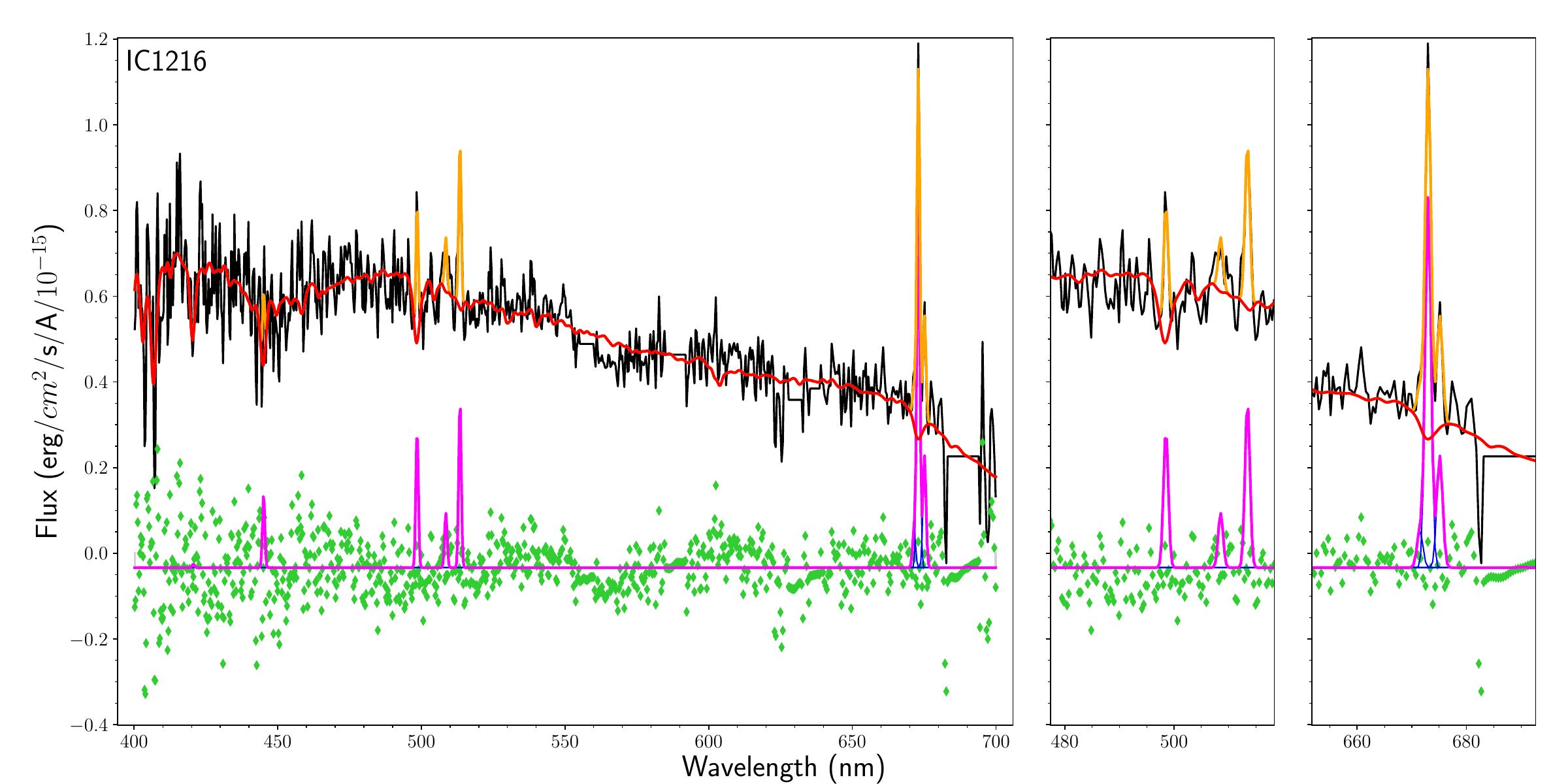}

\end{figure*}
\begin{figure*}
\includegraphics[width=7in]{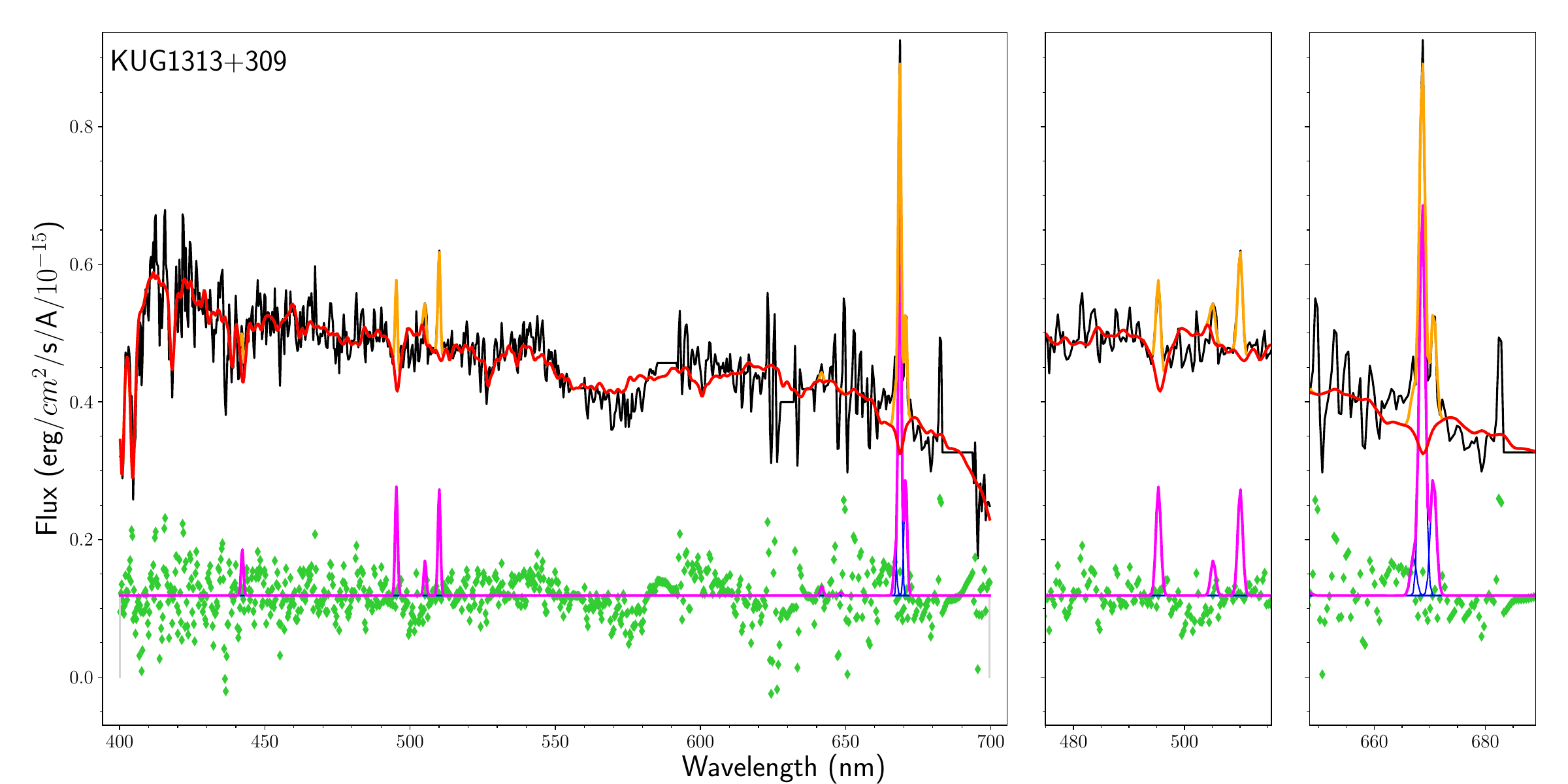}

\includegraphics[width=7in]{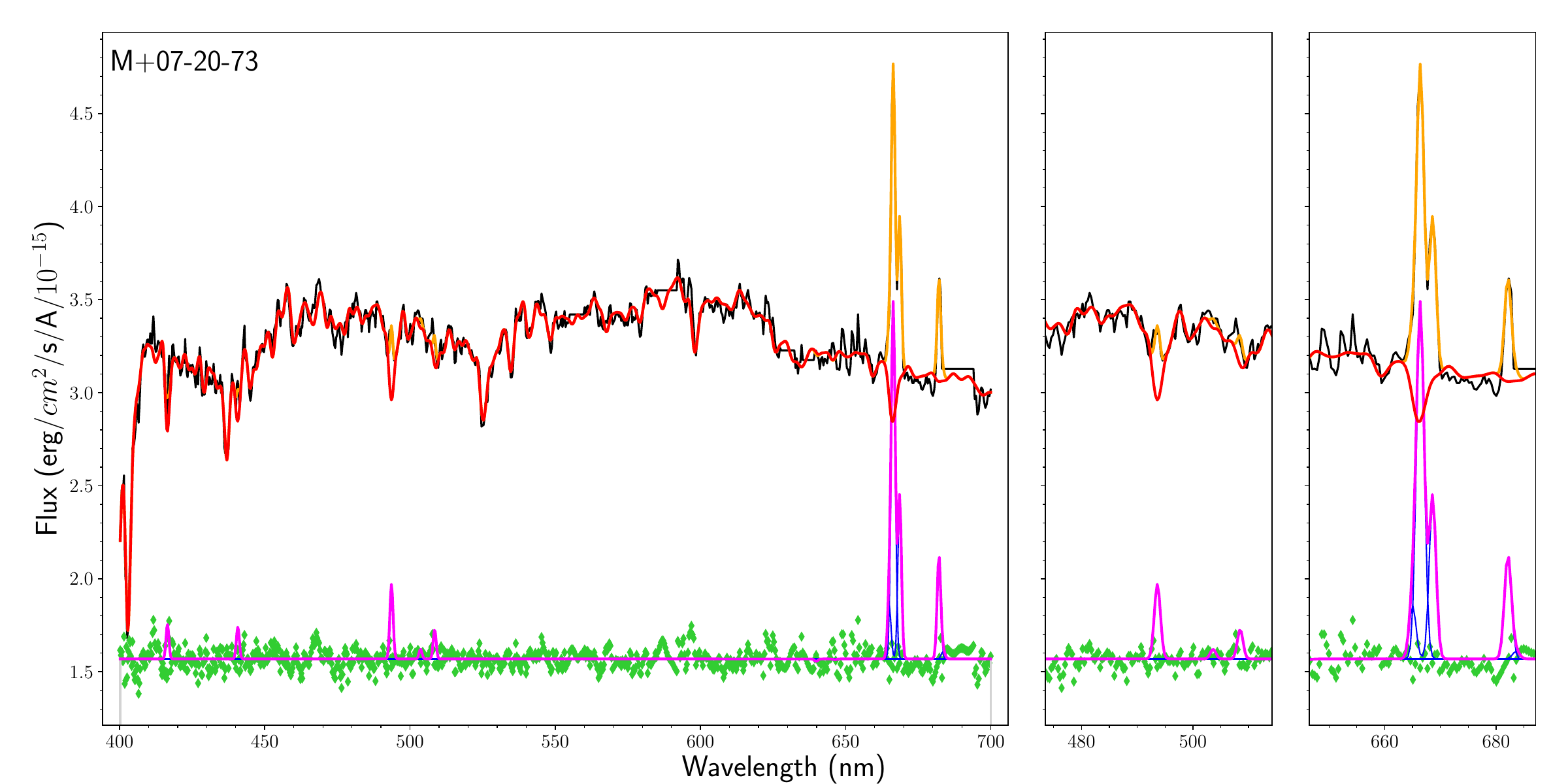}
\end{figure*}
\begin{figure*}
\includegraphics[width=7in]{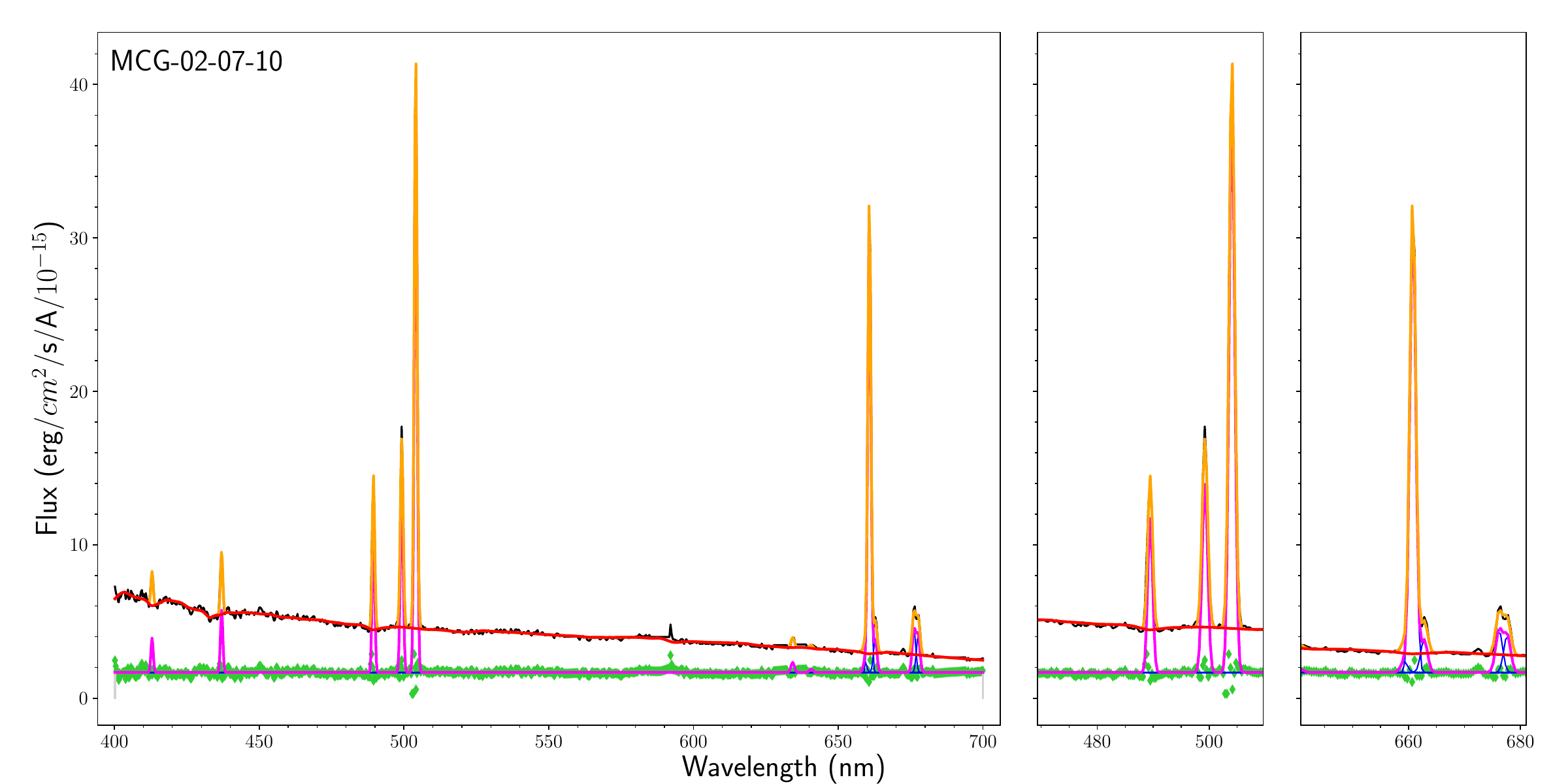}

\includegraphics[width=7in]{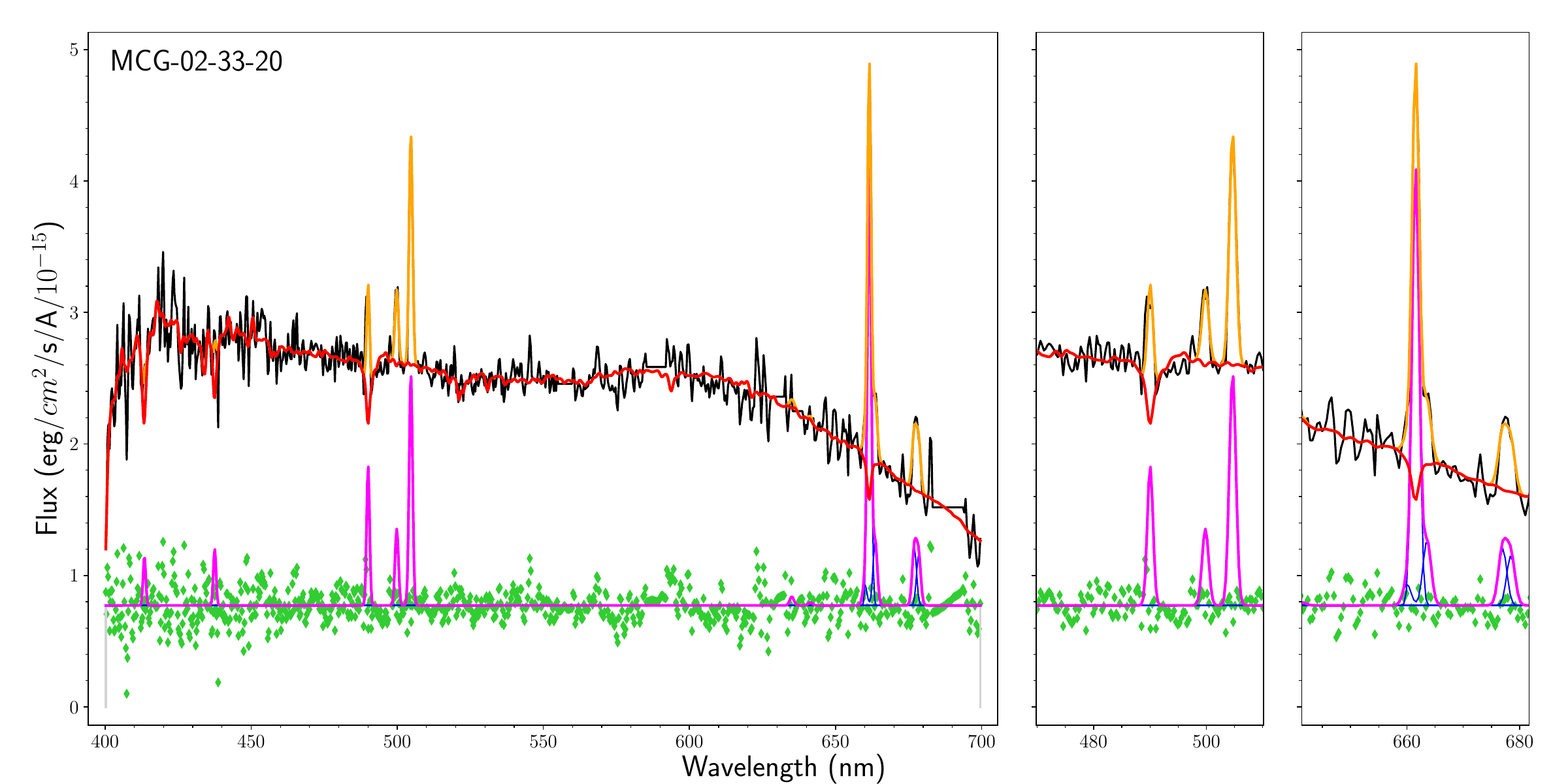}

\end{figure*}
\begin{figure*}
\includegraphics[width=7in]{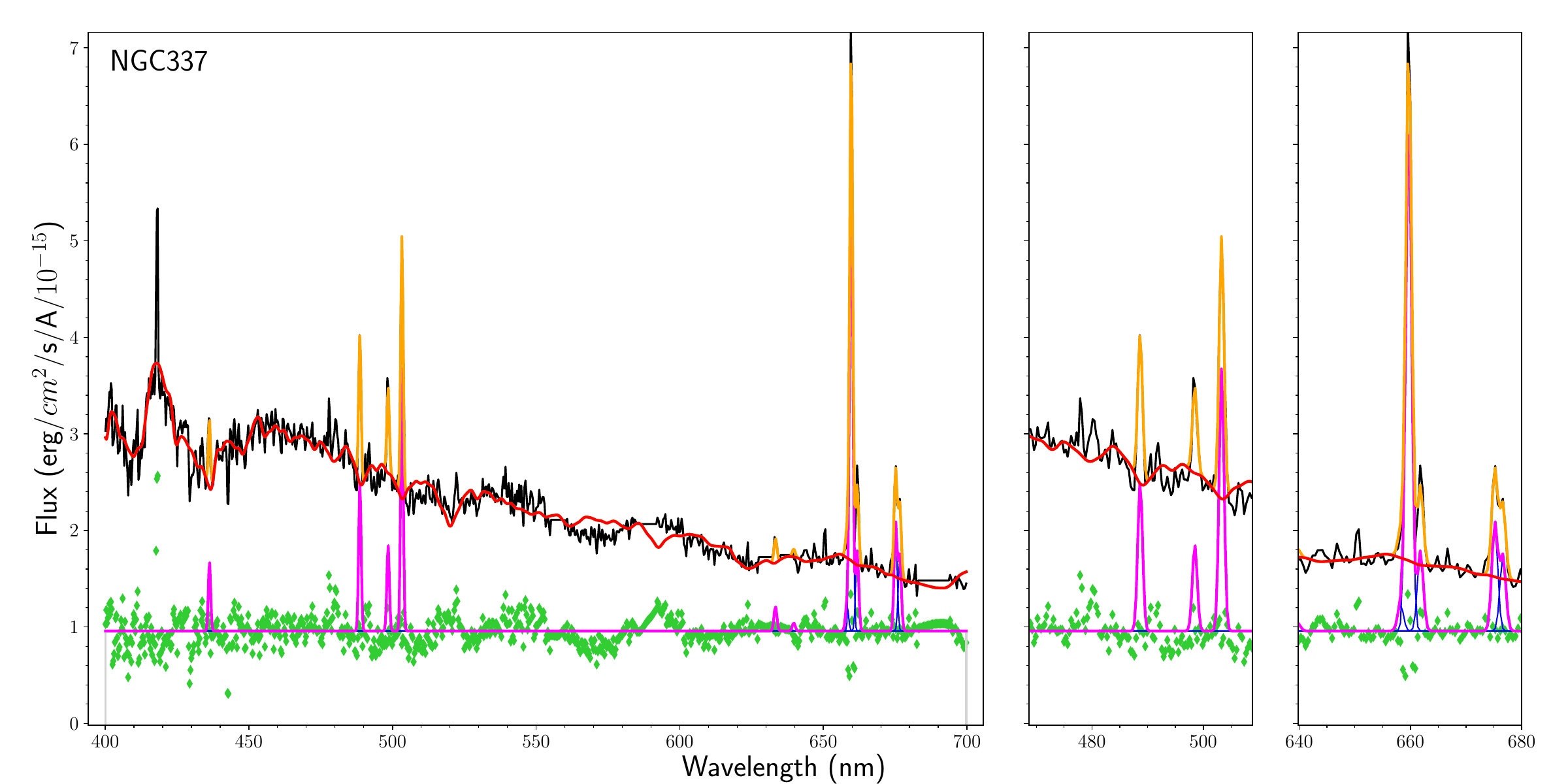}

\includegraphics[width=7in]{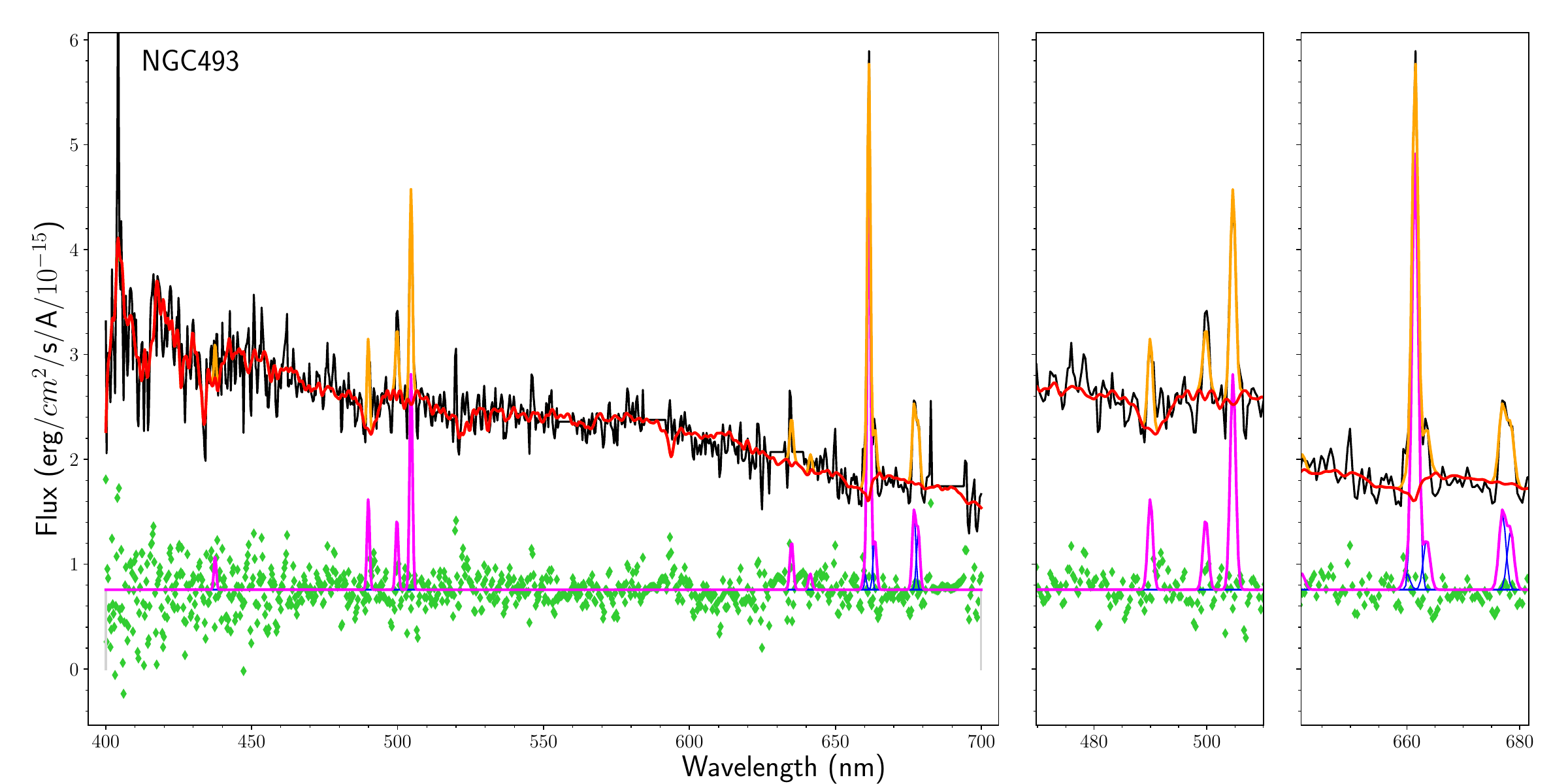}
\end{figure*}
\begin{figure*}
\includegraphics[width=7in]{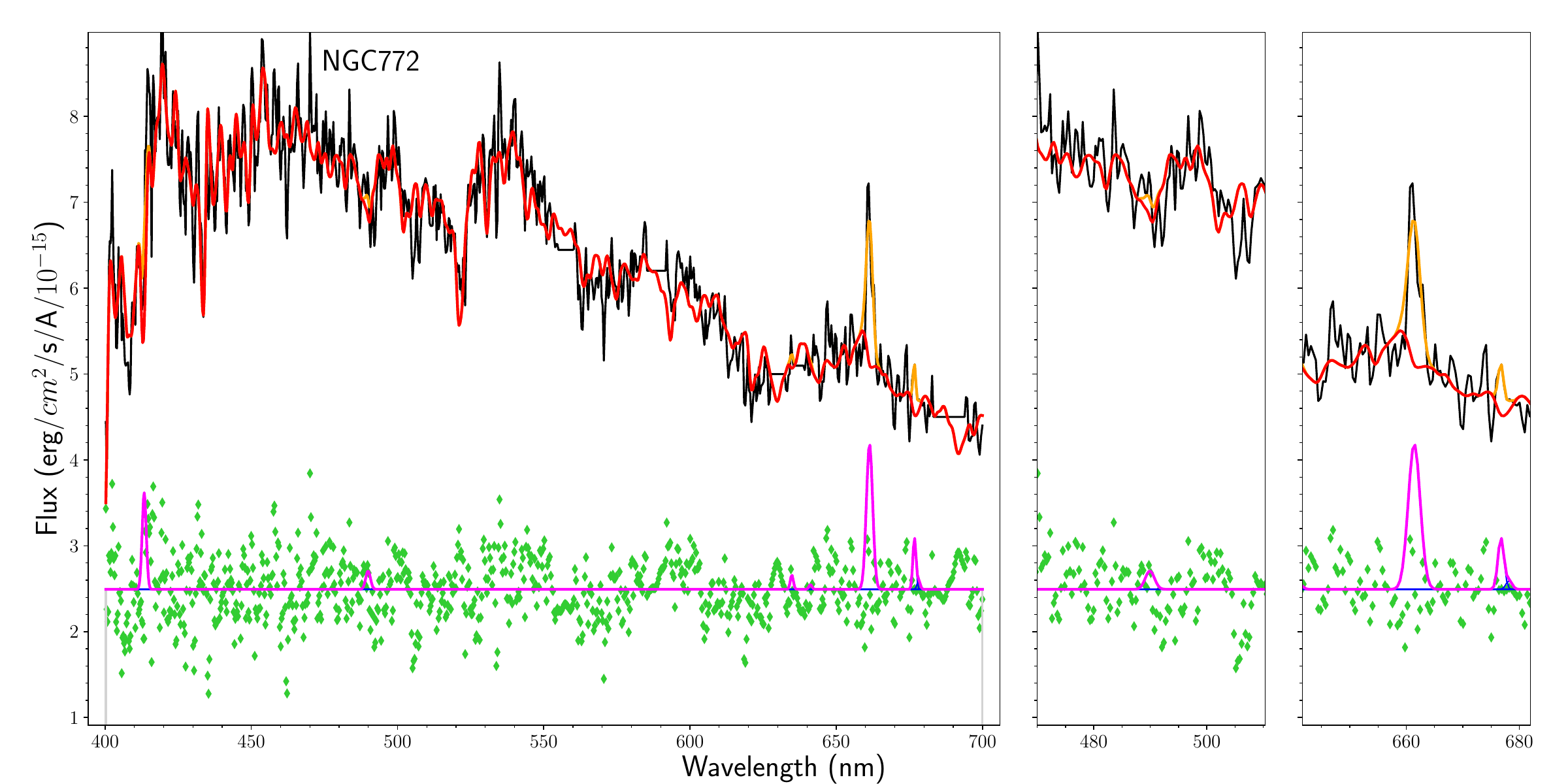}
  
\includegraphics[width=7in]{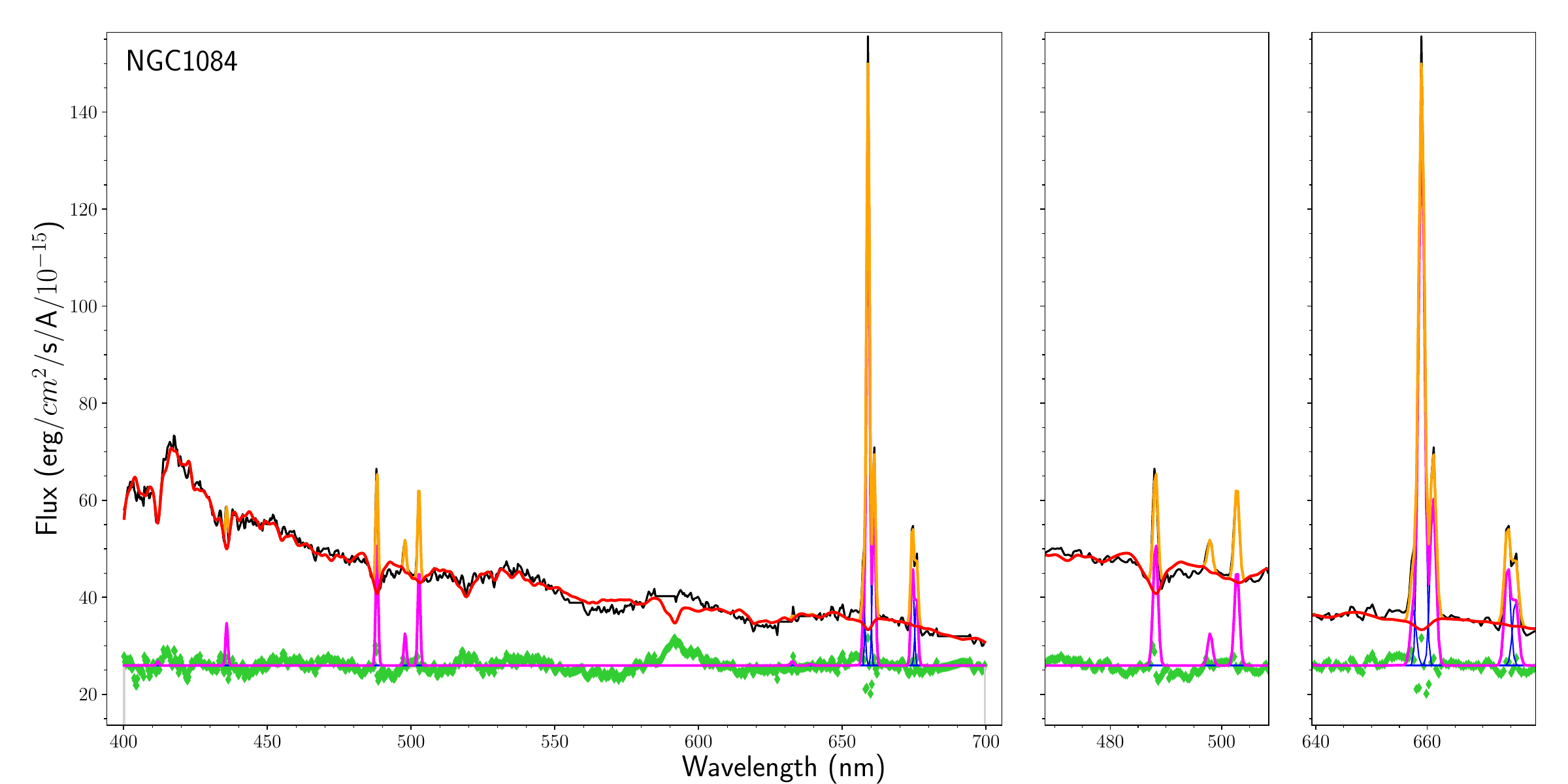}
 
\end{figure*}
\begin{figure*}
\includegraphics[width=7in]{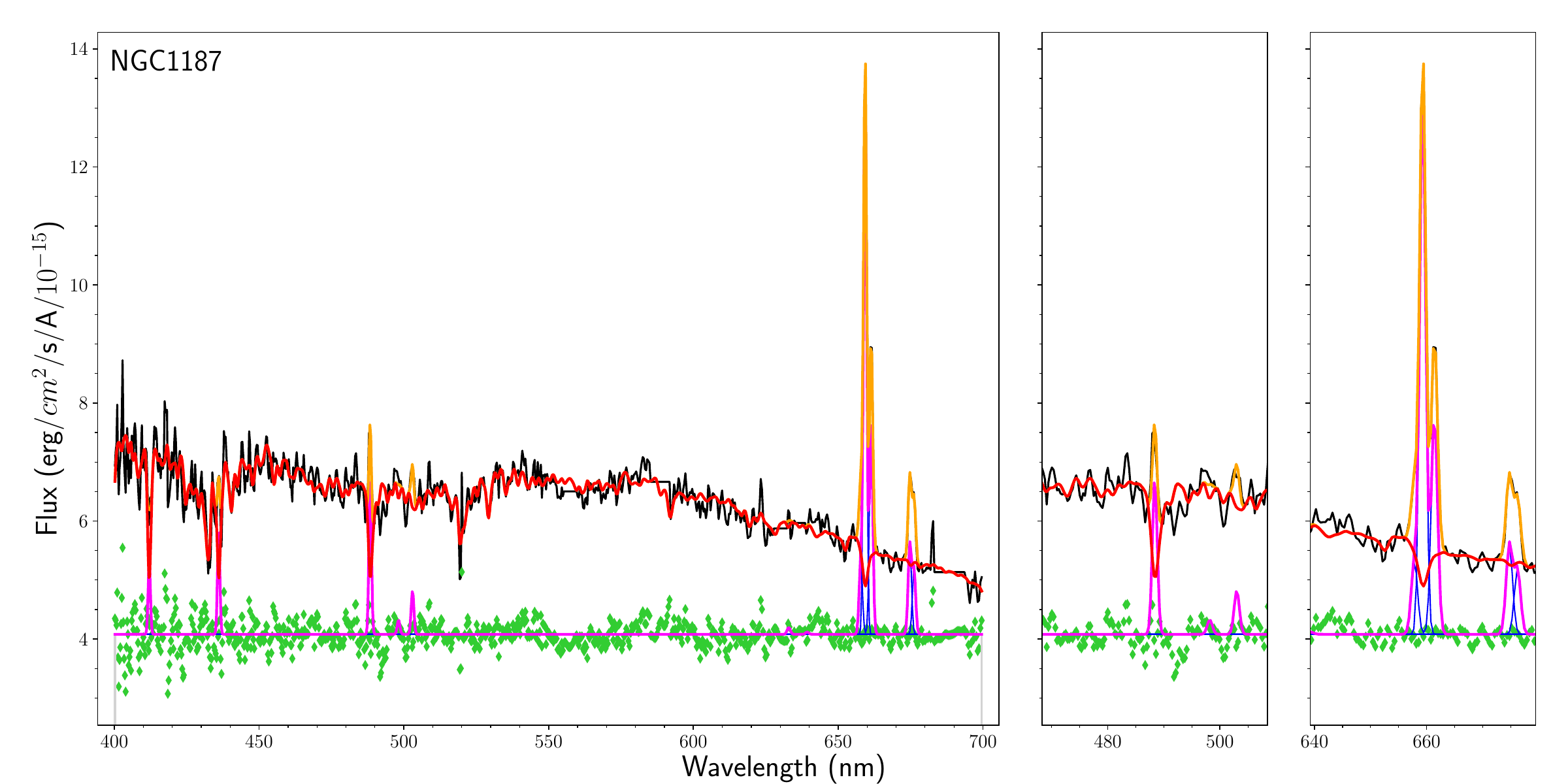}

\includegraphics[width=7in]{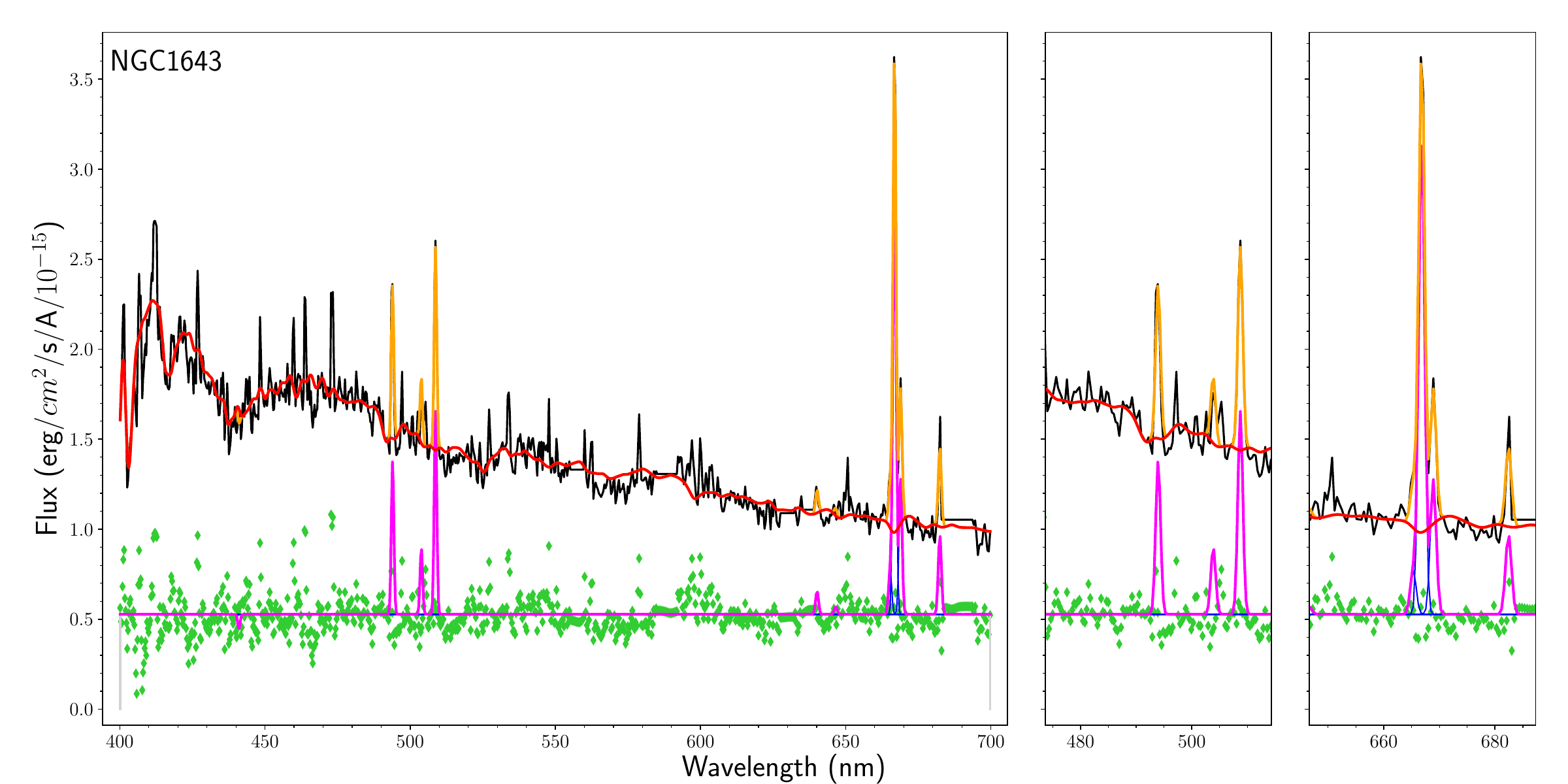}
\end{figure*}
\begin{figure*}
\includegraphics[width=7in]{image_files/NGC2276_Bok_ppxf_fit_orig.pdf}

\includegraphics[width=7in]{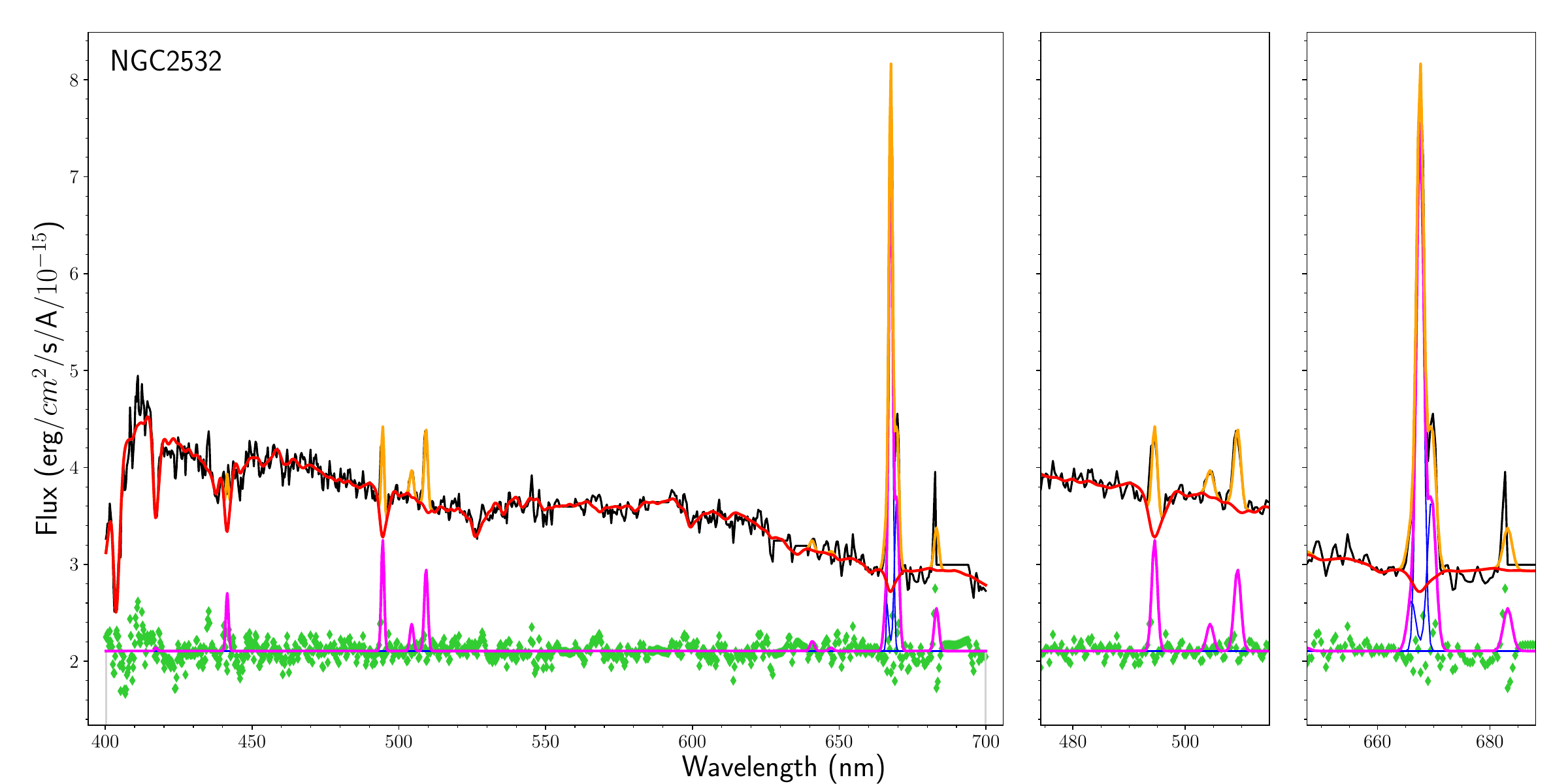}

\end{figure*}
\begin{figure*}
\includegraphics[width=7in]{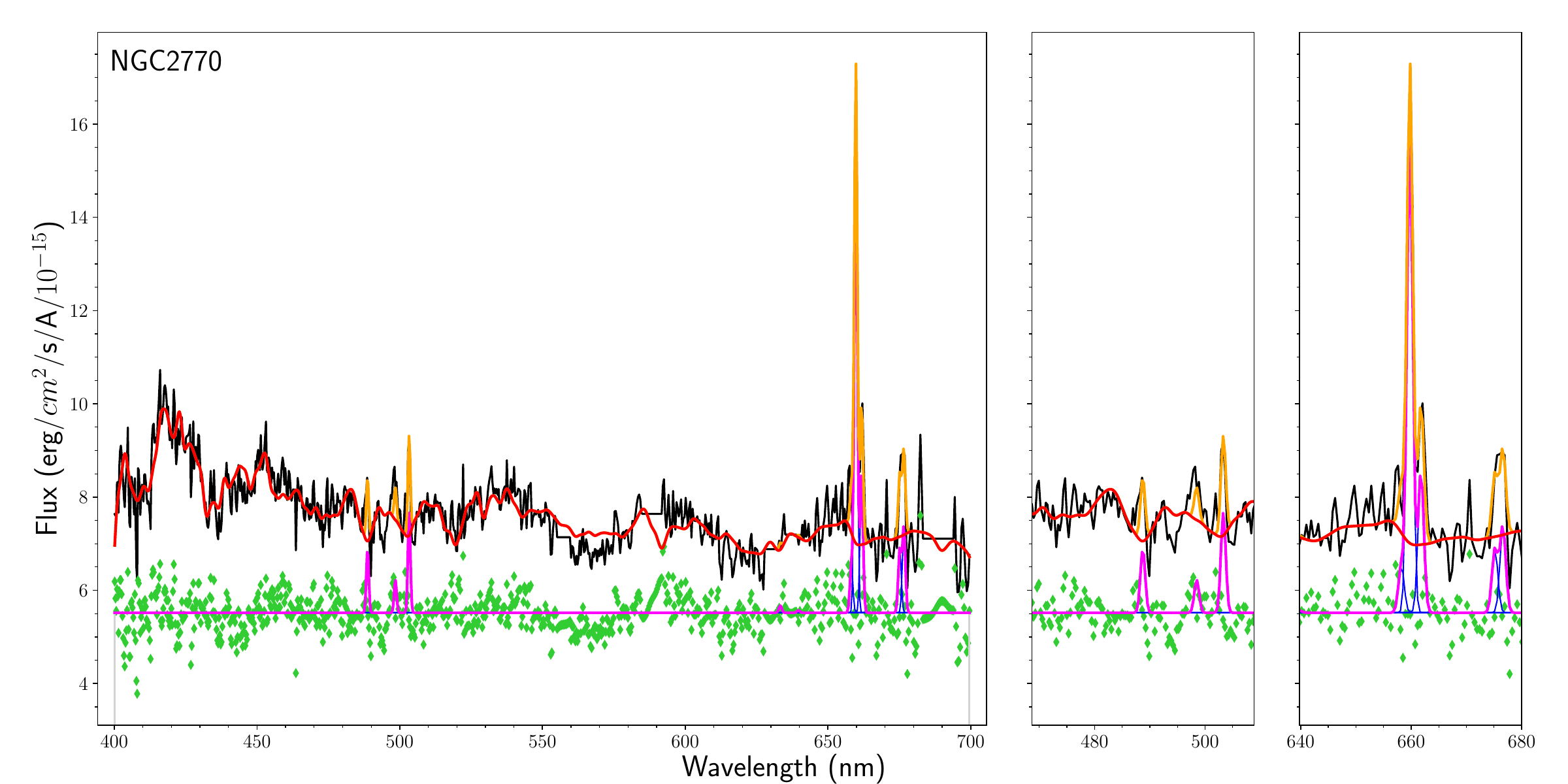}

\includegraphics[width=7in]{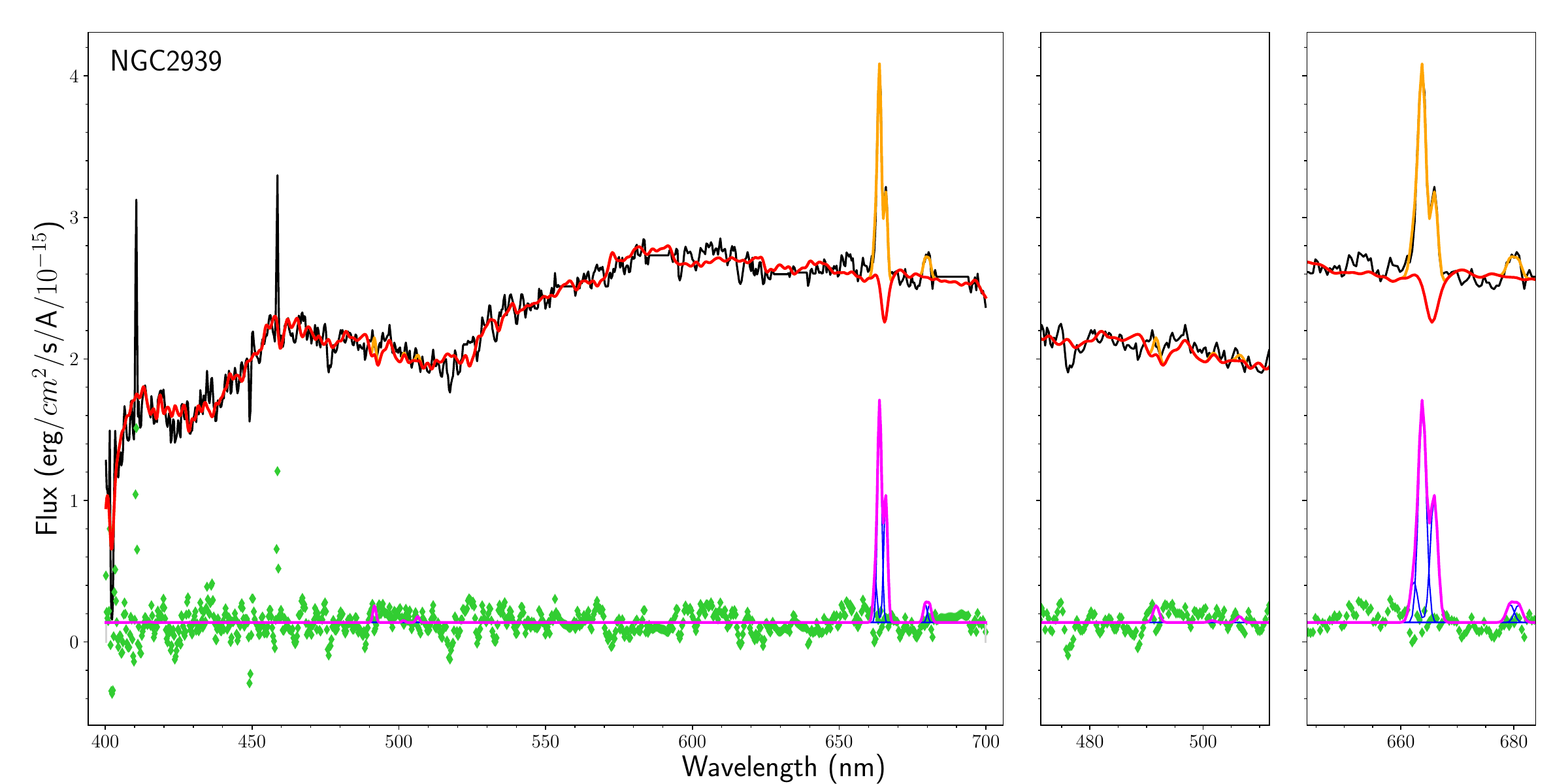}
\end{figure*}
\begin{figure*}

\includegraphics[width=7in]{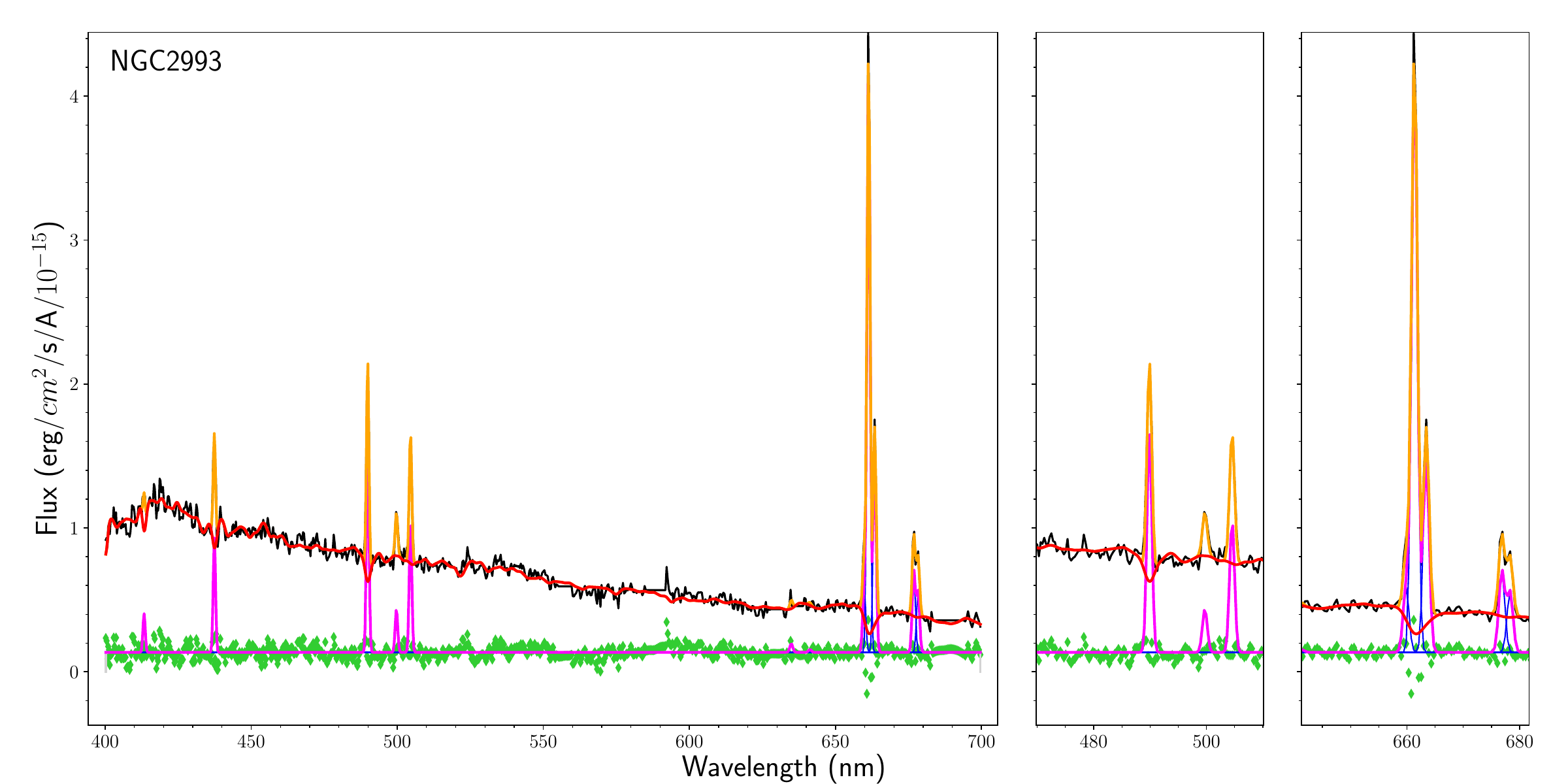}

\includegraphics[width=7in]{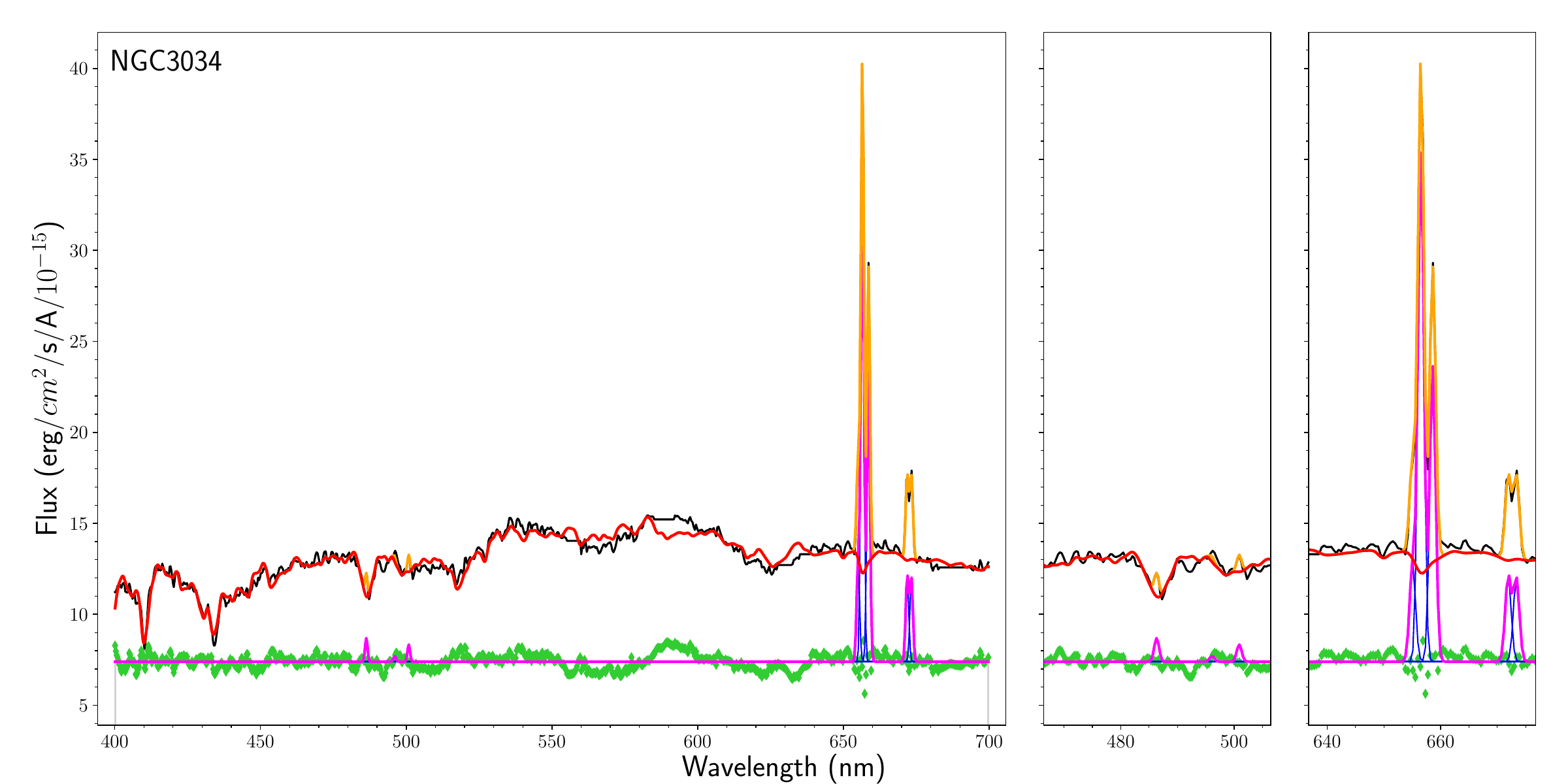}
\end{figure*}
\begin{figure*}
\includegraphics[width=7in]{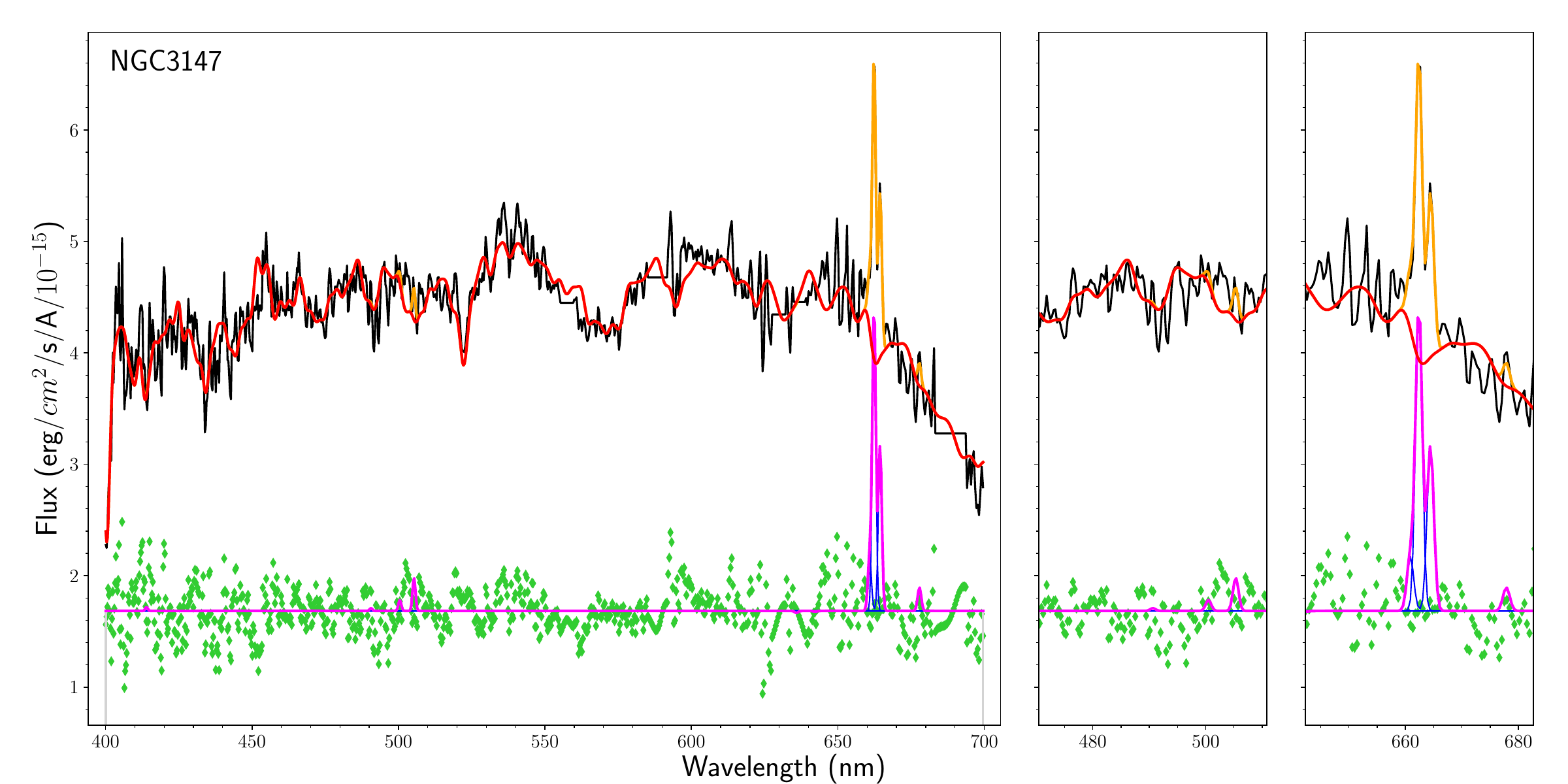}

\includegraphics[width=7in]{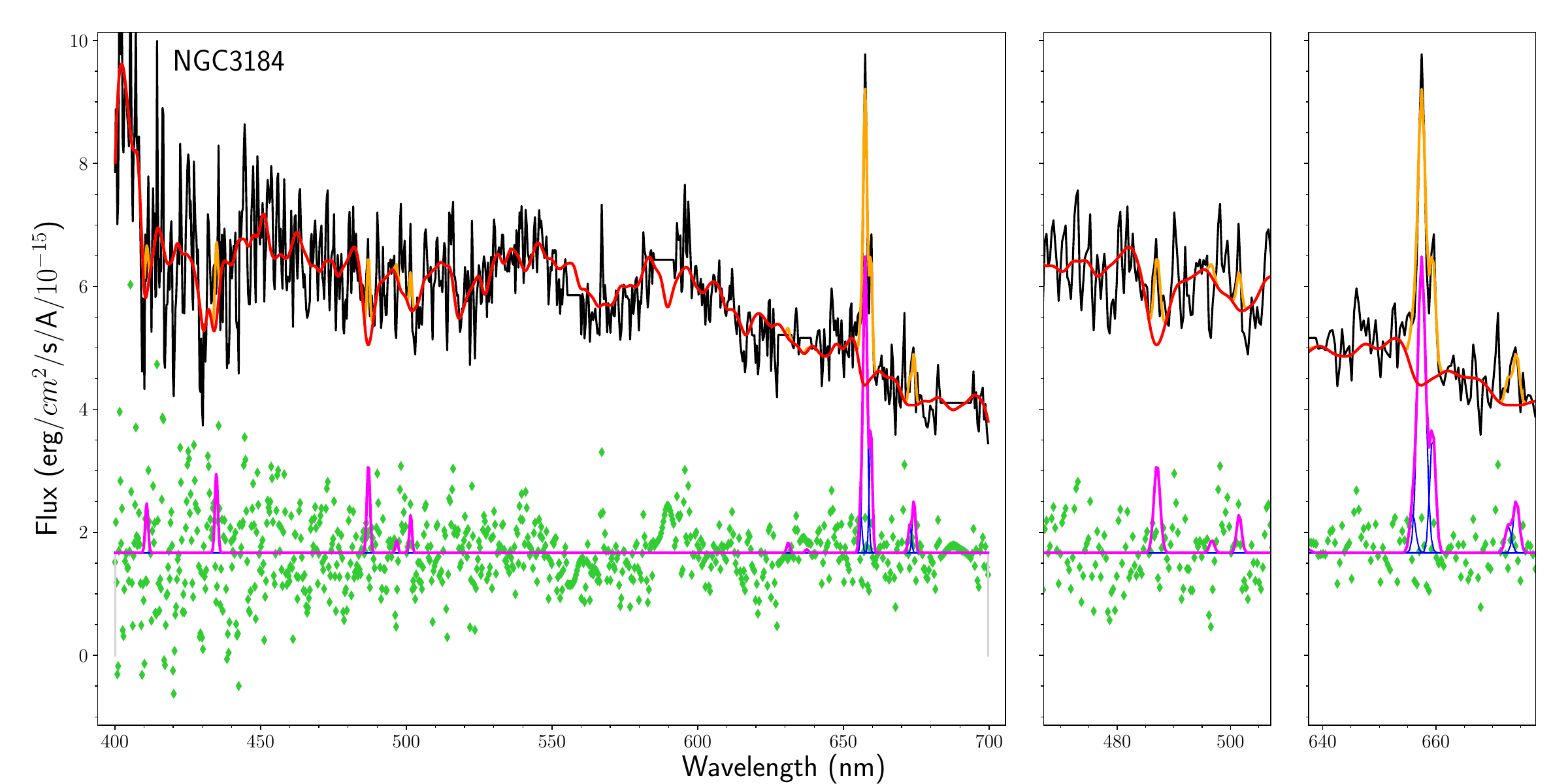}
\end{figure*}
\begin{figure*}
\includegraphics[width=7in]{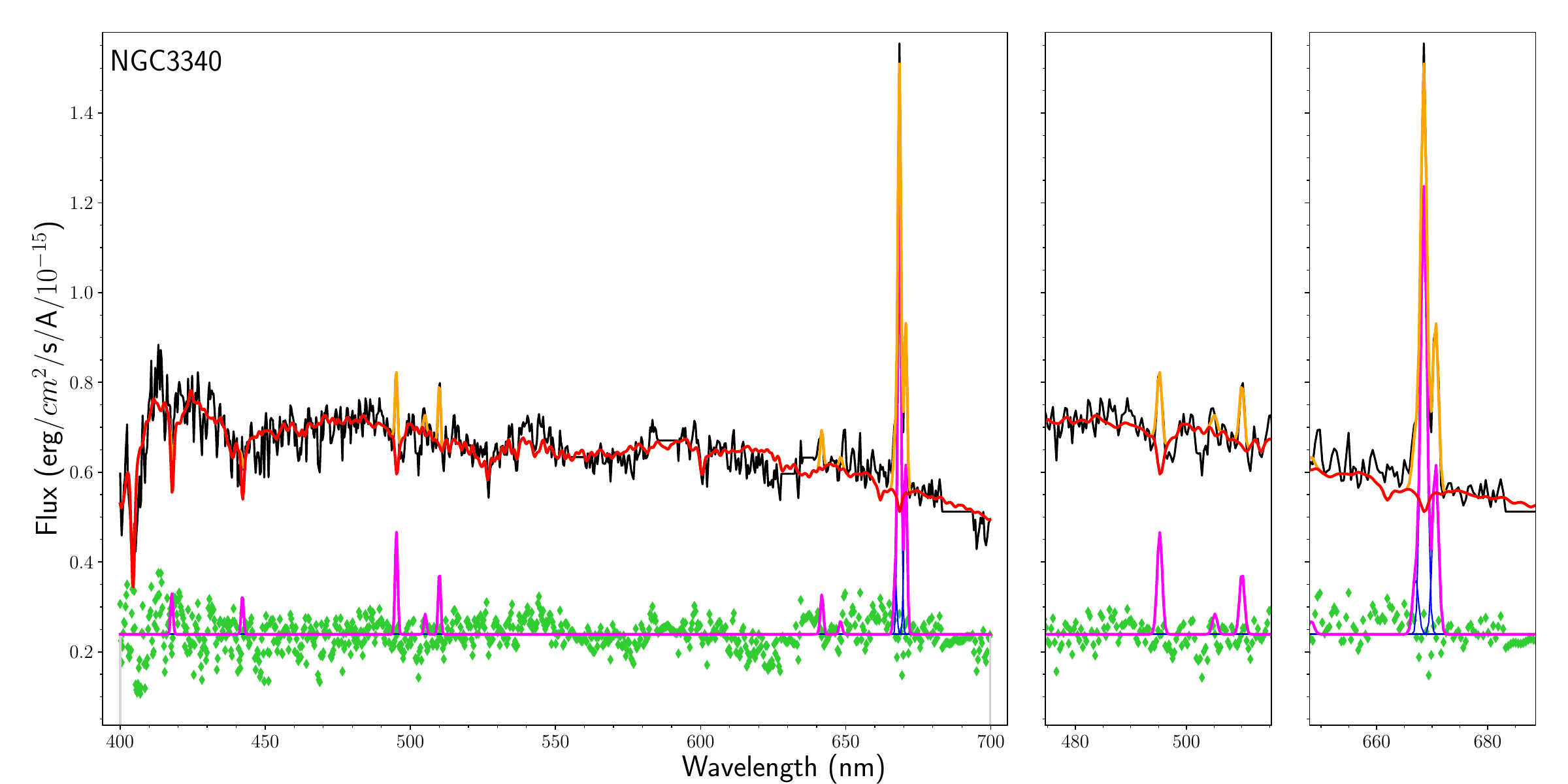}

\includegraphics[width=7in]{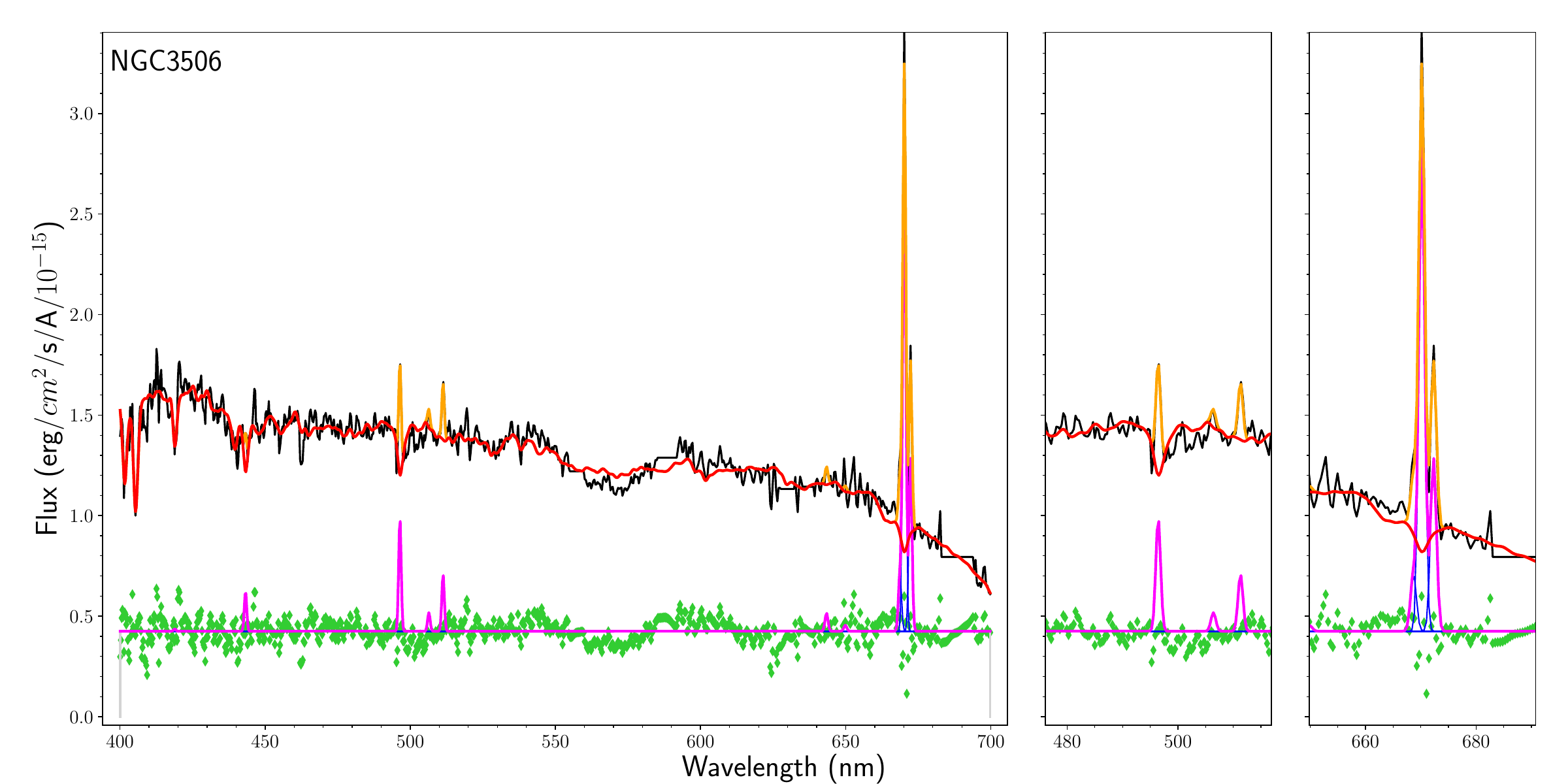}
\end{figure*}
\begin{figure*}
\includegraphics[width=7in]{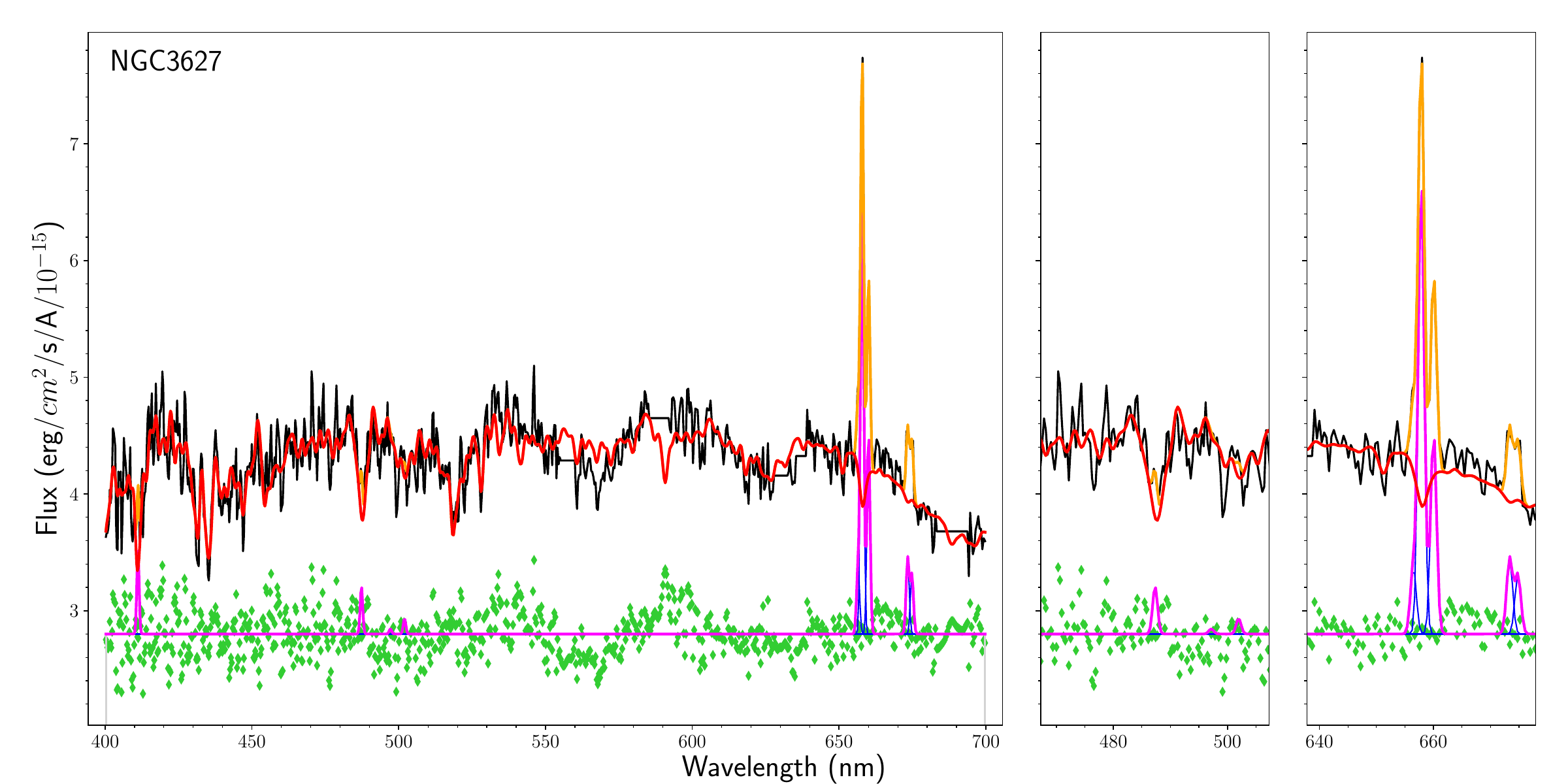}

\includegraphics[width=7in]{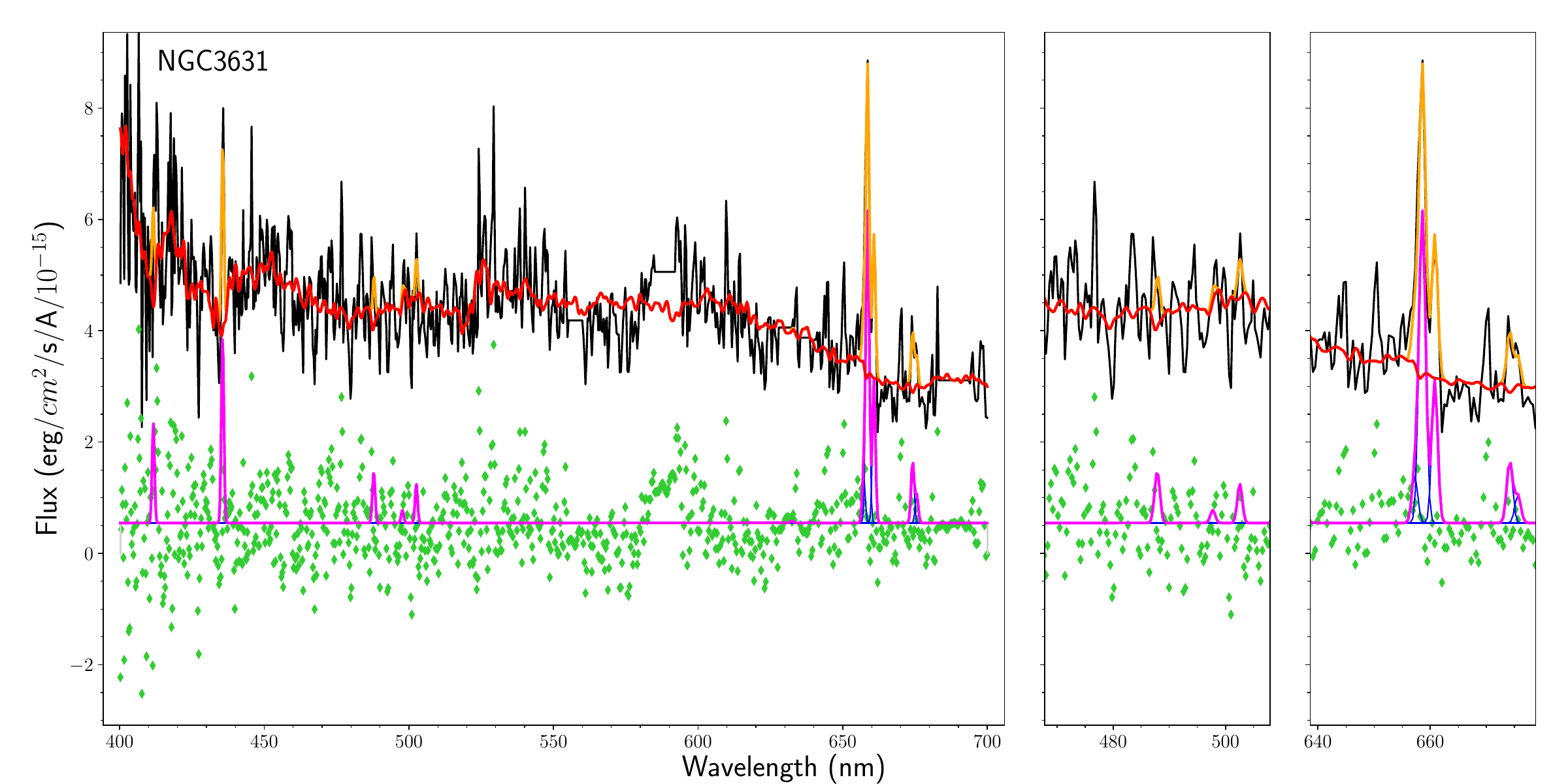}
\end{figure*}
\begin{figure*}
\includegraphics[width=7in]{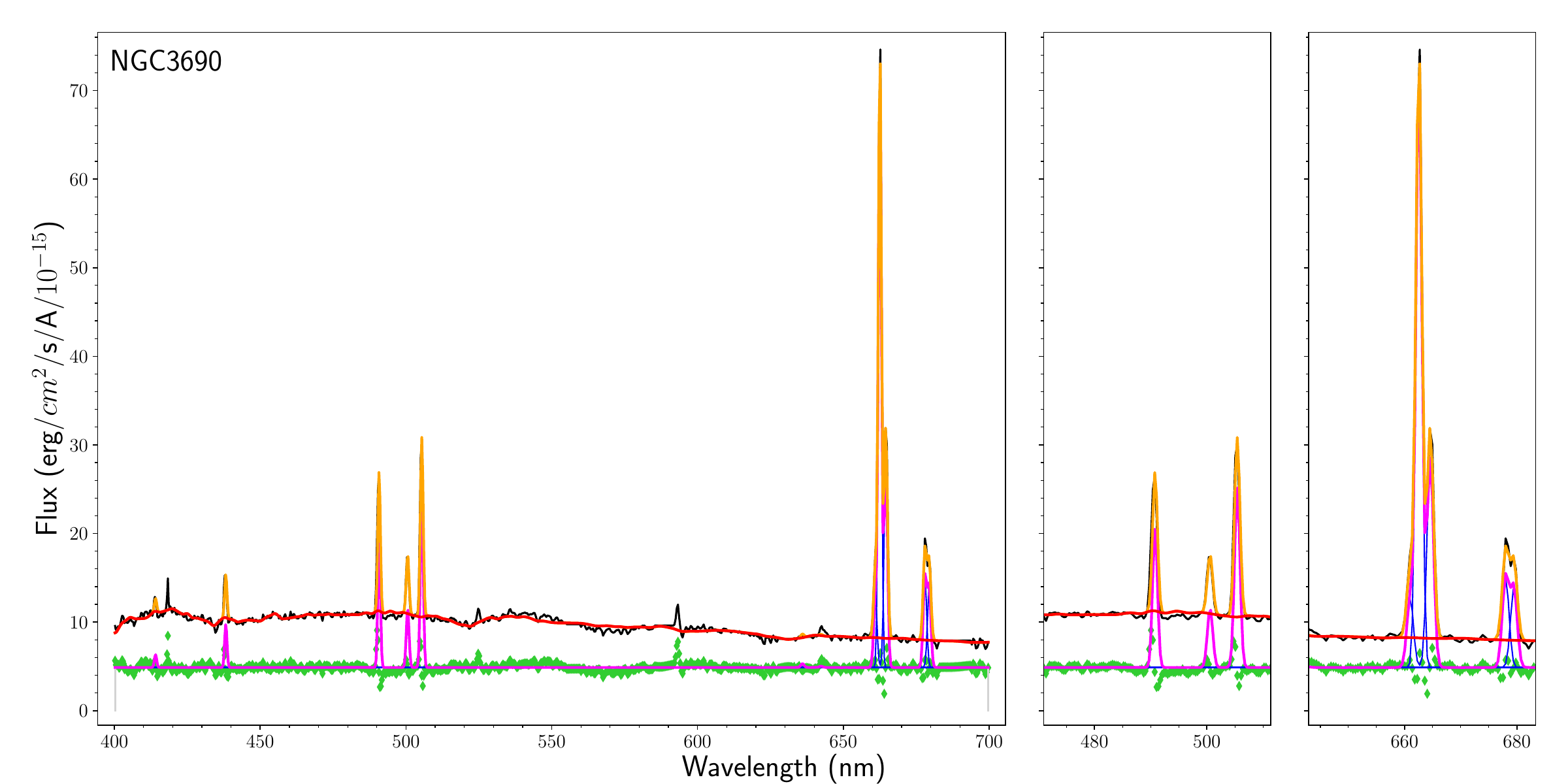}

\includegraphics[width=7in]{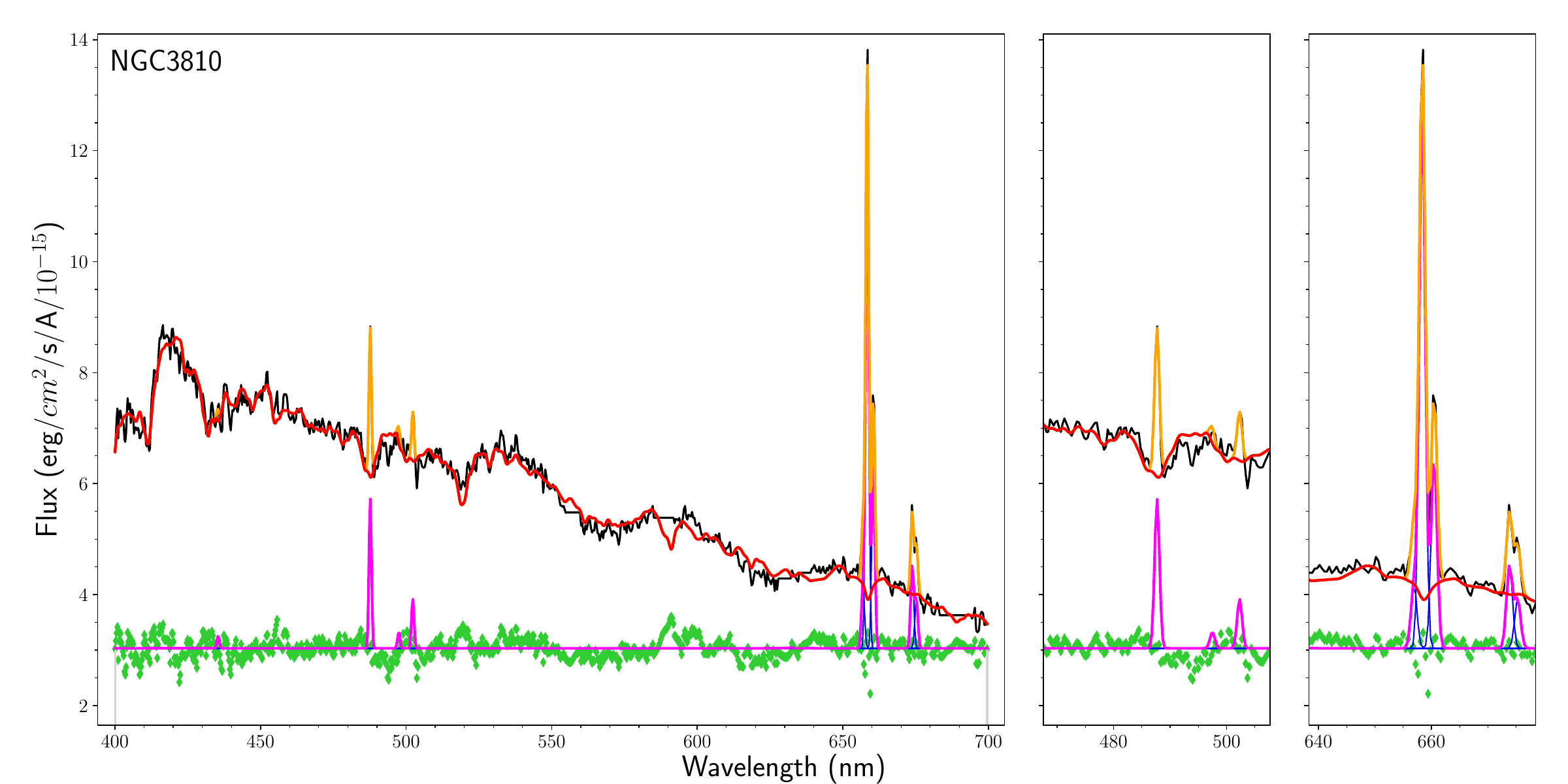}
\end{figure*}
\begin{figure*}
\includegraphics[width=7in]{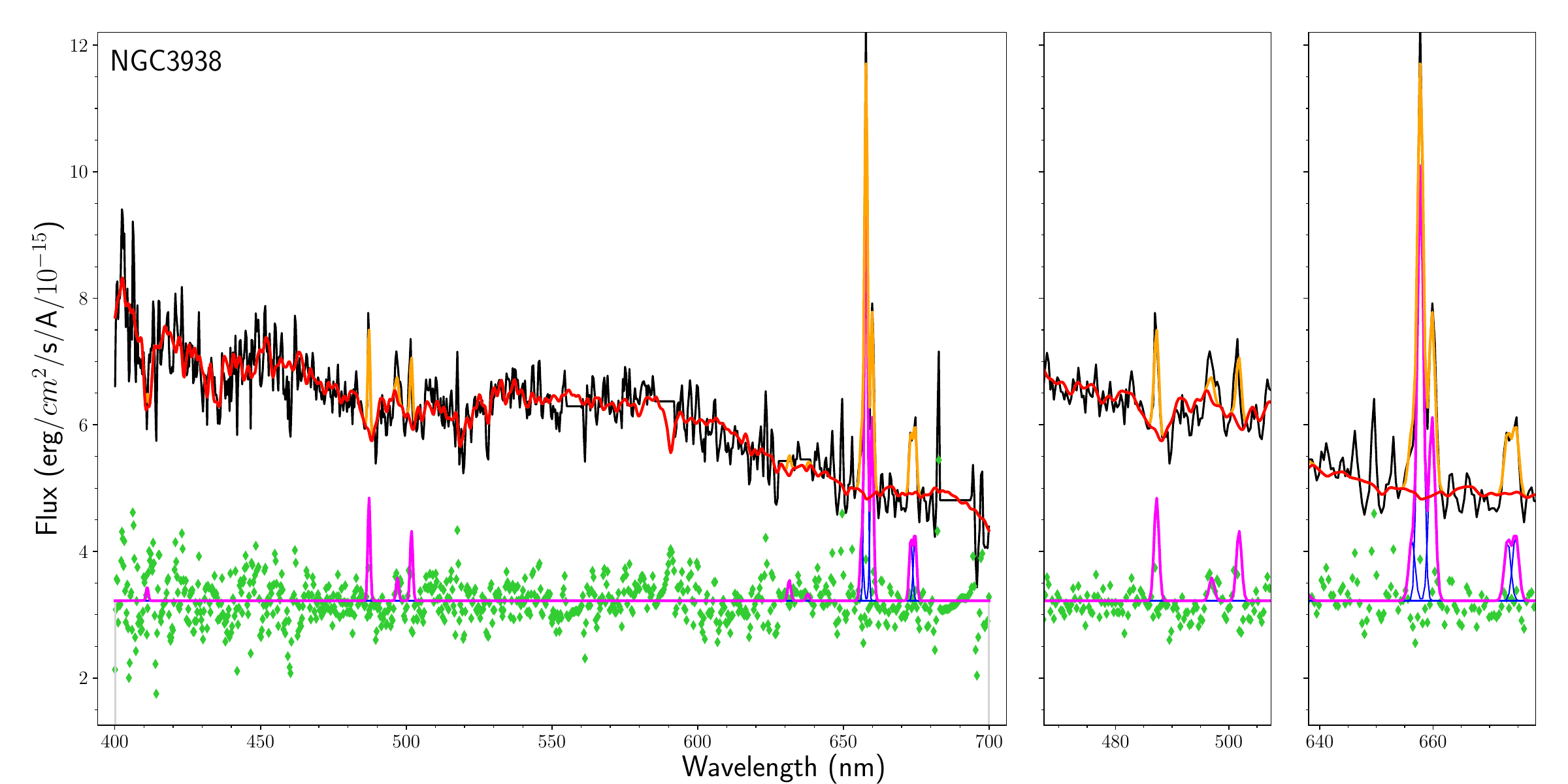}

\includegraphics[width=7in]{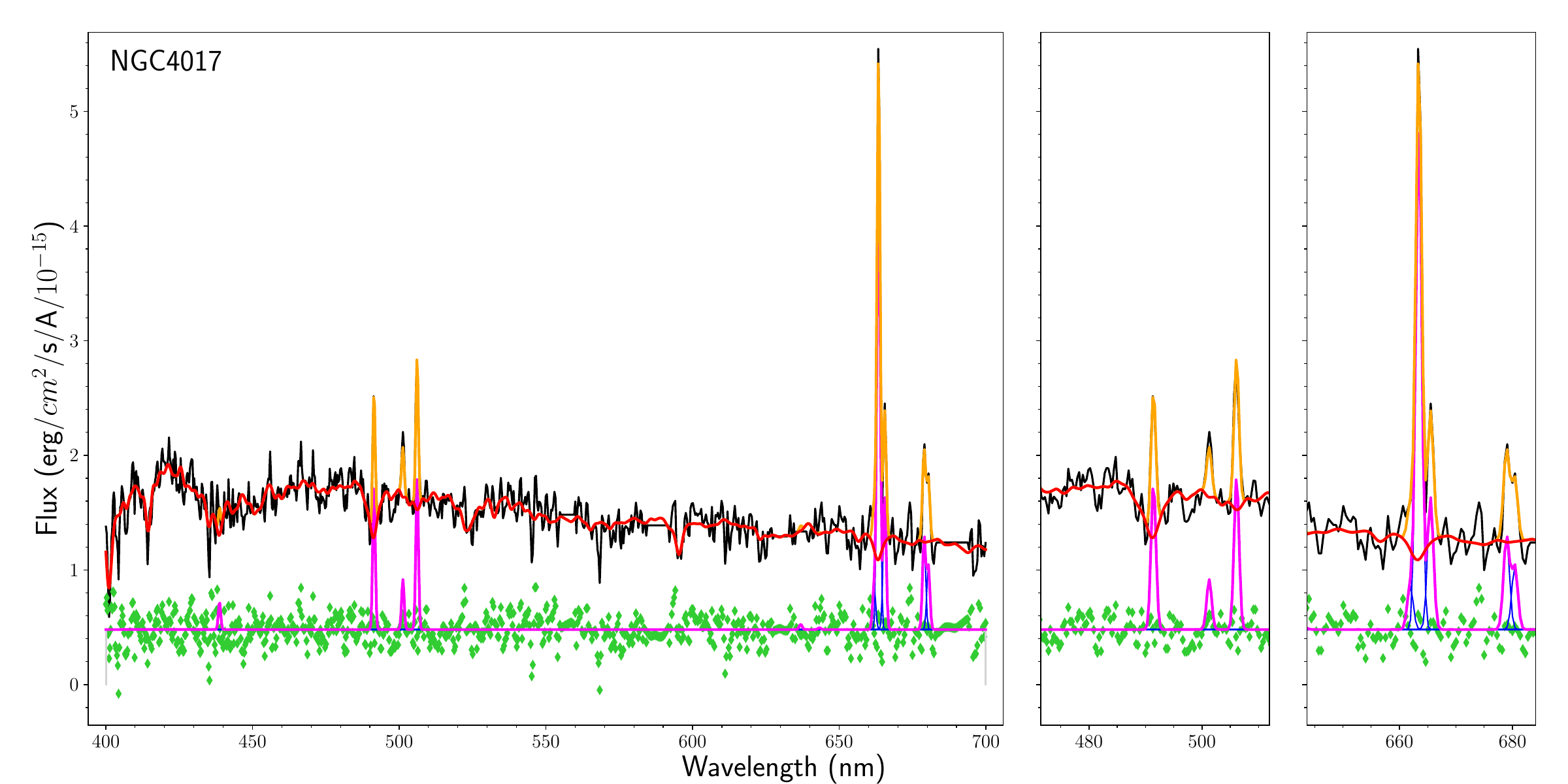}
\end{figure*}
\begin{figure*}
\includegraphics[width=7in]{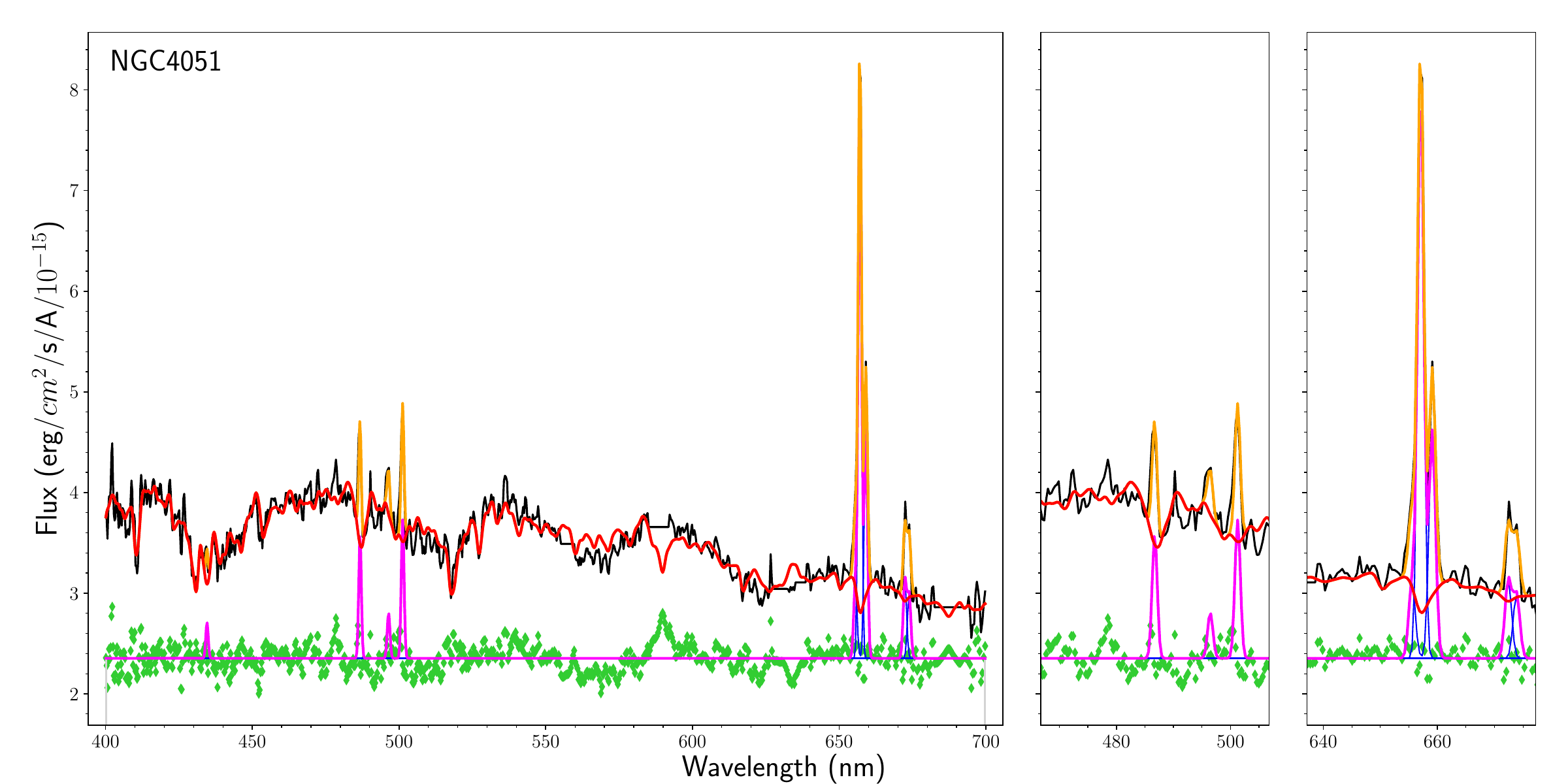}

\includegraphics[width=7in]{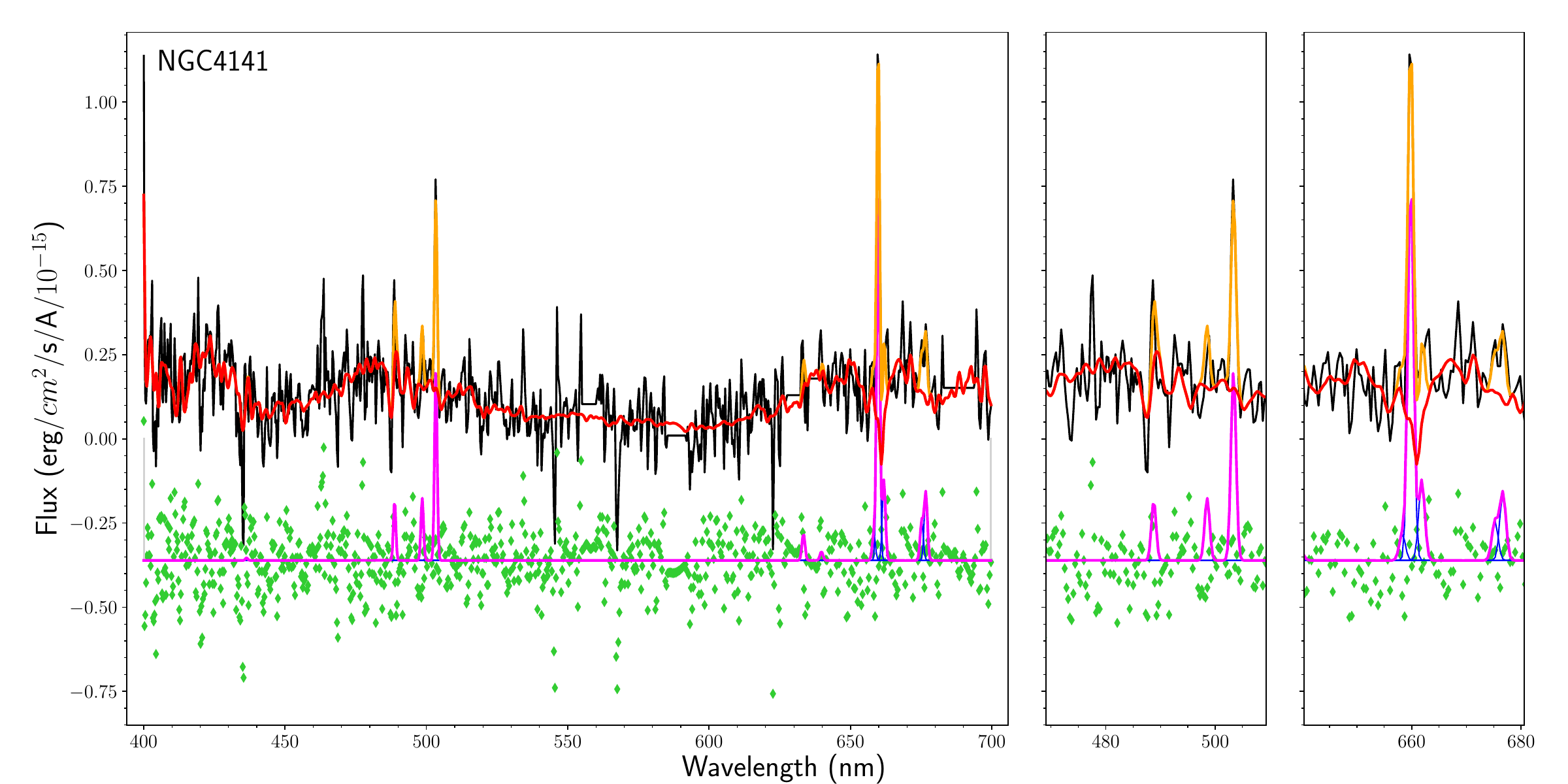}
\end{figure*}
\begin{figure*}
\includegraphics[width=7in]{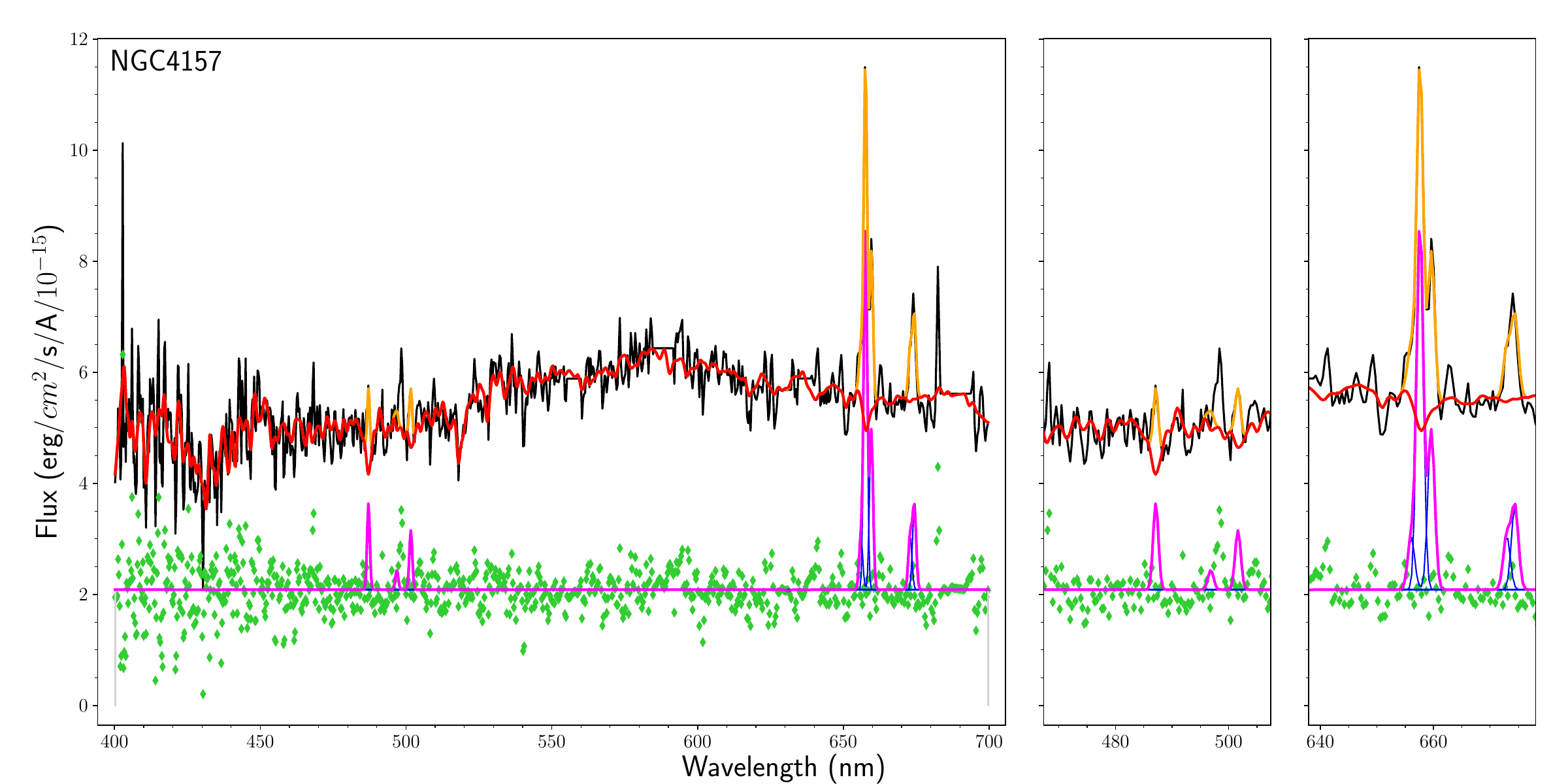}

\includegraphics[width=7in]{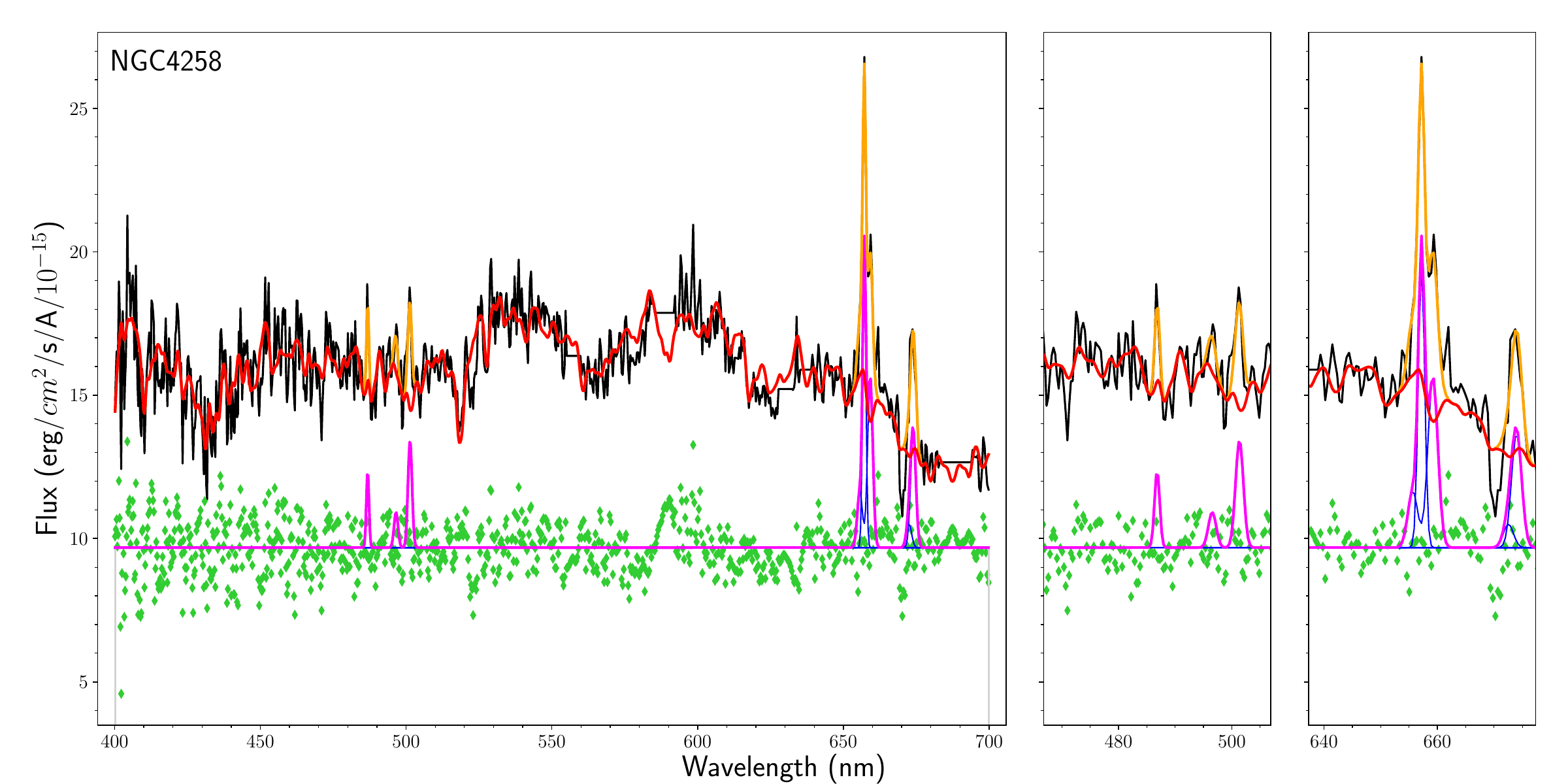}
\end{figure*}
\begin{figure*}
\includegraphics[width=7in]{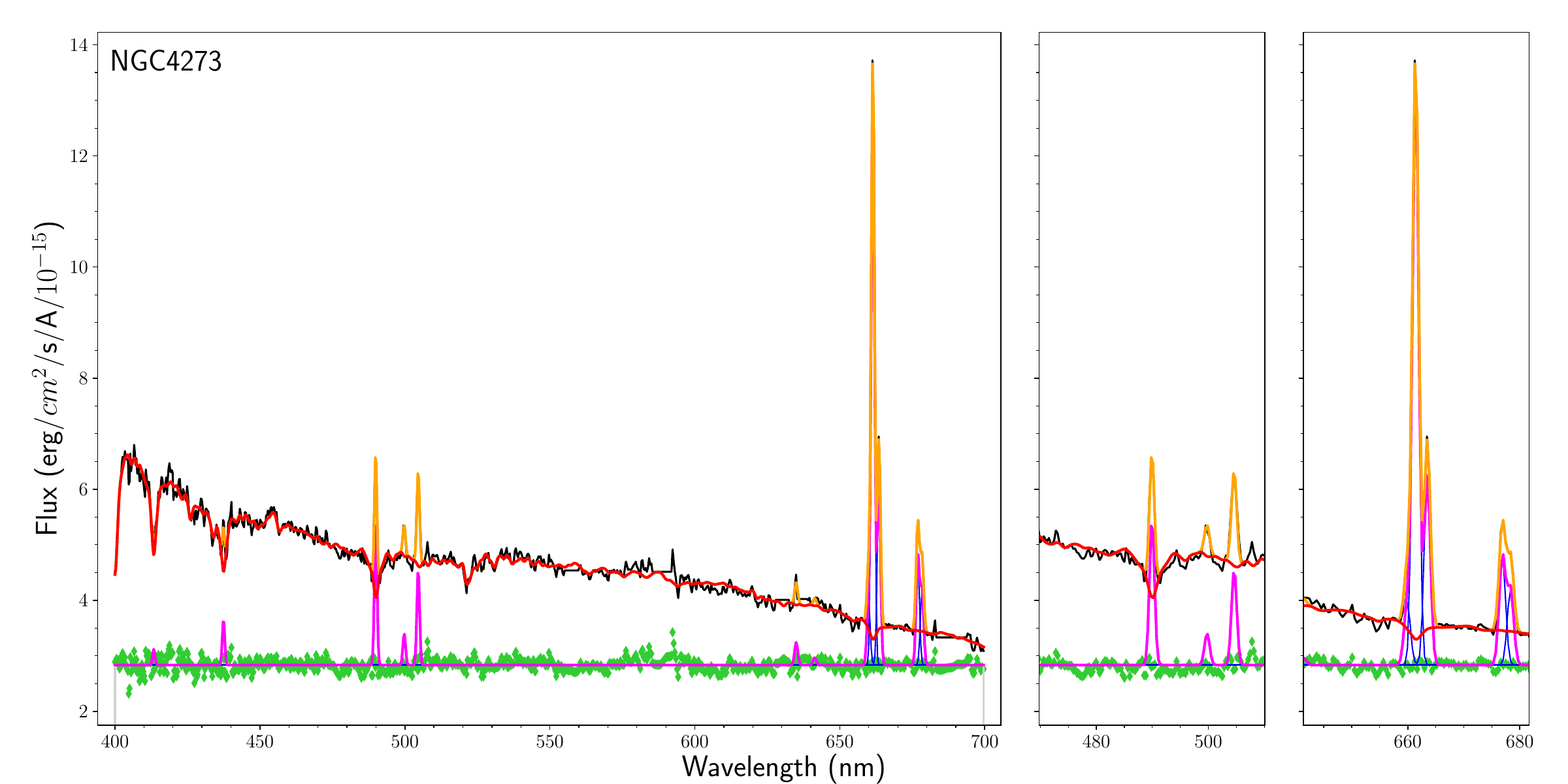}

\includegraphics[width=7in]{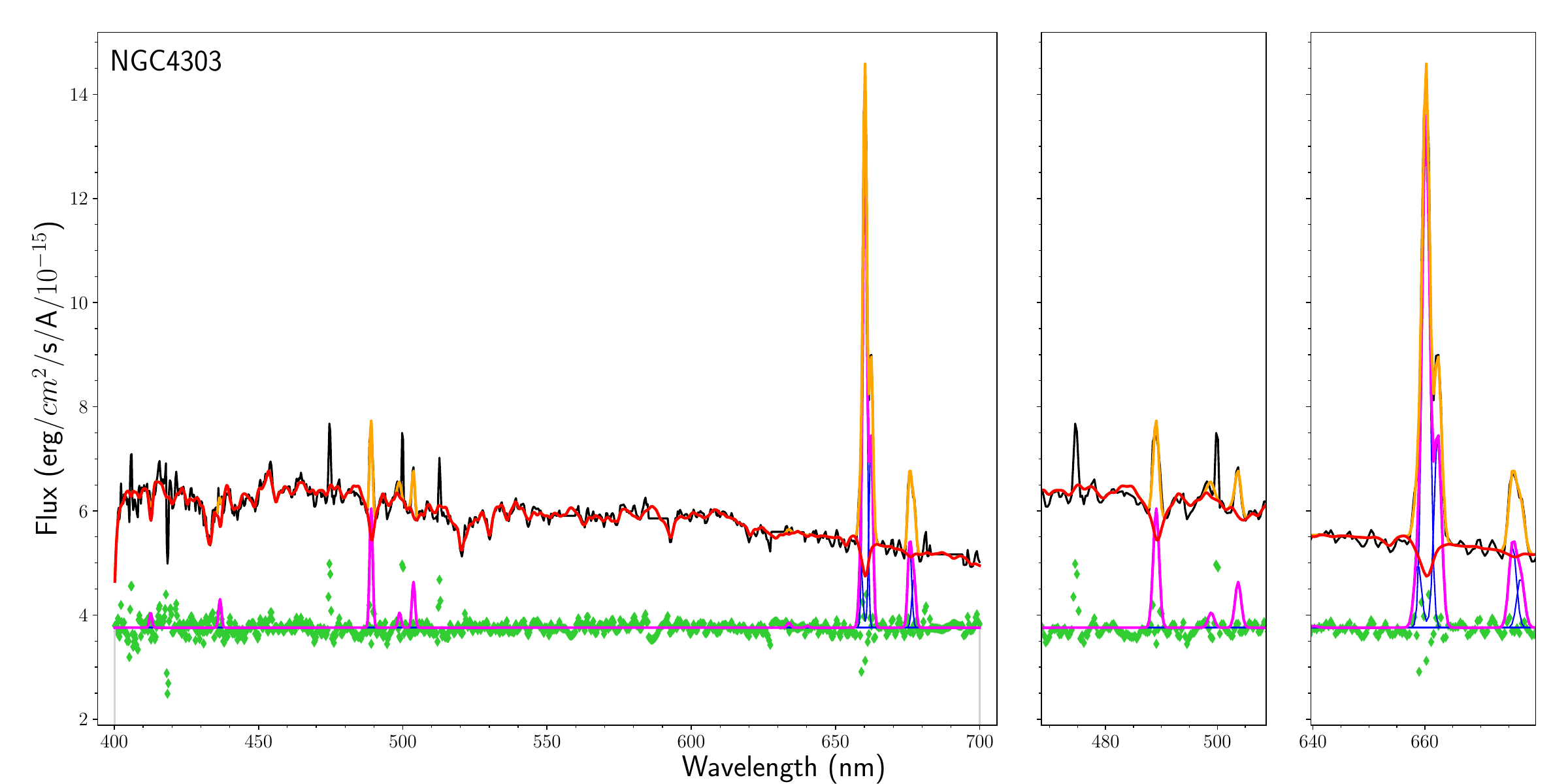}
\end{figure*}
\begin{figure*}
\includegraphics[width=7in]{image_files/NGC4490_Bok_ppxf_fit_orig.pdf}

\includegraphics[width=7in]{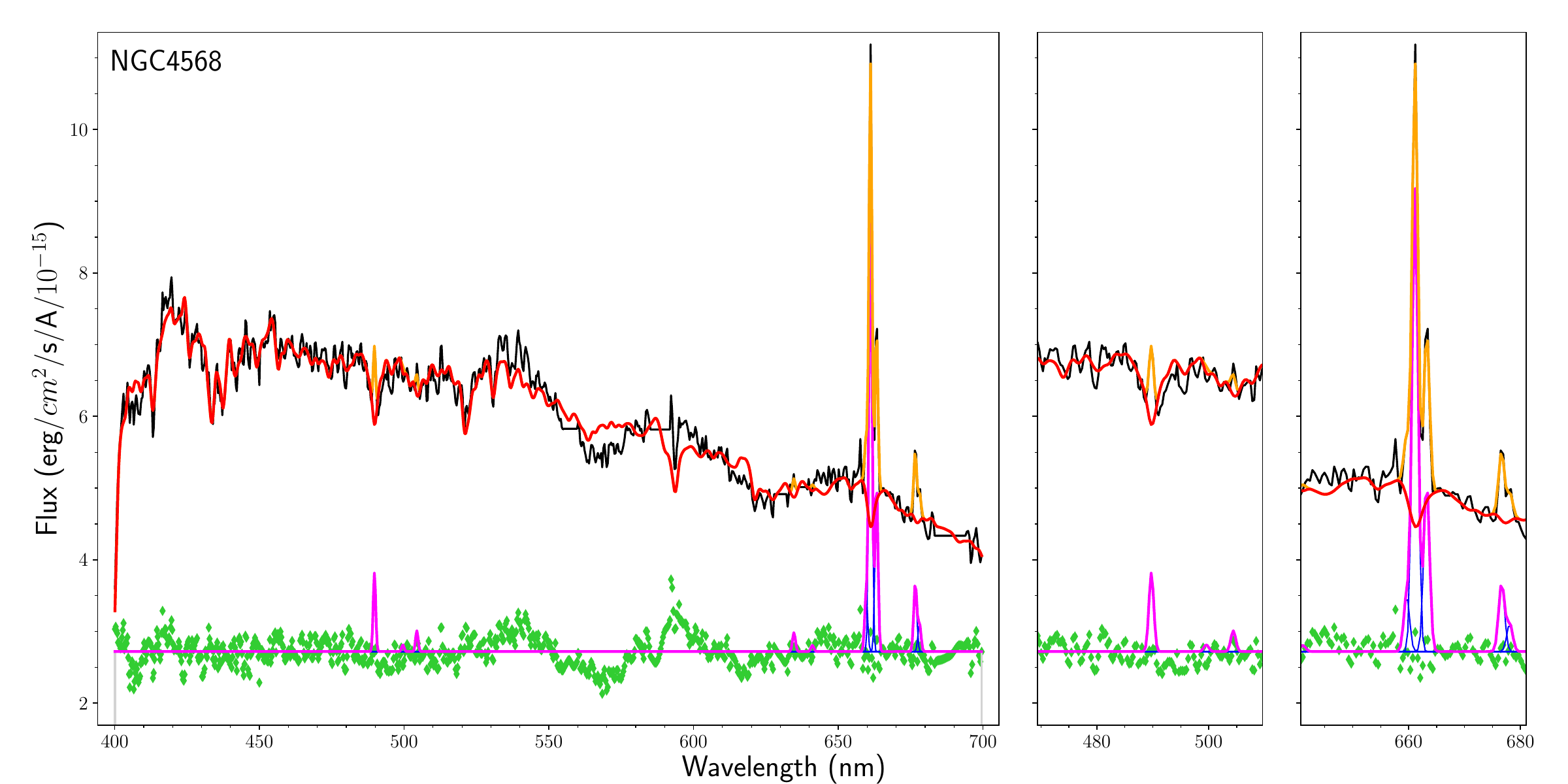}
\end{figure*}
\begin{figure*}
\includegraphics[width=7in]{image_files/NGC4666_Bok_ppxf_fit_orig.pdf}

\includegraphics[width=7in]{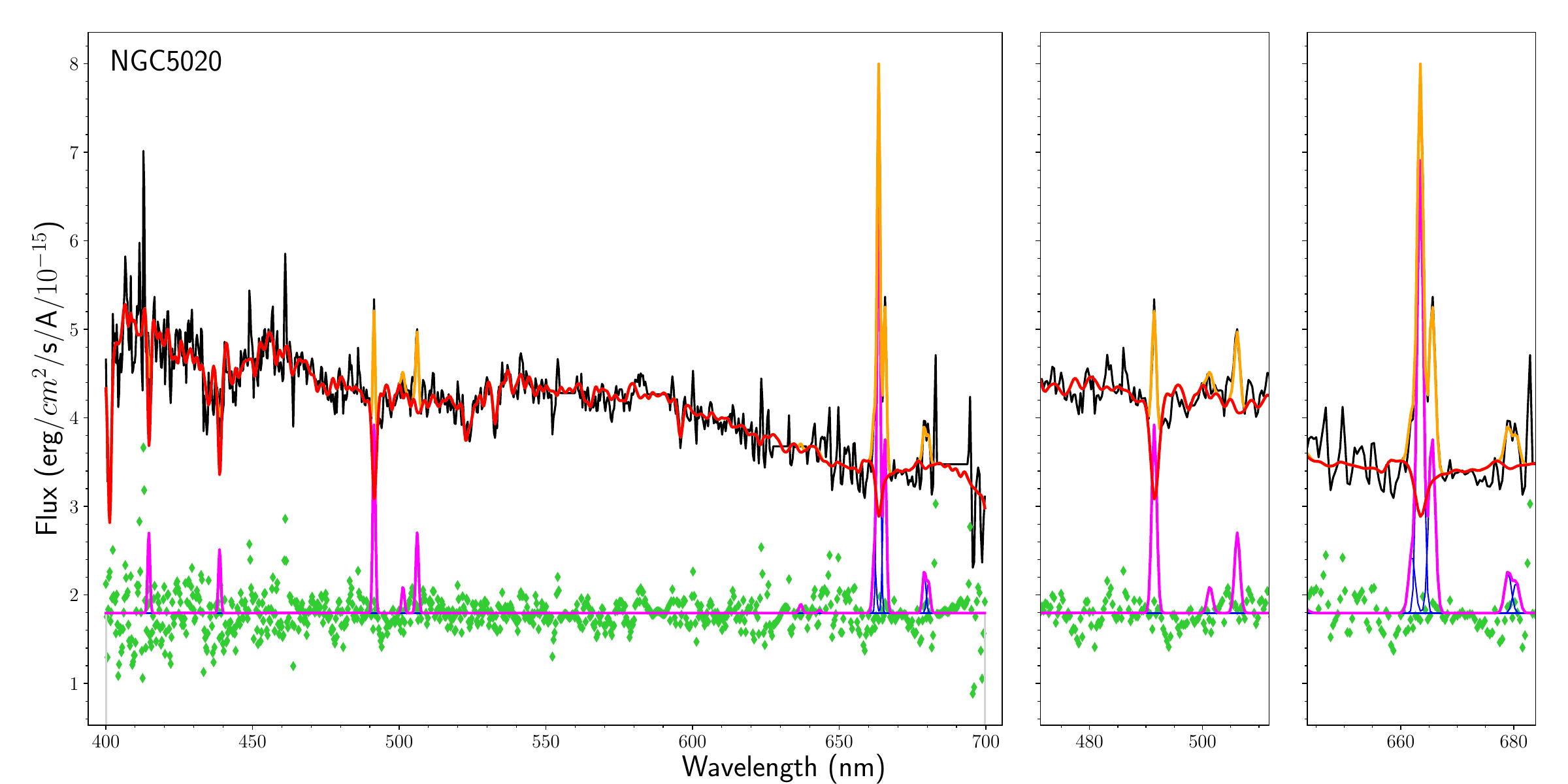}
\end{figure*}
\begin{figure*}
\includegraphics[width=7in]{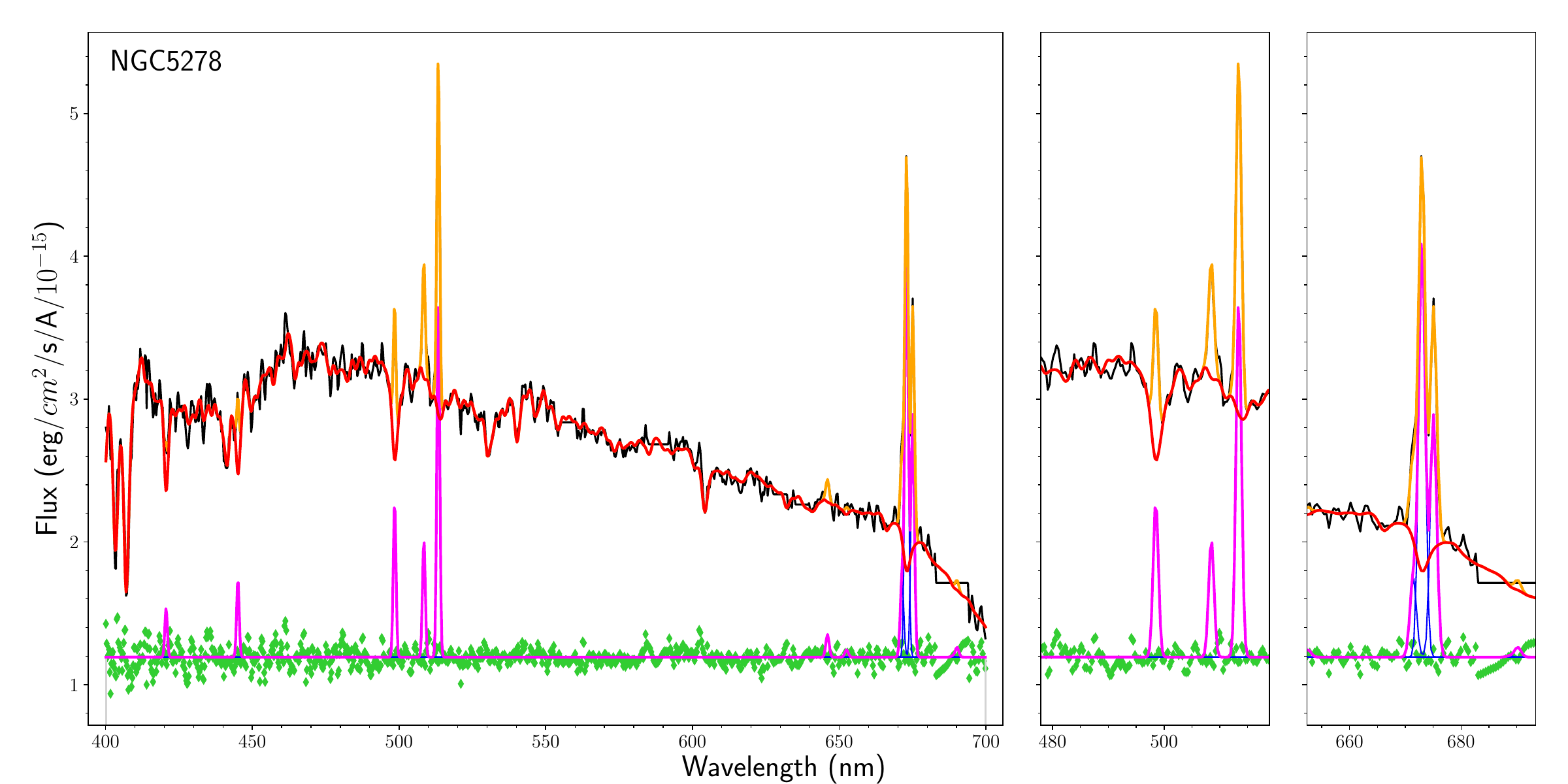}

\includegraphics[width=7in]{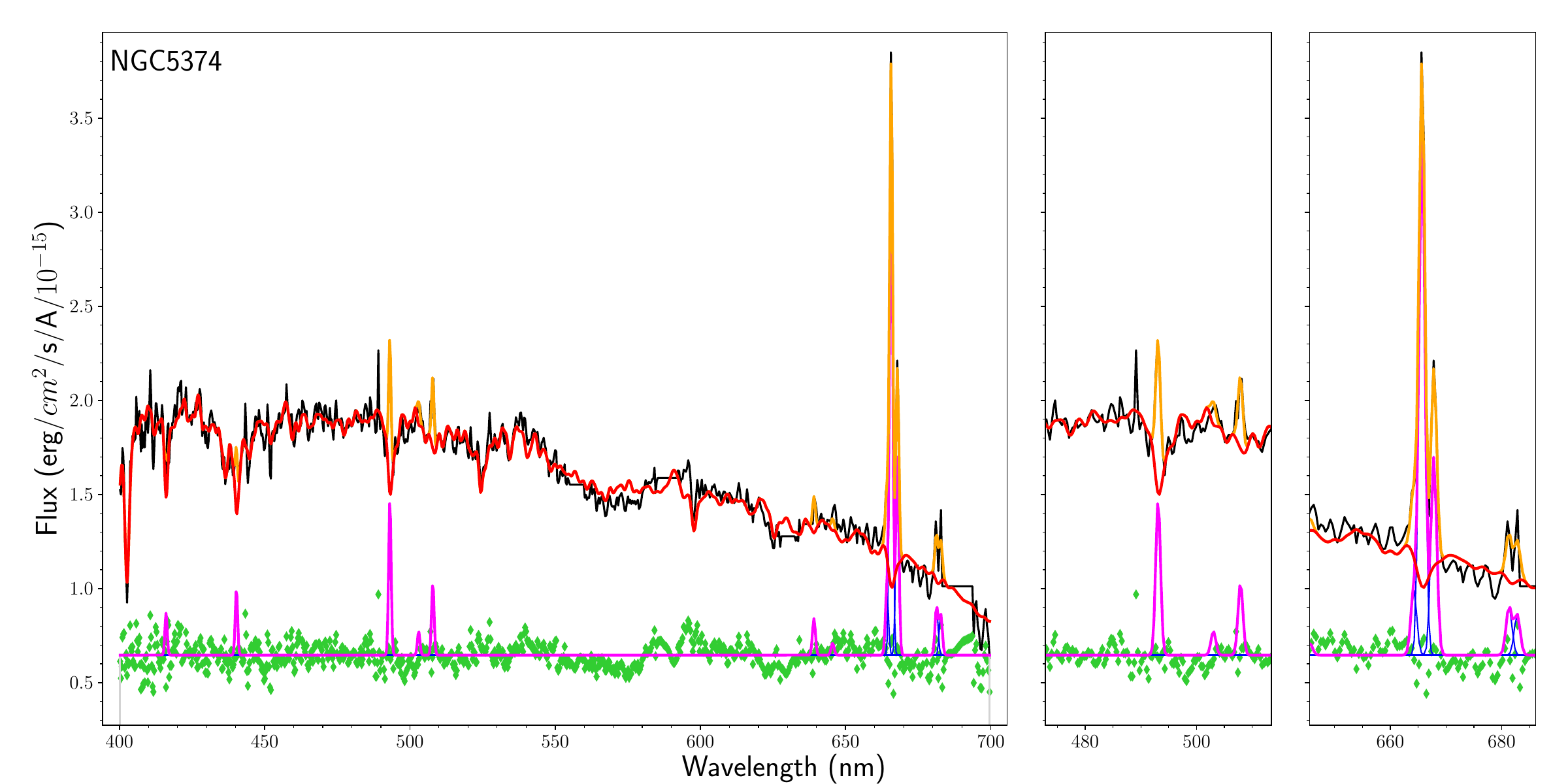}
\end{figure*}
\begin{figure*}
\includegraphics[width=7in]{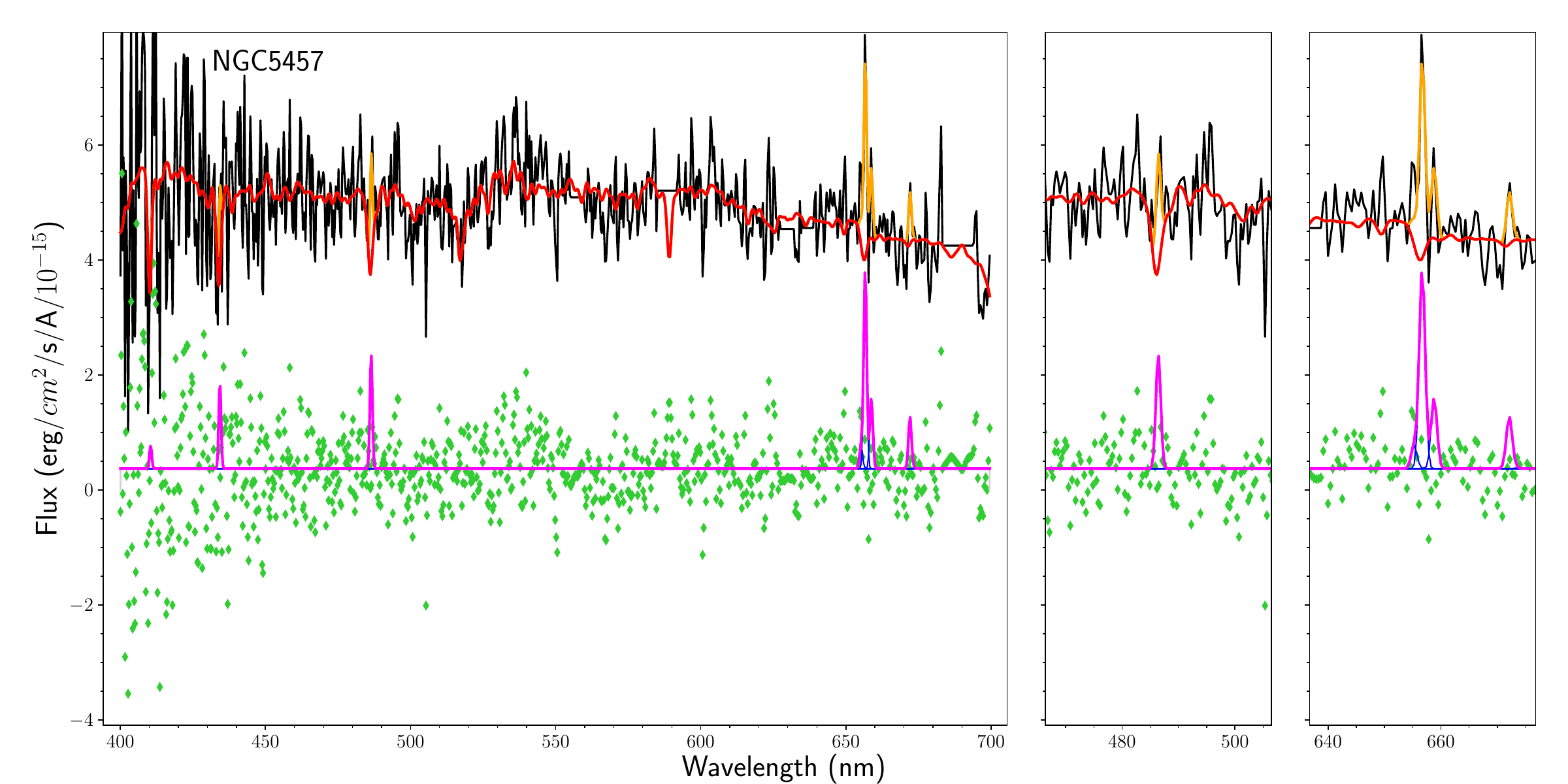}

\includegraphics[width=7in]{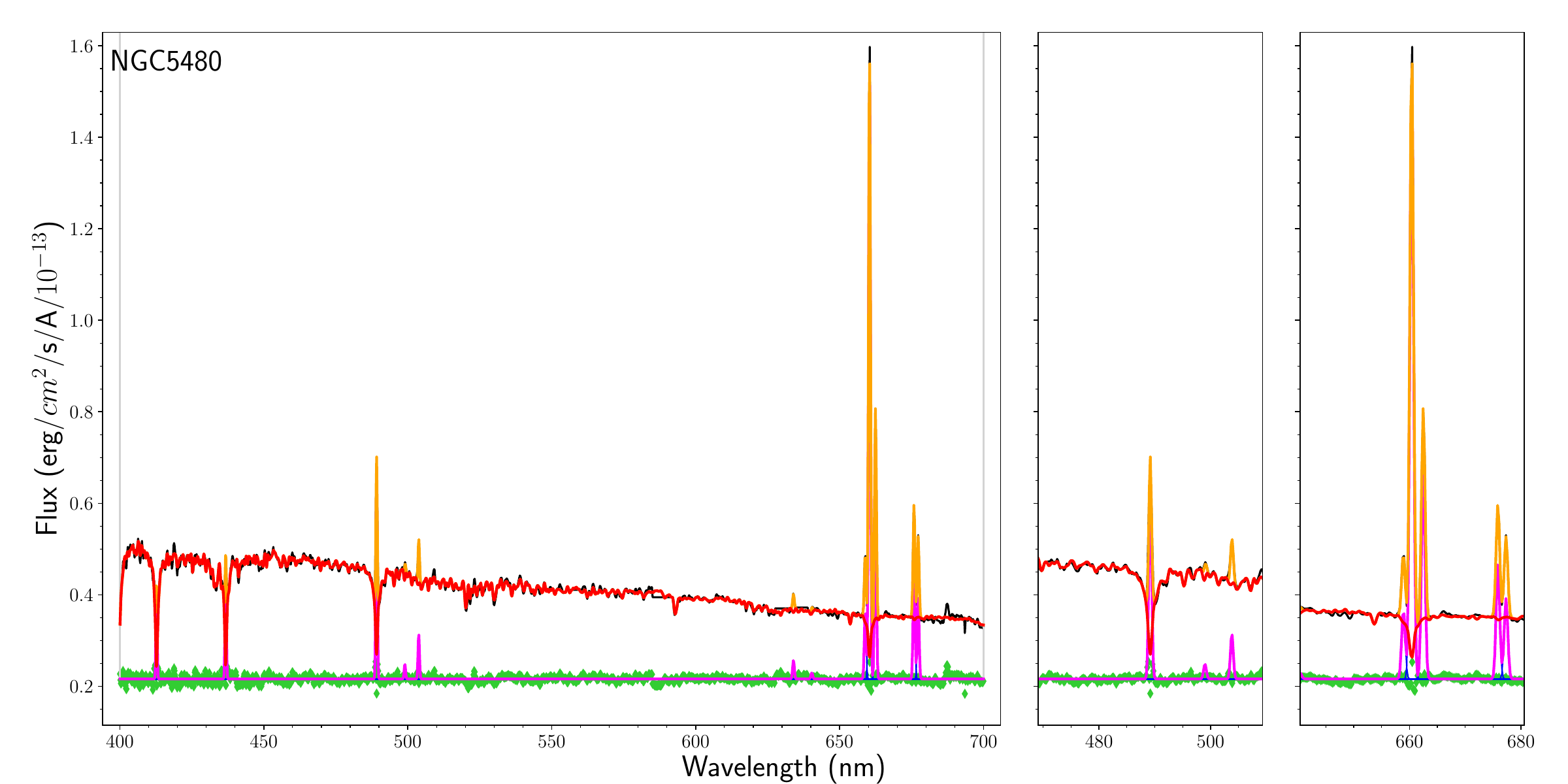}
\end{figure*}
\begin{figure*}
\includegraphics[width=7in]{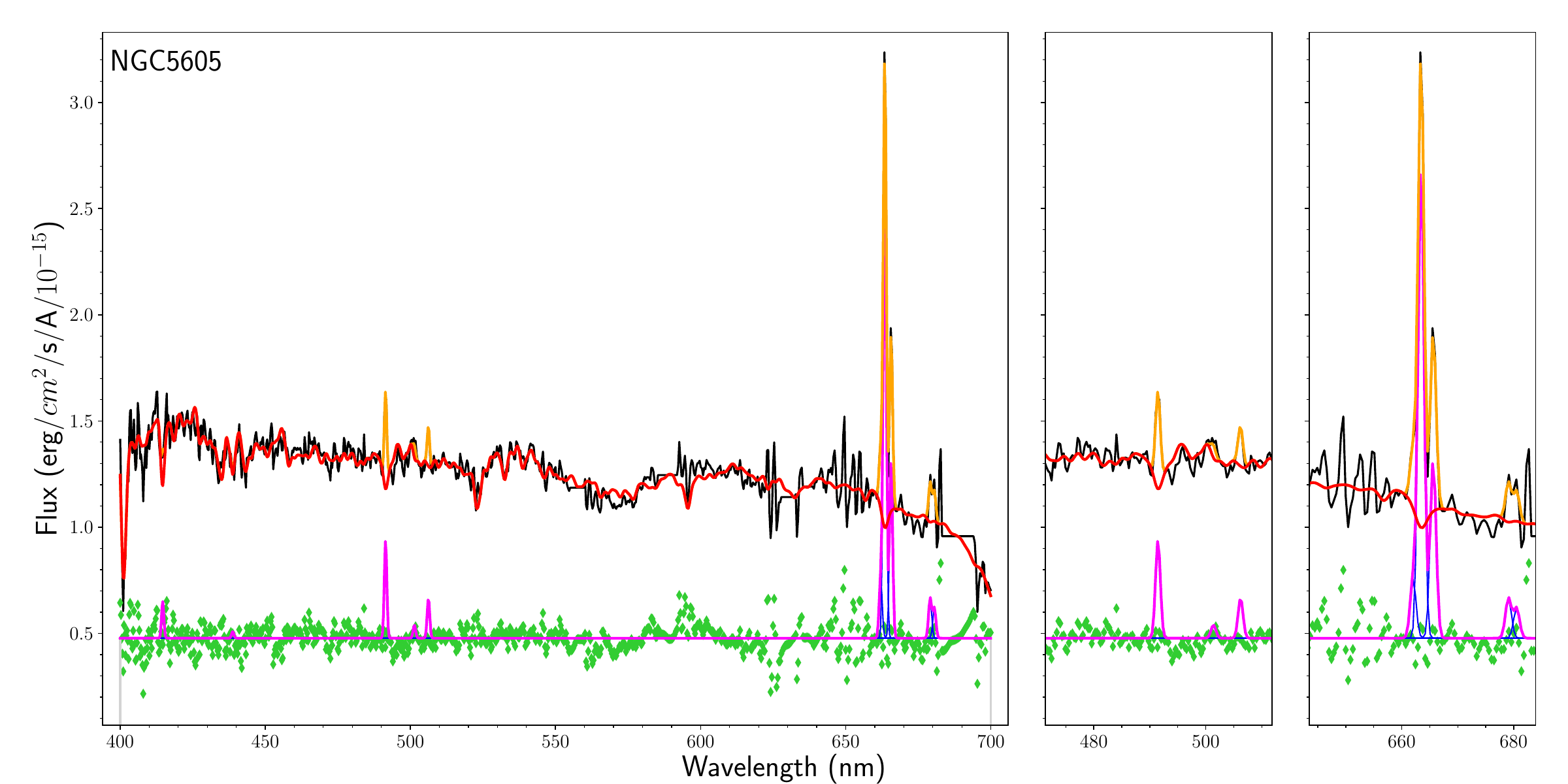}

\includegraphics[width=7in]{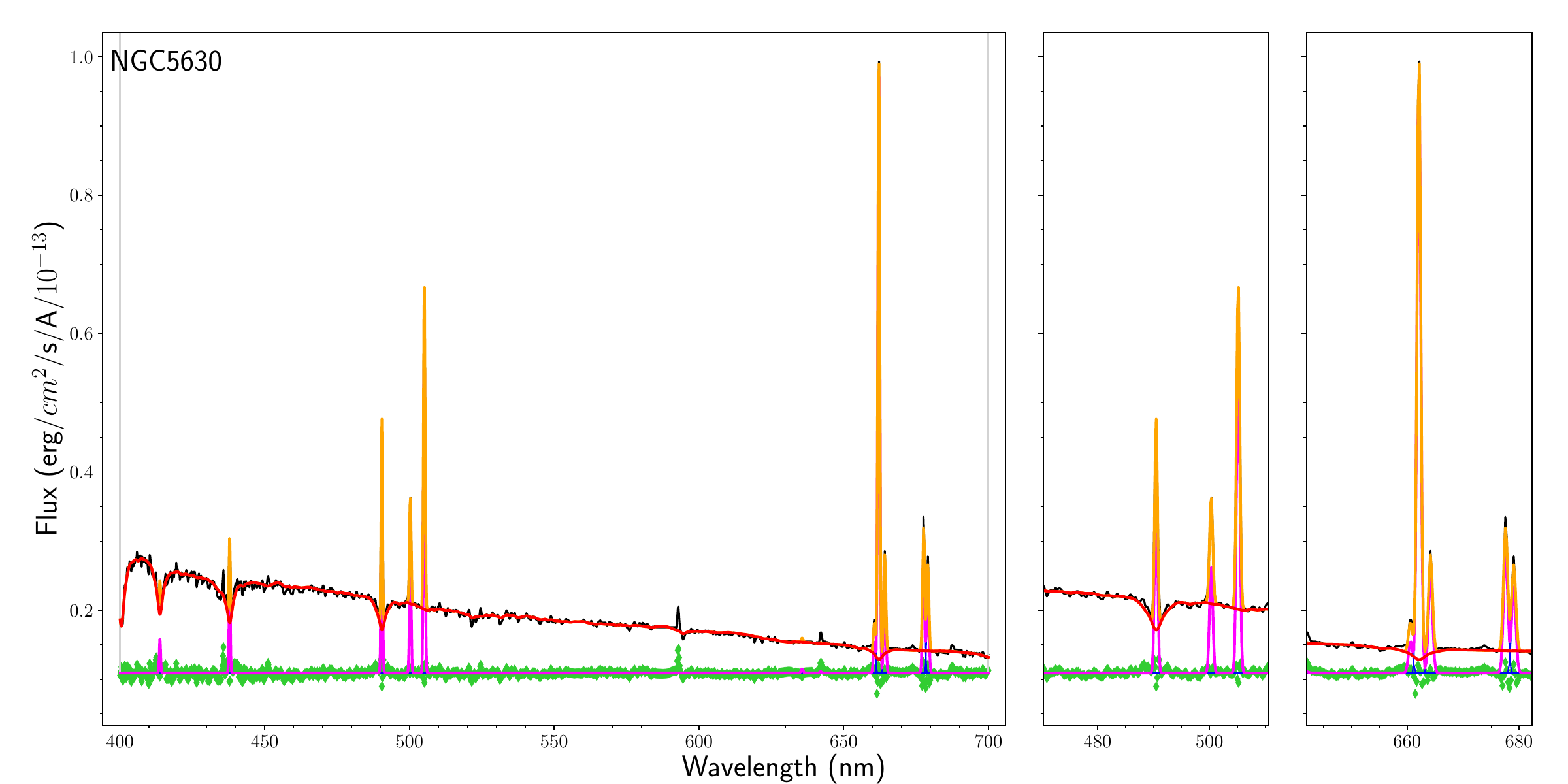}
\end{figure*}
\begin{figure*}
\includegraphics[width=7in]{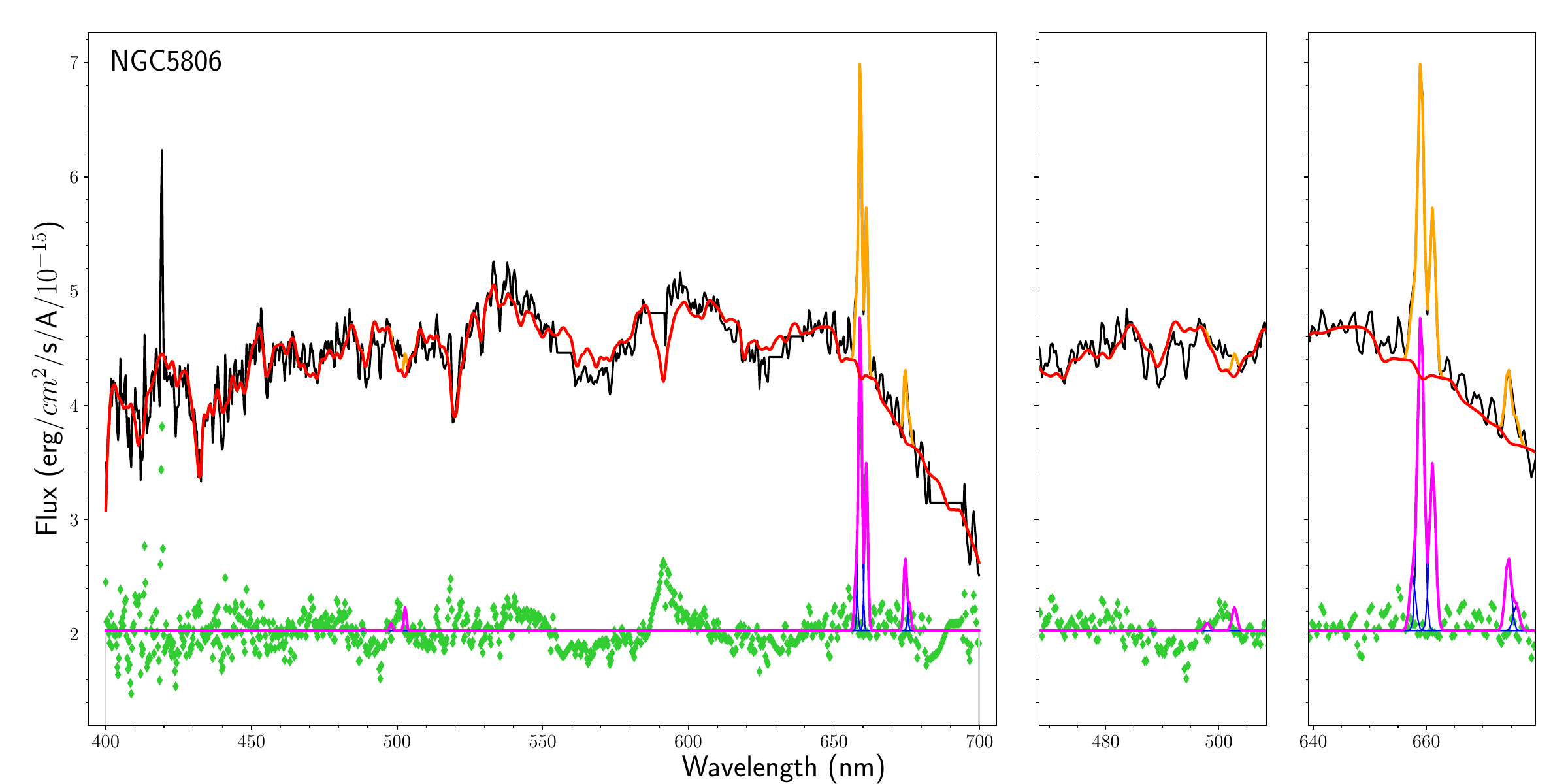}

\includegraphics[width=7in]{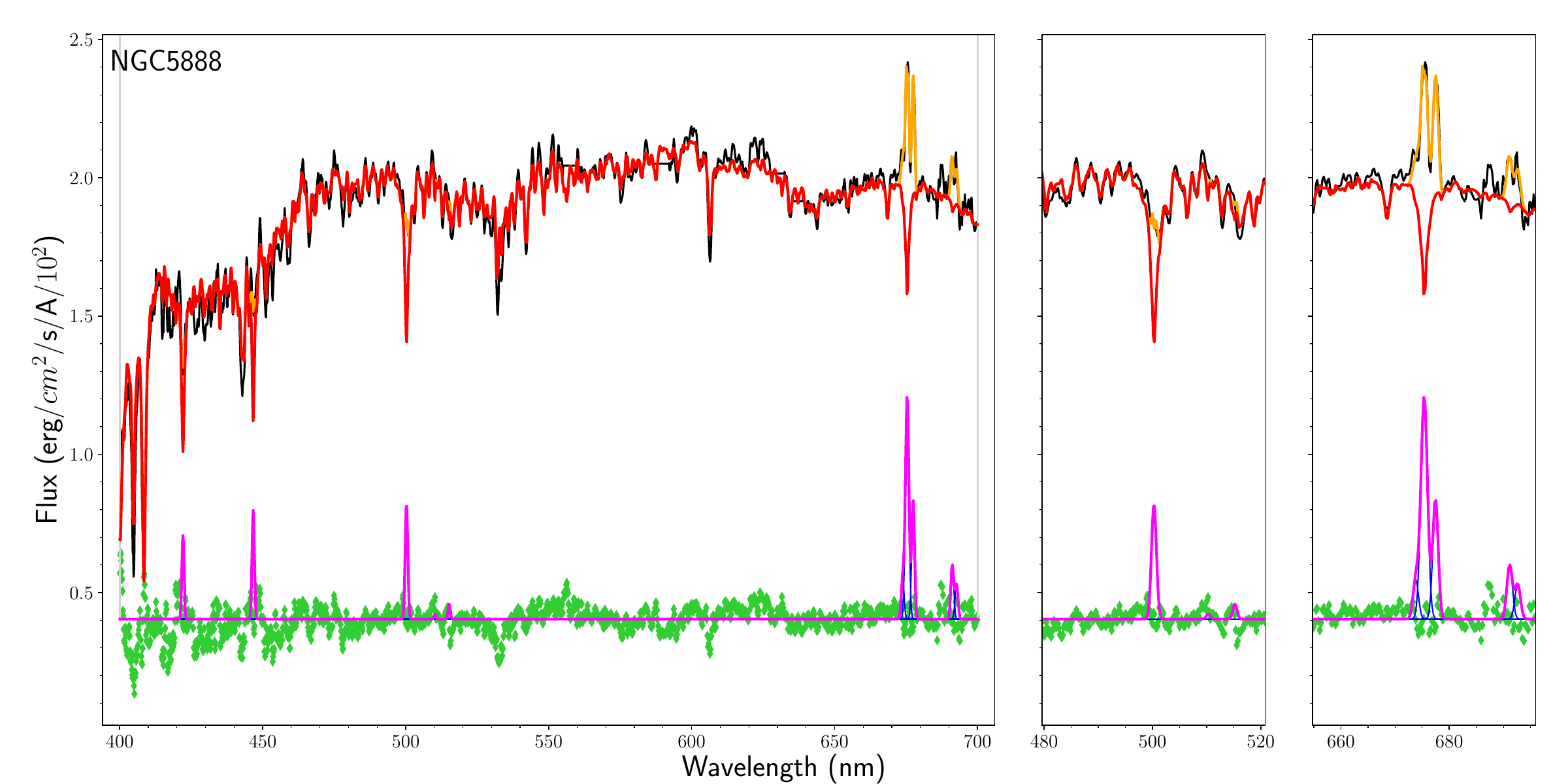}
\end{figure*}
\begin{figure*}
\includegraphics[width=7in]{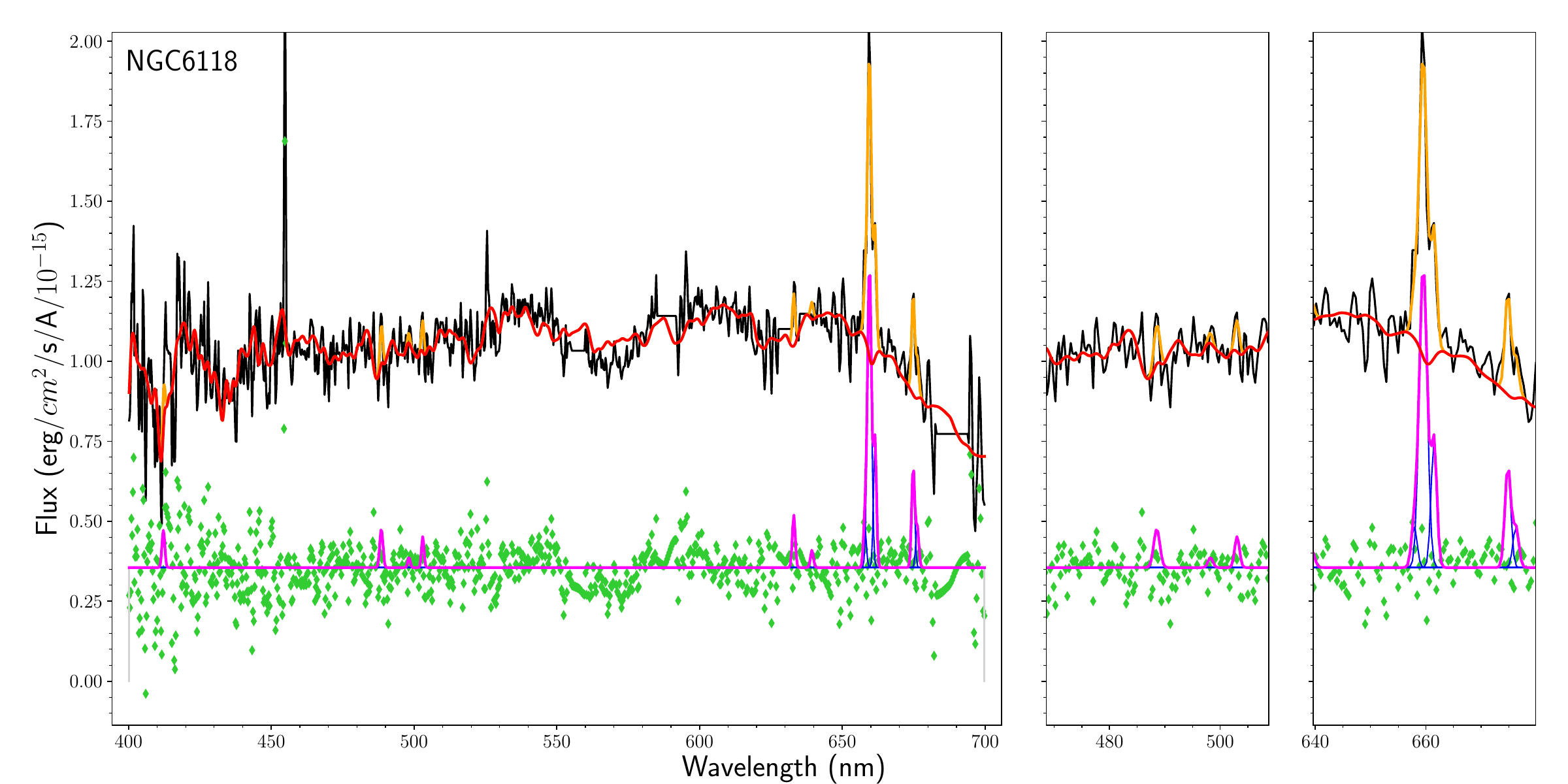}

\includegraphics[width=7in]{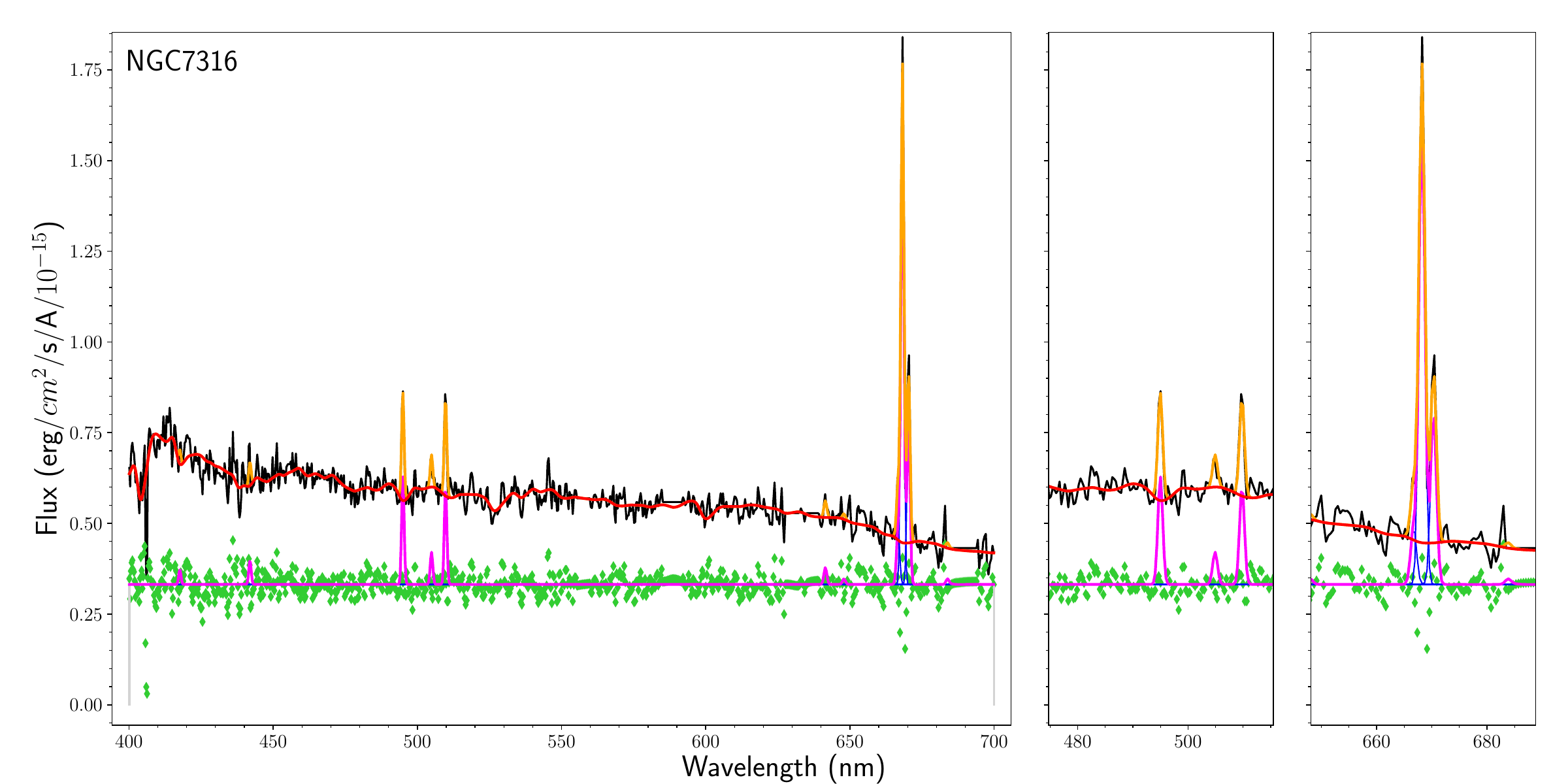}
\end{figure*}
\begin{figure*}
\includegraphics[width=7in]{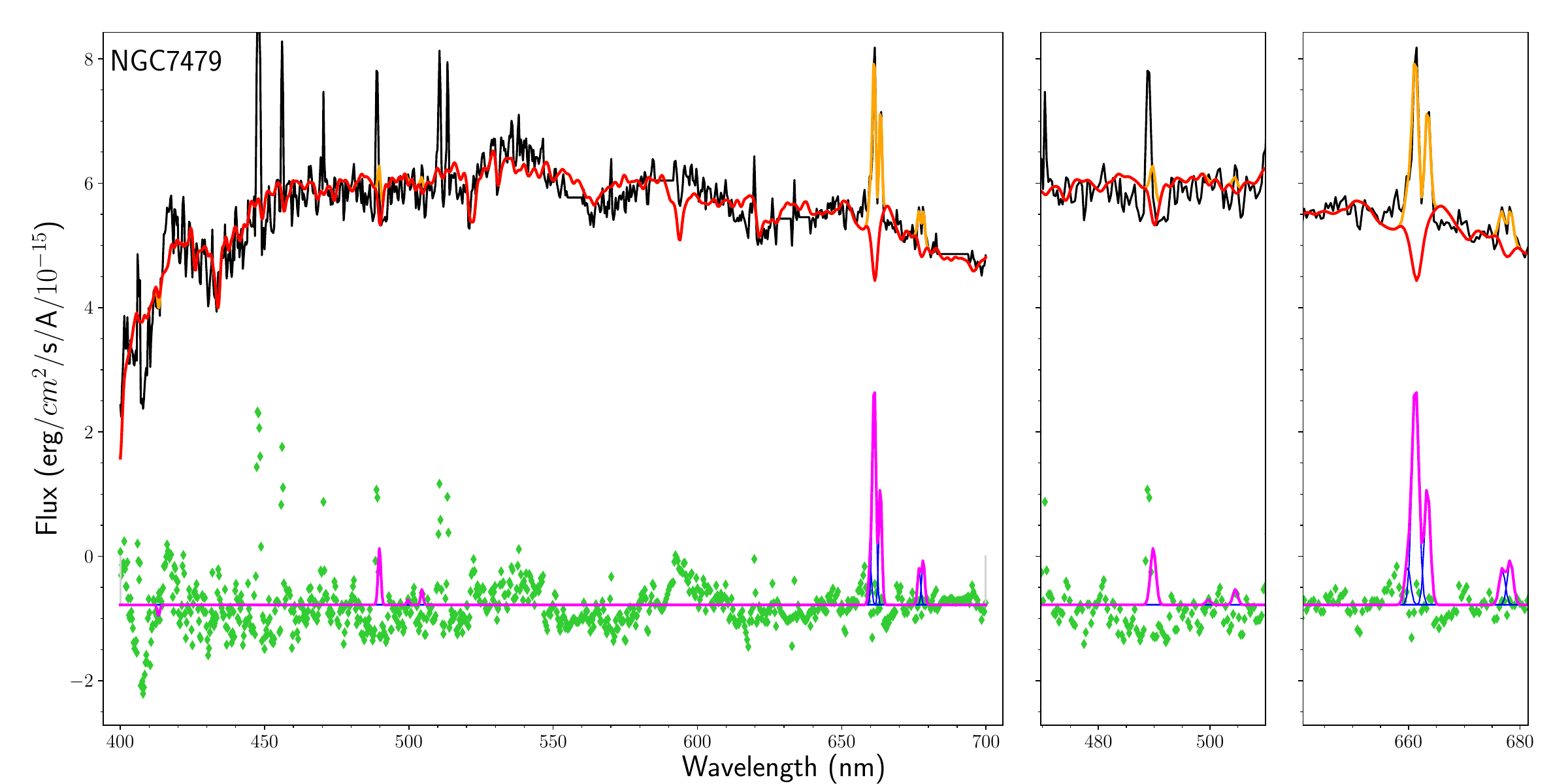}

\includegraphics[width=7in]{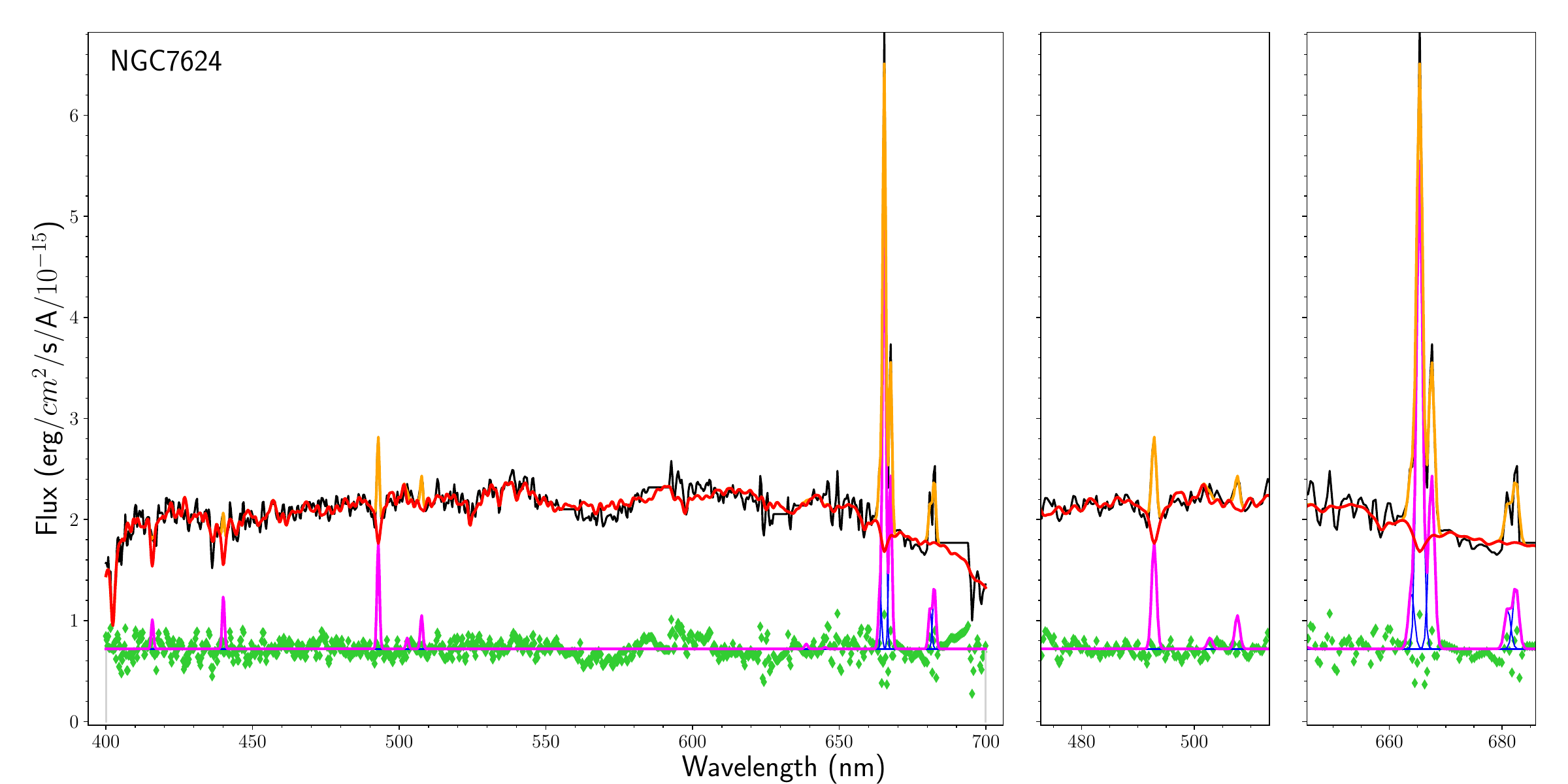}
\end{figure*}
\begin{figure*}
\includegraphics[width=7in]{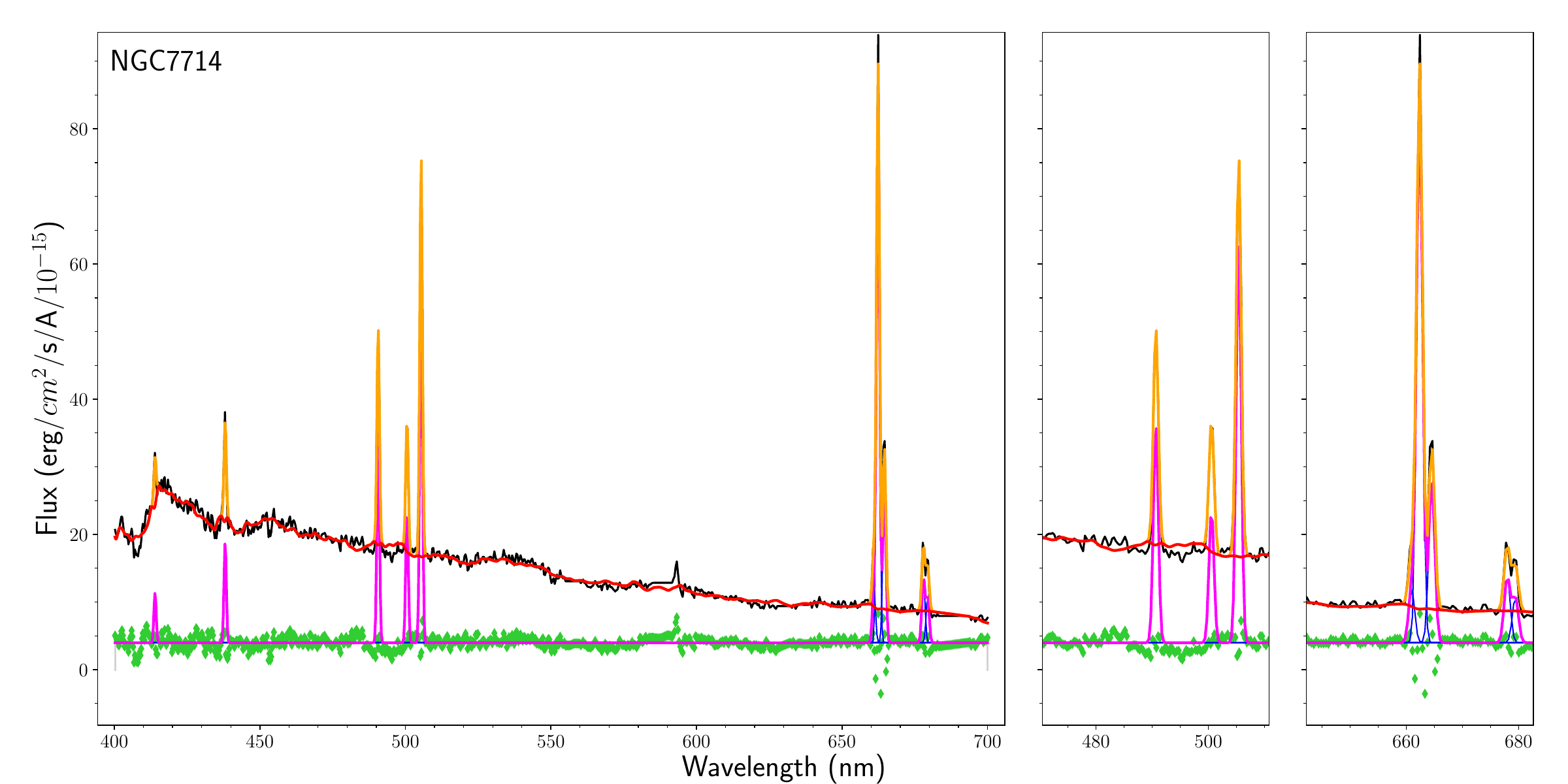}

\includegraphics[width=7in]{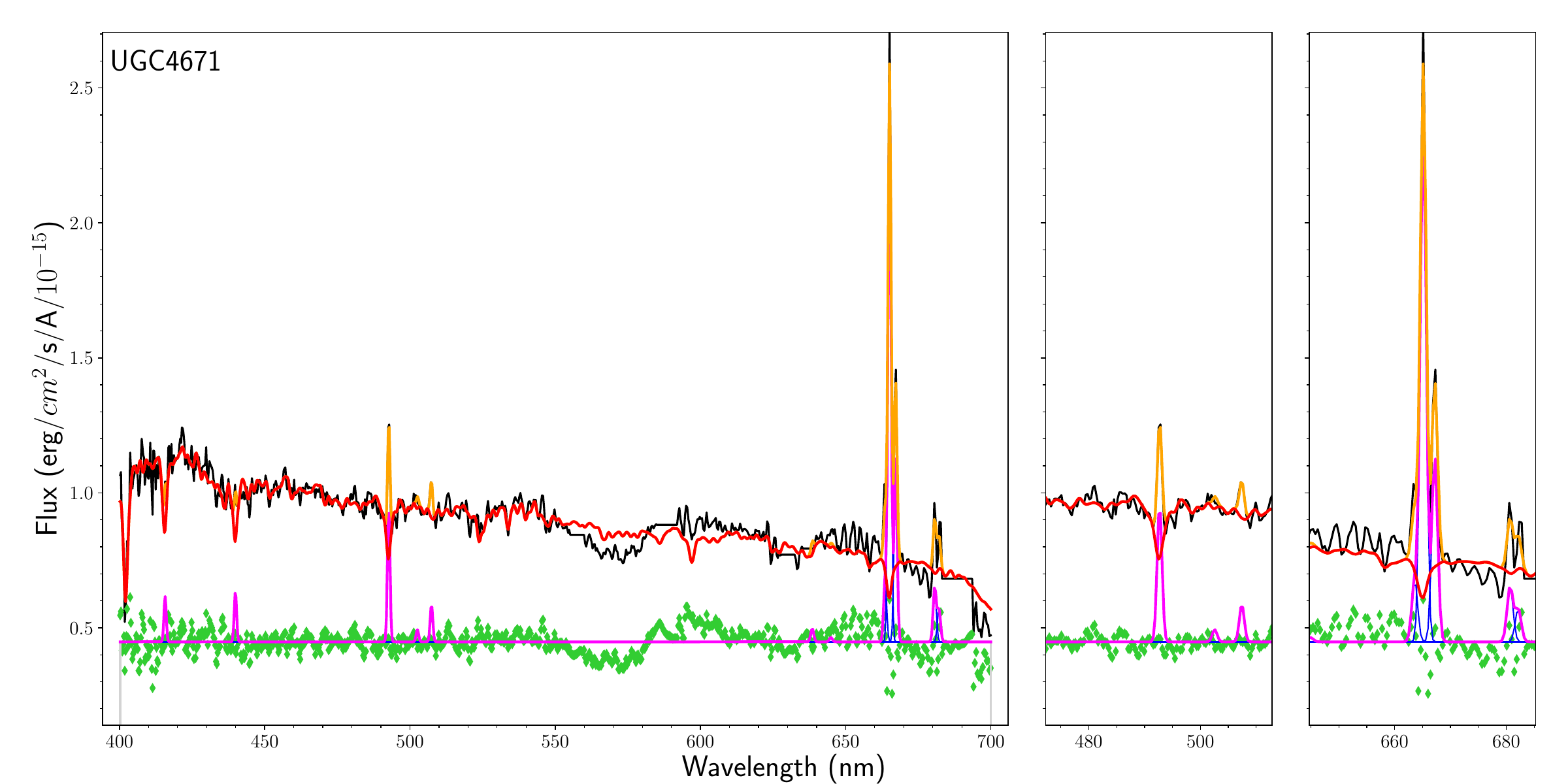}
\end{figure*}
\begin{figure*}
\includegraphics[width=7in]{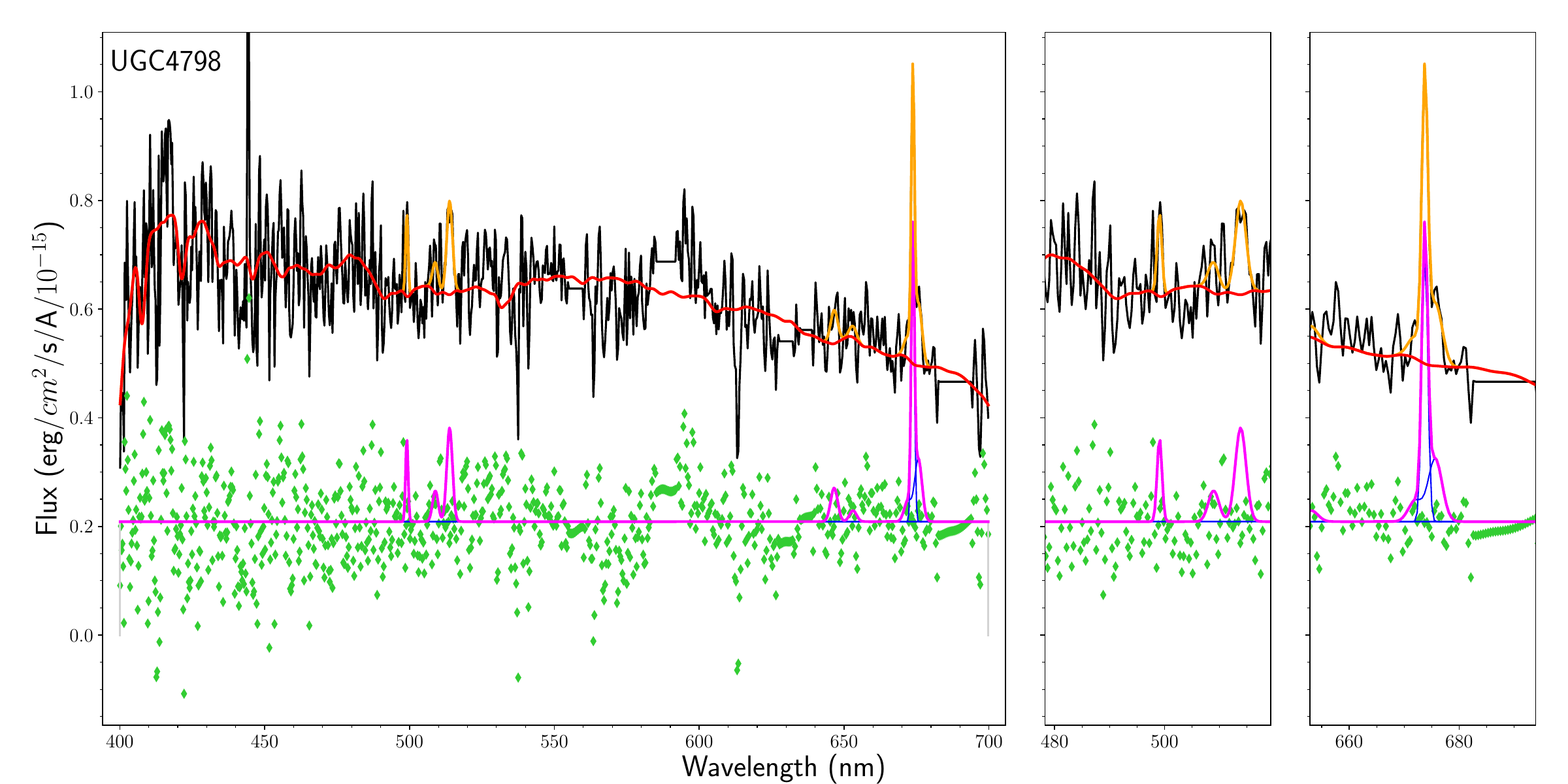}

\includegraphics[width=7in]{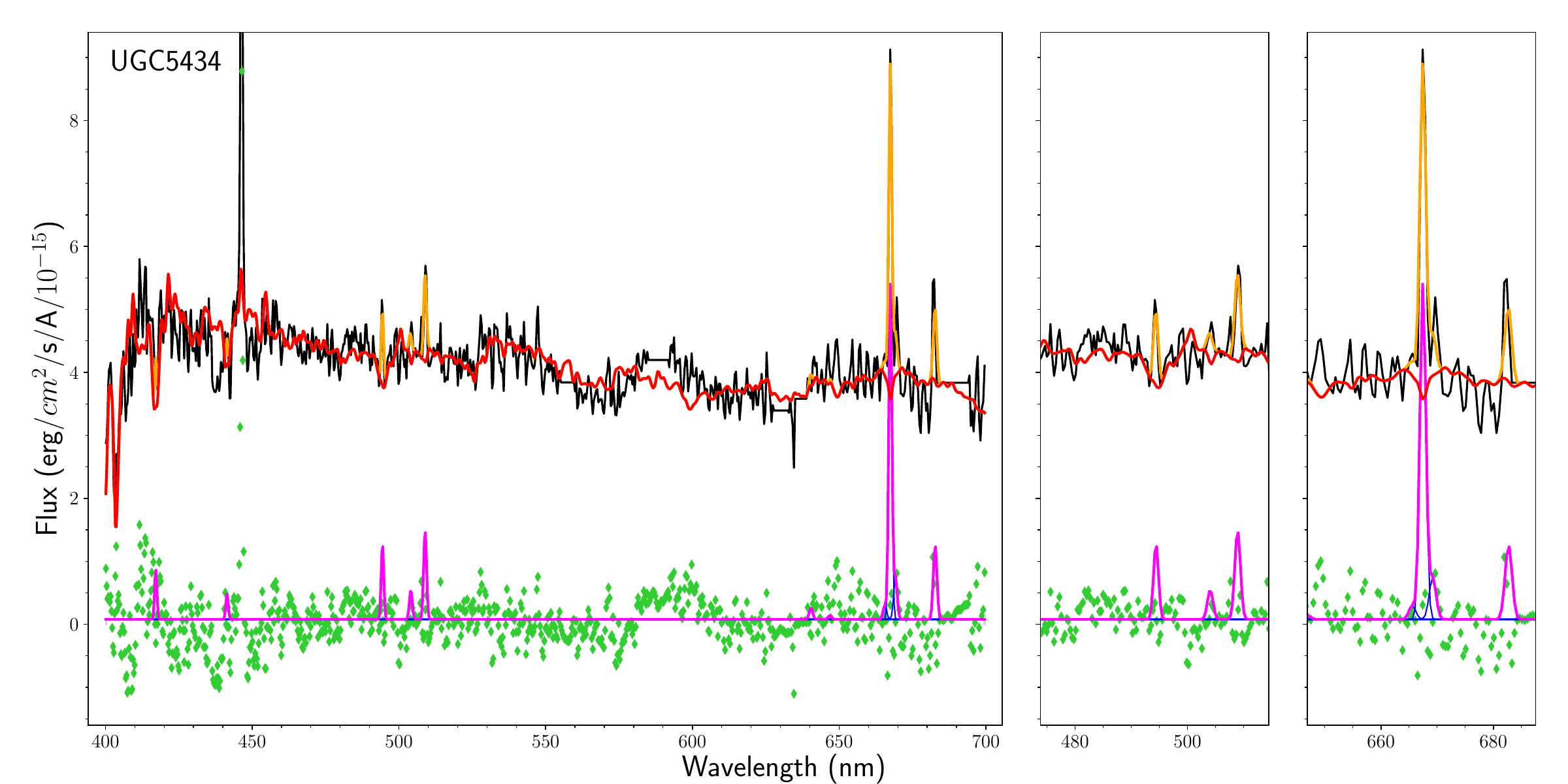}
\end{figure*}
\begin{figure*}
\includegraphics[width=7in]{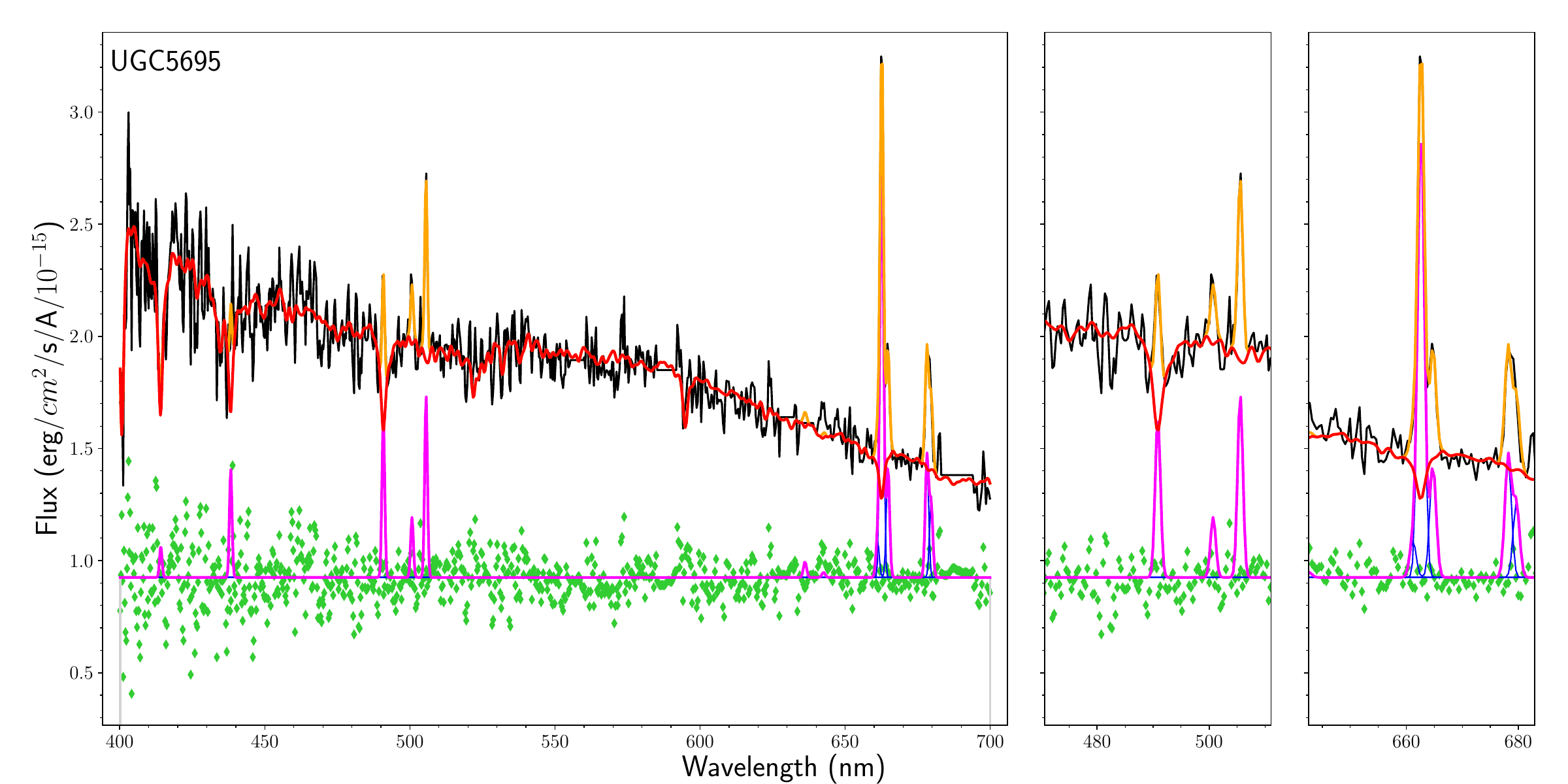}

\includegraphics[width=7in]{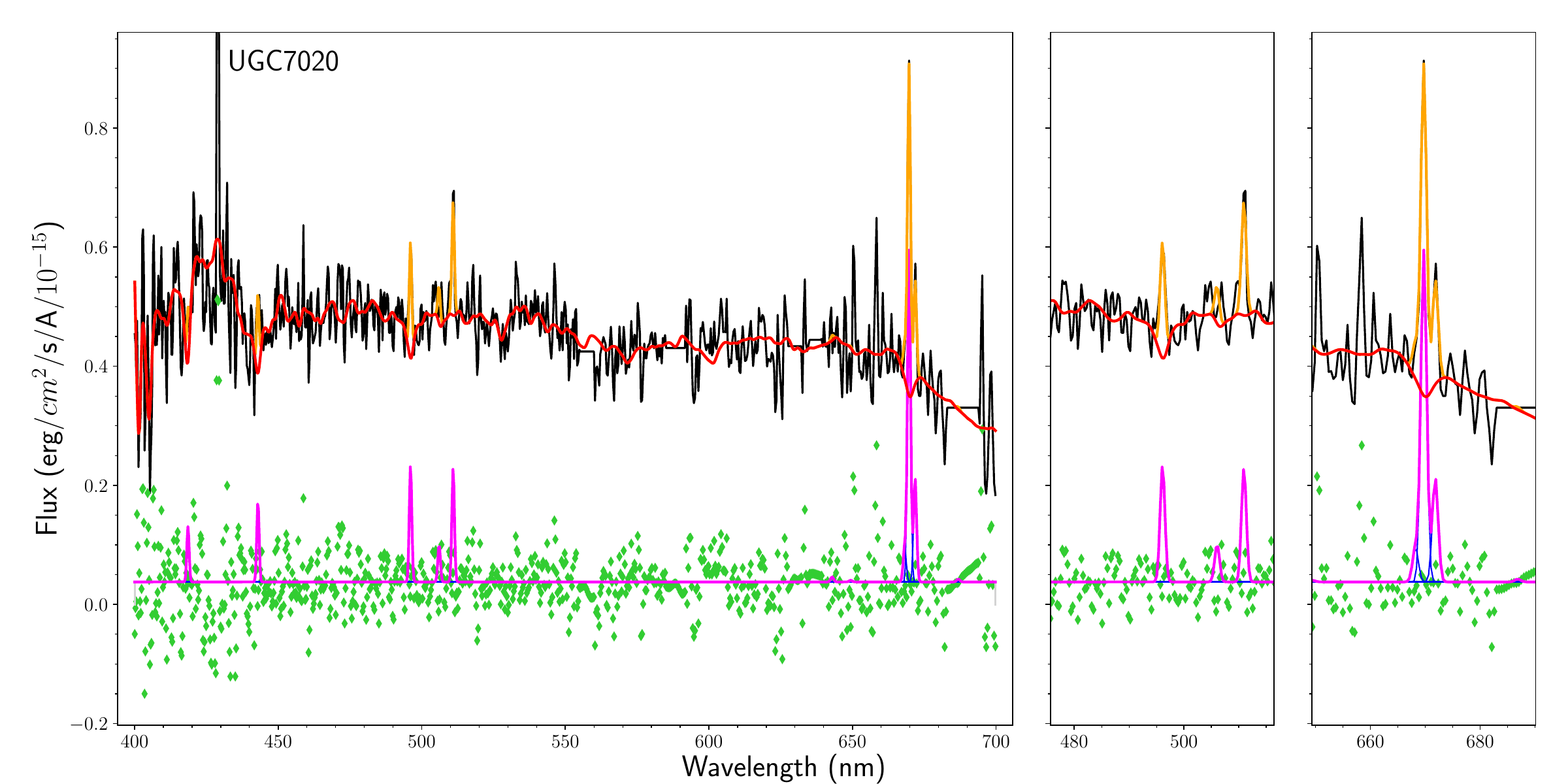}
\end{figure*}
\begin{figure*}
\includegraphics[width=7in]{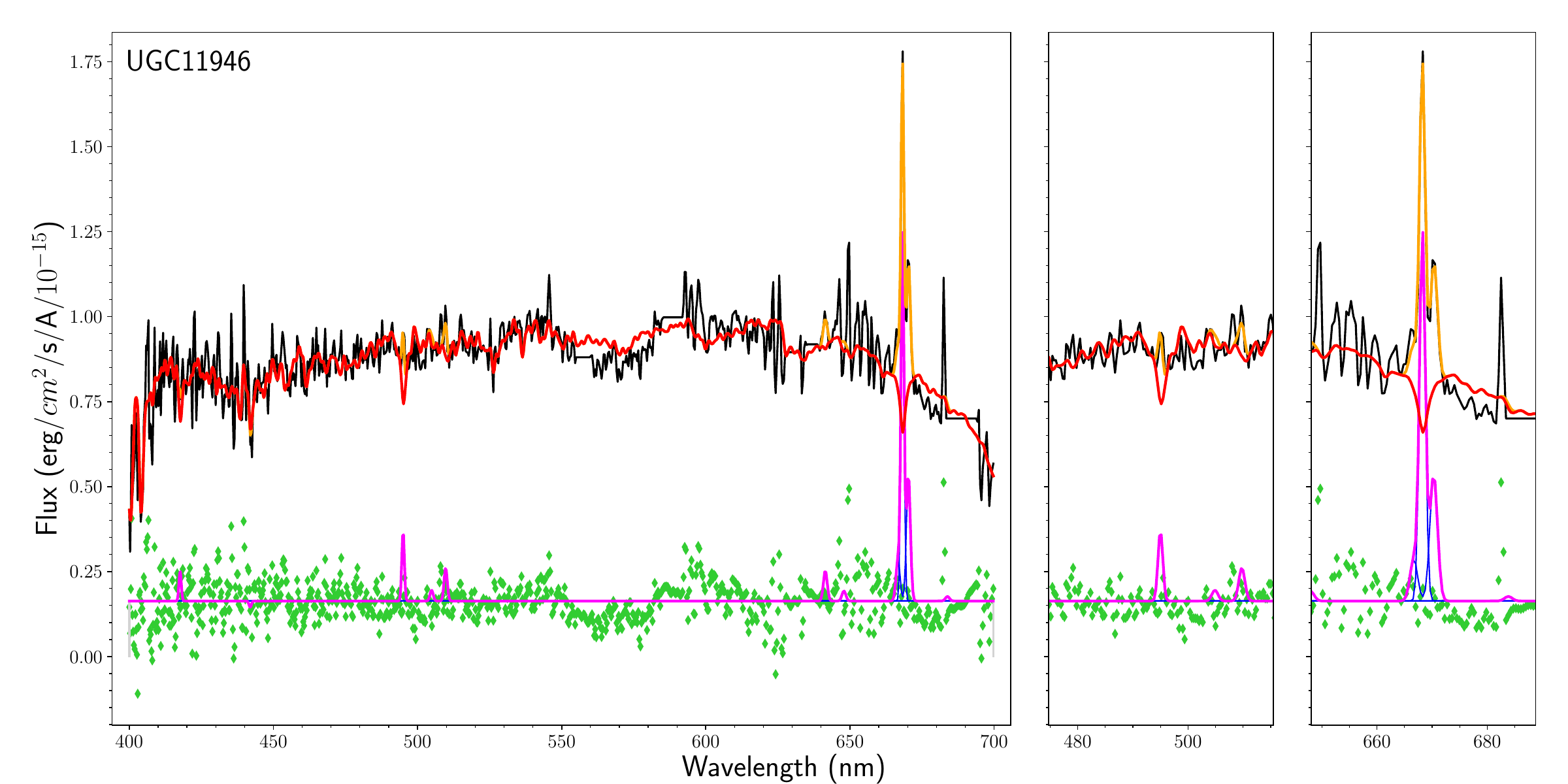}

  \caption{The above figures show the {\tt pPXF} fits for all galaxies observed, regardless of whether they were used in our analysis or not. The left plot shows the full spectrum. The middle plot show a zoom in around H$\beta$, and the right plot shows a zoom in around H$\alpha$. The black line is the flux density of the observed spectrum, the red line is the {\tt pPXF} model for the stellar component, and the orange line is the model of the nebular emission lines and the stellar continuum. The magenta line shows the nebular line model alone, and the blue line shows the individual lines that contribute to the nebular emission model. The green dots are the fit residuals. }
  \label{all_ppxf_fits}
\end{figure*}

\end{document}